\documentclass[%
 reprint,
 amsmath,amssymb,
 aps,
]{revtex4-2}
\usepackage{hyperref}
\usepackage{ragged2e}
\usepackage{graphicx}
\usepackage{dcolumn}
\usepackage{natbib} 
\setcitestyle{numeric}
\usepackage{bm}
\usepackage{tikz,graphics,color,fullpage,float,epsf,caption,subcaption}
\usepackage[font=small,labelfont=bf,
   justification=justified,
   format=plain]{caption} 
\usepackage{xcolor}
\usepackage[normalem]{ulem}
\usepackage{braket}

\makeatletter
\newcommand*{\rom}[1]{\expandafter\@slowromancap\romannumeral #1@}
\makeatother

\begin{document}
\preprint{APS/123-QED}
\title{On-chip microwave coherent source with in-situ control of the photon number distribution}
\author{P. Mastrovito$^{1,2}$}
\email{pasquale.mastrovito@unina.it}
\author{H.G. Ahmad$^{1,3}$}
\author{A. Porzio$^{4}$}
\author{M. Esposito$^{3}$}
\author{D. Massarotti$^{2,3}$}
\author{F. Tafuri$^{1,5}$}
\affiliation{%
$^1$ Dipartimento di Fisica “Ettore Pancini”, Università di Napoli “Federico II,” Monte S. Angelo, I-80126 Napoli, Italy \\
$^2$ Dipartimento di Ingegneria Elettrica e delle Tecnologie dell’Informazione, Università degli Studi di Napoli Federico II, via Claudio, I-80125 Napoli, Italy \\
$^3$ CNR-SPIN Complesso di Monte S. Angelo, 80126 Napoli, Italy \\
$^4$  Dipartimento di Ingegneria Civile e Meccanica (DICEM), Universita di Cassino e Lazio Meridionale, 03043 Cassino, Italy \\
$^5$ CNR–Istituto Nazionale di Ottica (CNR-INO), Largo Enrico Fermi 6, 50125 Florence, Italy
}
\date{\today}
\begin{abstract}
Coherent photon sources are key elements in different applications, ranging from quantum sensing to quantum computing. The possibility of designing and engineering superconducting circuits behaving like artificial atoms supports the realization of quantum optics protocols, including microwave photons generation. Here we propose and theoretically investigate a design that allows a tunable photon injection directly on-chip. Our approach enables control of the emission via an external knob while preserving, in principle, the stability and linewidth characteristics. This is achieved by replacing the conventional capacitive link between the source and the reservoir with a tunable coupler, with the advantage of avoiding direct dynamical manipulation of the photon source dynamics, providing pulsed control of the emission. We validate the dynamical control of the generated state under the effect of an external flux threading the tunable coupler and discuss the possibility of employing this scheme also in the context of multiple bosonic reservoirs.   
\end{abstract}
\maketitle
\section{\label{sec:level1}Introduction\protect}
The generation of tailored electromagnetic radiation is the foundational operation behind the control of quantum elements, such as natural or artificial atoms \cite{preskill2018quantum, saffman2019quantum,cirac1995quantum,gottesman2001encoding, gershenfeld1998quantum}. Precise control over these signals is essential for the performance of quantum computation and sensing processes \cite{houck2007generating,hofheinz2009synthesizing,clerk2010introduction,albert2016holonomic,degen2017quantum, blais2020quantum,krinner2022realizing}. In superconducting architectures, electromagnetic radiation is typically generated by room-temperature microwave electronics and delivered to the chip via dedicated transmission lines \cite{arute2019quantum,krinner2019engineering,bardin2021microwaves}. However, this approach introduces several drawbacks. Connections between the quantum components and room-temperature electronics can introduce unwanted thermal and electromagnetic noise, reducing the coherence of quantum elements \cite{houck2008controlling,clerk2010introduction}. Furthermore, such setups require numerous broadband connections with multiple stages of attenuation and filtering to mitigate noise at cryogenic stages \cite{krinner2019engineering}. As the number of connections scales linearly with the quantum elements, this results in a significant thermal load, necessitating larger refrigerators with increased cooling power \cite{reilly2019challenges}. These limitations indicate the need for alternative control strategies, particularly for large-scale quantum systems \cite{mcdermott2018quantum,DiPalma2023,joshi2023scaling}. \\
Cryogenic-integrated control electronics offer a promising alternative for addressing these challenges \cite{yan2021low,leonard2019digital,xue2021cmos}. Performing most control operations directly at the cryogenic stage eliminates the need for direct connections to room-temperature electronics, thereby improving thermal isolation and enhancing the coherence of controlled quantum elements. In this context, on-chip single-atom lasers are particularly notable \cite{meschede1985one,mu1992one,mckeever2003experimental}. These cryogenic-integrated coherent sources exhibit unique features \cite{pellizzari1994preparation,pellizzari1994photon} such as self-quenching and a pumping threshold for population inversion determined by circuit parameters rather than external pump power, unlike conventional multi-atom lasers \cite{lugiato1987connection,bjork1994definition,dubin2010quantum,astafiev2007single, cassidy2017demonstration}. Besides being interesting elements for the observation of different quantum optics phenomena, \cite{schawlow1958infrared,lugiato1987connection,pellizzari1994preparation,pellizzari1994photon,bjork1994definition,astafiev2007single,bartalini2010observing,dubin2010quantum}, these sources also present exceptional performance in terms of phase noise showing linewidth close to the standard quantum limit \cite{schawlow1958infrared, wiseman1999light,bartalini2010observing,liu2015semiconductor,cassidy2017demonstration,yan2021low,Liu_2021}. These features make them particularly well-suited as reference signals for microwave operations at cryogenic temperatures \cite{ball2016role} and for driving quantum systems such as qubits and oscillators in quantum computing applications \cite{simon2018theory,yan2021low}. Furthermore, their narrow linewidth enables high high-frequency resolution, making these sources highly desirable for applications in spectroscopy, communication, and sensing \cite{yuen1978optical,ma2017proposal,jerger2019situ,danilin2021quantum}.\\
For practical implementation, precise control over the power and frequency of emitted radiation is crucial. In most experimental realizations, the photon emitter is embedded in a reservoir, and its output power is controlled by direct manipulation of the population inversion dynamics \cite{astafiev2007single,chen2014realization,cassidy2017demonstration,yan2021low}. The emission frequency is typically tuned by modulating the energy associated with the transition characterized by population inversion \cite{cassidy2017demonstration}, while output power is regulated by varying the pumping rate that drives the population inversion dynamics \cite{astafiev2007single,yan2018tunable}.\\
An illustrative example is single-atom maser based on Josephson junctions \cite{astafiev2007single,cassidy2017demonstration,yan2021low}. Here, power tunability is achieved via an external DC voltage that determines the rate of Cooper pair breaking, which is the mechanism responsible for population inversion \cite{astafiev2007single}. However, emission control through this approach can impact the spectral features of the emissions \cite{yan2021low}. Fast and dynamical control of the output power is required to implement these cryogenic sources for the control of quantum elements, like qubits or cavities \cite{simon2018theory}. In this scenario, the direct control on the source and the finite and sharp time shape of control pulses directly for manipulating the inversion population dynamics can induce a broadening of the emissions, resulting in a degradation of the spectral stability and the control fidelity \cite{cassidy2017demonstration}.\\
In this work, we address these limitations by proposing a versatile scheme based on the introduction of a tunable coupler \cite{bialczak2011fast,yan2018tunable,heunisch2023tunable} between the photon source and its output reservoir. By incorporating a tunable coupler between the source and the reservoir, our scheme allows emission control without affecting their spectral content and the source stability. Within this scheme, the emissions are controlled by modulating the coupling strength, which enables power control via an external knob, without directly interfering with the photon source.  Such decoupled control allows to maintain and preserve emission stability and linewidth of these coherent sources, which are critical figures of merit for the successful implementation of these devices for practical applications in different quantum platforms \cite{gottesman2001encoding,gottesman2003secure,menicucci2014fault,marshall2016continuous,mcdermott2018quantum,pauka2019cryogenic,cai2021bosonic}. Furthermore, assuming the photon source itself is frequency-tunable \cite{cassidy2017demonstration}, the fan-out can be extended by inserting different elements resonating at different frequencies at the output, creating a practical microwave demultiplexer. Such a design eliminates the need for direct connections between the quantum elements on-chip and room-temperature electronics, enhancing thermal isolation and scalability for next-generation quantum devices.

\section{Results and discussions}

\subsection{Concept and implementation}

\begin{figure}[t]
\centering
\includegraphics[width=\columnwidth]{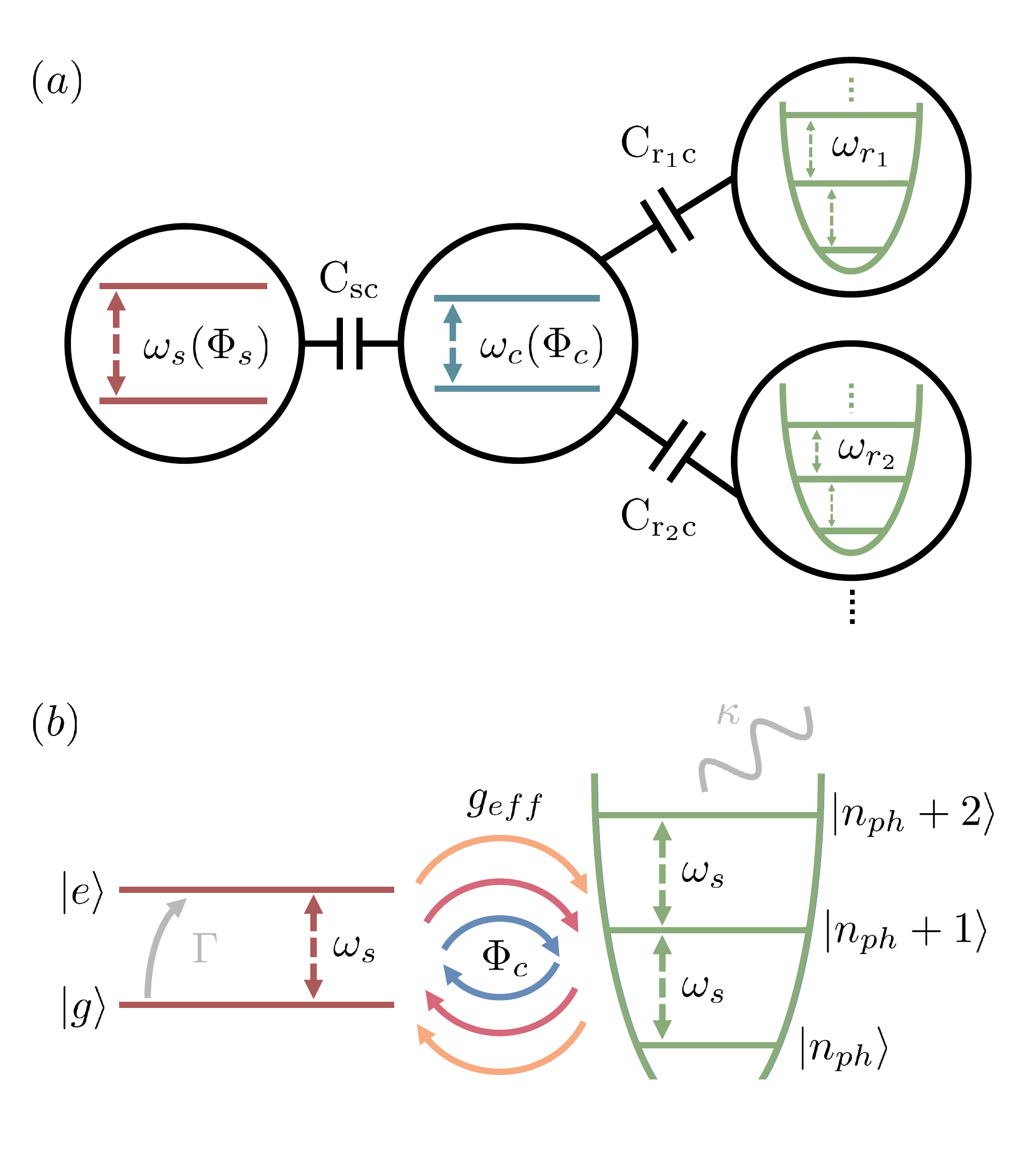}
  \caption{ \justifying \textbf{Concept and implementation sketches.} (a) Sketch of the general setup with the corresponding spectral characteristics of each element. The first element from the left is the photon source. We ignore the higher energy levels required for population inversion and consider only the tunable transition energy $\mathrm{\hbar \omega_s(\Phi_s)}$ that is set on resonance with the target reservoir. The middle element is the tunable coupler with energy $\mathrm{\hbar\omega_c(\Phi_c)}$, capacitively linked to various reservoirs with capacitors $\mathrm{C_{r_ic}}$ $\mathrm{i=1...N_r}$ and to the photon source with a capacitive coupling $C_{sc}$. (b) Description of the dynamics underlying photon injection. A counter-relaxation process represents population inversion in the system at a rate $\mathrm{\Gamma}$ from the ground state $\mathrm{\ket{g}}$ to the excited state $\mathrm{\ket{e}}$ of the system. The photon source is resonant with the reservoir to populate it with several photons depending on the external flux $\mathrm{\Phi_c}$. The tunable effective coupling between the source and the reservoir is labeled $g_{eff}$.} 
\label{fig:setup}
\end{figure}

A schematic of the setup for the implementation of the proposed system is shown in Fig. \ref{fig:setup}(a). The system consists of $\mathrm{N_r + 2}$ elements, where $\mathrm{N_r}$ corresponds to the number of elements on the output, here represented as harmonic reservoirs. 
The latter have resonance frequencies equal to $\mathrm{\omega_{r_i}}$, where $\mathrm{i}$ ranges from 1 to $\mathrm{N_r}$ and $\mathrm{\omega_{r_i} \neq \omega_{r_j}}$ for any $\mathrm{i \neq j}$. The remaining two elements correspond to the tunable photon source and the tunable coupler. These two elements are considered as two-level systems with frequencies $\mathrm{\omega_s}(\Phi_s)$ and $\mathrm{\omega_c}(\Phi_c)$, respectively. Differently from the tunable coupler, the photon source possesses an inner counter-relaxation rate $\Gamma$, indicated in Fig. \ref{fig:setup}(b), representative of the population inversion dynamics from which the spontaneous emission from the source originates. By adequately engineering the system, the tunable coupler can be dispersively coupled to the other elements, recreating an analog system of $\mathrm{N_r+1}$ elements, similar to its implementation in quantum computation for the realization of two-qubit gates \cite{yan2018tunable, sung2021tunablecoupler}. In this scenario, the effect of the coupler is to renormalize the characteristic frequencies and coupling rates of the system by an amount that depends on its frequency $\mathrm{\omega_c}(\Phi_c)$. The ability to modulate the frequency of the coupler defines a dynamical knob that gives control over the interaction between the photon source and a specific target reservoir, as depicted in Fig. \ref{fig:setup}(b).\\
The presented scheme can be easily implemented with typical elements of cQED. The tunable coupler corresponds to a flux tunable superconducting qubit with a frequency adequately far from the other elements to fall under the dispersive regime. The photon source can be any generic system capable of initiating population inversion between two energy levels in its spectra, showing the universality of the proposed scheme to any photon source based on masing.  
In the following section, we review the physics underlying the masing regime and explore feasible approaches to achieve population inversion using elements of cQED. Subsequently, we discuss the methodology to effectively engineer the proposed scheme in the case of a single target resonator. We outline the key ingredients to build an experimentally
 feasible system that enables precise control over the photon population in the resonator via an external flux controlling the frequency of the tunable coupler. In this framework, we discuss different strategies to prevent detrimental effects coming from flux noise and faulty fabrication processes. We asses all these properties through a study of the normalized second-order correlation function denoted as $\mathrm{g^{(2)}(0)}$, which we use as an indicator of the emission performance of the photon source.
We conclude the analysis of this scheme by simulating its behavior and discussing the extension to a scheme with multiple harmonic reservoirs, focusing on the capability to control each reservoir while avoiding cross-population individually.

\subsection{Masing regime}

Masing is a physical phenomenon that governs the dynamics of masers and occurs when emissions prevail over absorption processes. Traditional masers typically involve multiple atoms and operate within a regime characterized by weak coupling \cite{fowles1989introduction}. In the realm of cQED, the strong coupling achievable between an artificial Josephson atom and photons \cite{wallraff2004strong} enables the observation of the masing regime even at the level of a single artificial atom. There have been multiple demonstrations of single atom masing in cQED, employing a variety of techniques \cite{astafiev2007single,chen2014realization,cassidy2017demonstration,Oelsner2018,thomas2020thermally,yan2021low}. All these techniques leverage on the atom-like behavior of superconducting circuits and the possibility to engineer their spectra. The first experimental realization of masing in cQED involved a voltage-biased charge qubit \cite{astafiev2007single}. In this setup, population inversion was achieved by connecting a drain electrode to one arm of the tunnel junction. A sufficiently high voltage bias breaks Cooper pairs, triggering sequential single-electron tunneling events. Under proper charge offset, these events will trigger a population inversion mechanism that can be represented as a counter-relaxation rate $\Gamma$ between the first two energy levels. The counter-relaxation rate depends on the circuit parameters of the qubit which can be engineered according to the required output power from the source. Other experimental validations are based on the same phenomena but employ different circuital designs \cite{chen2014realization,cassidy2017demonstration, yan2021low}. A second approach consists in employing a transmon qubit coupled to an auxiliary low-Q cavity \cite{sokolova2021single}. In this scheme, population inversion is initiated between the first two energy levels of the transmon through a two-photon excitation that drives the qubit to the second excited state. Coupling the low-Q cavity to the second transition $\ket{f} \leftrightarrow \ket{e}$ increases its decay rate allowing it to significantly exceed the decay rate of the $\ket{e} \rightarrow \ket{g}$ transition. This creates a population inversion between the first two energy levels with a counter-relaxation rate $\Gamma$ that depends on the quality factor of the cavity and the coupling strength with the transmon qubit. Other approaches are also based on the coupling between the qubit and resonator but rather achieve population inversion through the engineering of the transition rate between dressed states \cite{jaewoo2010electroinduc,marthaler2011lasing,oelsner2013dressed}.\\
In all mentioned methods, the photon source can always be modeled as a two-level system with an effective counter-relaxation rate $\Gamma$, as depicted in Fig. \ref{fig:setup}(b). The dynamics of a single-atom maser can thus be characterized by studying the emission of a two-level system with a counter-relaxation rate $\Gamma$ capacitively coupled to a resonant harmonic reservoir with a loss rate $\kappa$. The full system dynamics is well described by the Jaynes-Cummings Hamiltonian \cite{blais2021circuit}:

\begin{equation}
    \mathrm{\hat{H}_{\text{JC}} = \hbar\omega_r \hat{a}^\dagger \hat{a} + \frac{\hbar\omega_s}{2} \hat{\sigma}_z + \hbar g_{eff} (\hat{a}^\dagger \hat{\sigma}_- + \hat{a} \hat{\sigma}_+)} \, ,
\end{equation}
where $\mathrm{\hat{a}^\dagger}$ and $\mathrm{\hat{a}}$ are the creation and annihilation operators for the reservoir mode, $\mathrm{\hat{\sigma}_z}$ is the Pauli-Z operator, $\mathrm{\hat{\sigma}_-}$ and $\mathrm{\hat{\sigma}_+}$ are the lowering and raising operators for the artificial atom, $\mathrm{\omega_r}$ is the reservoir frequency, $\mathrm{\omega_s}$ is the artificial atom transition frequency and $\mathrm{g_{eff}}$ represents the coupling strength between the atom and the reservoir.
The resonant coupling establishes a photon-mediated interaction between the states $\mathrm{\ket{N,e} \leftrightarrow \ket{N+1,g}}$. Here, $\mathrm{\ket{N,g}}$ ($\mathrm{\ket{N,e}}$) represents the system when the resonator is populated by $\mathrm{N}$ photons and the artificial atom is in the ground (excited) state.
Assuming that the counter-relaxation process in the artificial atom transition is characterized by a rate $\mathrm{\Gamma \gg g_{eff}}$, the dominant transition becomes $\mathrm{\ket{N,e} \rightarrow \ket{N+1,g}}$. This predominance is directly linked to the counter-relaxation rate overcoming the atom-reservoir coupling strength, thus, during the typical interaction time $\mathrm{(\sim 2\pi/g_{eff})}$, the reservoir interacts with the artificial atom in its excited state $\mathrm{\ket{e}}$.
Consequently, we can assume that the artificial atom is initially in the excited state. In this hypothesis, if the reservoir contains $\mathrm{N}$ photons, the  coupling $\mathrm{g_{eff}\sqrt{N+1}}$ induces dynamic transitions between the states $\mathrm{\ket{N,e}}$ and $\mathrm{\ket{N+1,g}}$. Since $\mathrm{\Gamma \gg g_{eff}}$, the transition will take the form of an incoherent process where any population that starts to accumulate in the state $\mathrm{\ket{N,g}}$ will quickly relax to the state $\mathrm{\ket{N,e}}$. These two steps complete the transition from the state $\mathrm{\ket{N,e}}$ to the state $\mathrm{\ket{N+1,e}}$, which allows to populate the reservoir. This process will inevitably stop when the coupling strength $g_{eff}\sqrt{N}$ becomes comparable or larger than the counter relaxation rate $\mathrm{\Gamma}$. In this last situation the transition between $\mathrm{\ket{N,e}}$ and $\mathrm{\ket{N+1,g}}$ corresponds to the typical coherent oscillations of dressed state systems. \\
The scaling of the Fock state ladder can also be limited by two other phenomena: photon blockade \cite{lang2011observation, sokolova2021single} and the quantum Zeno effect \cite{itano1990quantum, de1997quantum}. The photon blockade effect primarily arises from the proportional relationship between the splitting of the transition energy levels $\mathrm{\Delta}$ and the number of photons in the reservoir $\mathrm{(\Delta \propto \sqrt{N})}$. As the number of injected photons increases, the resonant transition progressively deviates from the resonance condition. It is worth noting that a technique proposed in \cite{sokolova2021single}, involving an auxiliary reservoir linked to the artificial atom, can mitigate the photon blockade effect. Hence, we will not consider this effect in our subsequent analysis. The quantum Zeno effect, instead, occurs for large values of $\mathrm{\Gamma}$ and results from the decoherence associated with the pumping process, which affects the quantum behavior of the system and, in turn, the masing regime \cite{de1997quantum,ashhab2009single}. 
Considering a system where these last two phenomena can be neglected, we study the condition for which the masing regime takes place. Following the model used above for describing the dynamics under the masing regime, we can define the loss rate from the reservoir to be $\mathrm{\Gamma_l = N \kappa}$, while the emission rate linked to the atom-reservoir transition will be $\mathrm{\Gamma_e = 4 N g_{eff}^2 / \Gamma }$ \cite{ashhab2009single}. Connecting these rates, the emissions overcome the losses when the following condition is satisfied \cite{ashhab2009single}:

\begin{equation}
    \label{eq:las_ratio}
    \lambda = \frac{\Gamma_l}{\Gamma_e} = \mathrm{\frac{\Gamma \kappa}{4 g_{eff}^2} < 1} \, ,
\end{equation}
where $\mathrm{\lambda}$ corresponds to the masing ratio, a fundamental parameter that regulates the emissions in a masing photon source.
Under this condition the interaction between the artificial atom and the reservoir leads to photon population of the reservoir through the mechanism described before, allowing the observation of the masing regime.
Moreover, using a mean-field approximation, it is possible also to extract an analytical expression of the average number of steady-state photons that can be stored in the reservoir \cite{ashhab2009single}:

\begin{equation}
\label{eq:n_photons}
\mathrm{n_{ph} = \frac{\Gamma}{2\kappa} \left(1 - \lambda \right)}  \, .
\end{equation}
Eq. \ref{eq:n_photons} shows that a system characterized by well-defined values of $\mathrm{\Gamma}$ and $\mathrm{\kappa}$ can attain a maximum number of photons, determined by the ratio $\mathrm{\Gamma/2\kappa}$, when the masing ratio is far from unity $\mathrm{(\lambda \ll 1)}$. 
\subsection{Control through tunable coupling}
\label{sec:cavity_control}

\begin{figure}
\includegraphics[width=\columnwidth]{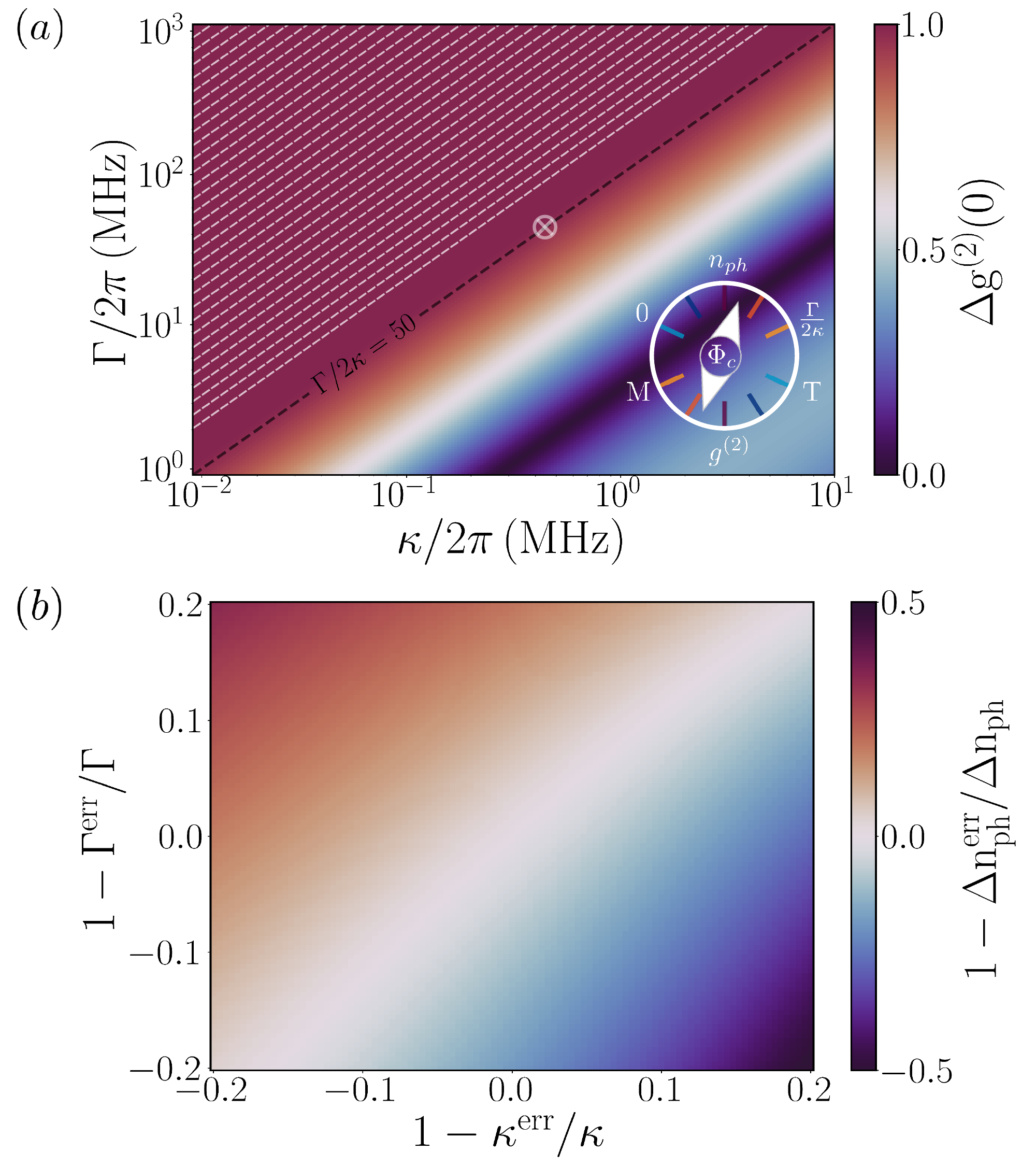}
  \caption{ \justifying \textbf{Efficiency and robustness of the tunable scheme}(a) Variation of the second-order correlation function $\mathrm{g^{(2)}(0)}$ as a function of the counter-relaxation rate $\mathrm{\Gamma}$ and the loss rate in the reservoir $\mathrm{\kappa}$ for a dynamic range of coupling strengths spanning from $\mathrm{g_{eff} = \sqrt{\Gamma\kappa/8}}$ to $\mathrm{g_{eff} = \sqrt{\Gamma\kappa/0.4}} $. The area enclosed by dashed white lines corresponds to working points where $\mathrm{\Gamma/2\kappa > 100}$, which are not computed in this analysis. The black line tracks working points characterized by a maximum emission $\mathrm{\Gamma/2\kappa = 50}$, while the white circle indicates the specific working point $\mathrm{\left(\Gamma, \kappa\right) = \left(50,0.5\right) \, 2 \pi \, MHz} $ used in this study. The role of the flux $\mathrm{\Phi_c}$ tuning the coupling $\mathrm{g_{eff}}$ at this working point is depicted as a flux-dependent knob that allows transitions between deep Masing (M) and Thermal (T) regime. This control mechanism regulates accordingly the average number of steady-state photons injected into the reservoir. The second-order correlation function is computed using QuTip \cite{johansson2012qutip}, considering the Hamiltonian in Eq. \ref{eq:dispersive_hamiltonian}. The Hilbert subspace dimension $\mathrm{N}$ for the reservoir is dynamically adjusted according to the working point and always satisfies the condition $\mathrm{N = \Gamma/2\kappa + 50}$. (b) Relative error in the emission range $\mathrm{\Delta n_{ph}}$ for different values of $\mathrm{\Gamma^{err}}$ and $\mathrm{\kappa^{err}}$ up to a distance of the $\mathrm{20 \%}$ around the reference working point $\mathrm{\left(\Gamma, \kappa \right) = \left( 50, 0.5 \right)2\pi \, MHz} $ discussed in Sec. Control through tunable coupling. The emission range $\mathrm{\Delta n_{ph}}$ is calculated as the difference in the emitted photon number between the maximum coupling strength $\mathrm{g_{eff} = \sqrt{\Gamma \kappa /0.4}}$ and the minimum coupling strength $\mathrm{g_{eff}=\sqrt{\Gamma \kappa / 8}}$ associated to the white point in (a). The photon number $\mathrm{n_{ph}}$ is computed using QuTip considering the steady-state dynamics of the system defined by the Hamiltonian in Eq. \ref{eq:dispersive_hamiltonian}. The dimension $\mathrm{N}$ of the Hilbert subspace changes according to the working point to satisfy the condition $\mathrm{N = \Gamma/2\kappa + 50}$.}
\label{fig:las_phasespace}
\end{figure}

The model described above reveals that the reservoir population is strongly dependent on the specific choices of the parameters $\mathrm{\Gamma}$, $\mathrm{\kappa}$, and $\mathrm{g_{eff}}$ in a single-atom maser. Consequently, the tunability of one of these parameters provides complete control over the steady-state photon population in the reservoir.\\
In this study, we explore the implementation of a tunable coupler scheme \cite{yan2018tunable} to manipulate the coupling strength $\mathrm{g_{eff}}$ between the photon source and a target reservoir. The proposed configuration, depicted in Fig. \ref{fig:setup}(a-b), comprises a two-level system coupled to a superconducting reservoir through a flux-tunable qubit. By employing a broadband coaxial line inductively coupled to the SQUID in the tunable coupler, flux can be delivered on-chip, enabling control over the effective coupling strength between the photon source and the target reservoir without directly interfacing with either element. 
 The circuit can also be integrated into a Single Flux Quantum (SFQ) architecture \cite{mukhanov2011energy}, in which flux is provided via dedicated voltage-biased superconducting circuits, removing the need for a broadband coaxial line.
We model the dynamics of this system by using a Hamiltonian that incorporates the free evolution terms of its constituent elements and the exchange interactions until the next nearest neighbors. Taking into account just the first two levels of the tunable coupler, the Hamiltonian takes the following form:
\begin{equation}
\label{eq:hamiltonian}
\begin{aligned}
\hat{H}/\hbar &= \left(\hat{H}_s + \hat{H}_c + \hat{H}_r + \hat{H}_{rc} + \hat{H}_{sc} + \hat{H}_{sr}\right)/\hbar \\
&= \sum_{i=s,c} \frac{\omega_i \hat{\sigma}_i^z}{2} + \omega_r \hat{a}^\dagger \hat{a} + g_{rc}(\hat{a}^\dagger \hat{\sigma}_c^- + \hat{a} \hat{\sigma}_c^+) \\
&\quad + g_{sc}(\hat{\sigma}_s^+ \hat{\sigma}_c^- + \hat{\sigma}_s^- \hat{\sigma}_r^+) + g_{sr}(\hat{\sigma}_s^+ \hat{a} + \hat{\sigma}_s^- \hat{a}^\dagger) \, .
\end{aligned}
\end{equation}
Here, $\mathrm{\omega_j}$, $\mathrm{\hat{\sigma}_{j}^z}$, $\mathrm{\hat{\sigma}_{j}^+}$, and $\mathrm{\hat{\sigma}_{j}^-}$ for $\mathrm{j=s,c}$, represent the transition frequency and the Z-Pauli, raising and lowering operators for the photon source and the tunable coupler, respectively. Additionally, $\mathrm{\hat{a}}$ and $\mathrm{\hat{a}^\dagger}$ are the ladder operators associated with the target resonator modes. The terms $\mathrm{g_{ij}}$ for $\mathrm{i \neq j=s,r,c}$ represent the coupling strengths governing the interactions between different elements, defined in terms of the circuital parameters of the scheme:

\begin{equation}
\label{eq:coupling_strengths}
\begin{aligned}
\mathrm{g_{ic}} &\approx \frac{1}{2}\frac{C_{ic}}{\sqrt{C_i C_c}} \sqrt{\omega_i \omega_c} \, , \\
\mathrm{g_{sr}} &\approx \frac{1}{2}\frac{C_{sc} C_{rc}}{\sqrt{C_s C_r C_c^2}} \sqrt{\omega_s \omega_r} \, ,
\end{aligned}
\end{equation}
where $\mathrm{C_{ij}}$ for $\mathrm{i \neq j=s,r,c}$ correspond to the coupling capacitor between the different elements in the scheme, while $\mathrm{C_{i}}$ for $\mathrm{i=s,r,c}$ denote their capacitance to ground. The coupling strengths $\mathrm{g_{rc}}$ and $\mathrm{g_{sc}}$ associated with the interactions involving the tunable coupler can be controlled via the magnetic flux $\mathrm{\Phi_c}$, which modulates the frequency of the tunable coupler $\mathrm{\omega_c\left( \Phi_c \right)}$. In contrast, the coupling strength $\mathrm{g_{sr}}$ related to the next-nearest neighbor interaction between the source and the reservoir, remains fixed.
Flux tunability of the coupling between the photon source and the reservoir can be realized by engineering a dispersive coupling between the tunable coupler and both the photon source and the reservoir when $\mathrm{g_{jc} \ll \Delta_{jc} = \omega_j - \omega_c}$ for $\mathrm{j=s,r}$. Under these assumptions, the Hamiltonian in Eq. \ref{eq:hamiltonian} can be simplified using a Schrieffer-Wolff transformation through the unitary operator $\mathrm{\hat{U} = \exp\left[ \sum_{j=s,r} \frac{g_{jc}}{\Delta_{jc}}(\hat{\sigma}_j^+\hat{\sigma}_c^- - \hat{\sigma}_j^-\hat{\sigma}_c^+)\right]}$. This transformation effectively decouples the tunable coupler from the photon source and reservoir to the second order in $\mathrm{g_{sc}/\Delta_{sc}}$ and $\mathrm{g_{rc}/\Delta_{rc}}$. This transformation results in a Jaynes-Cummings Hamiltonian with a flux-tunable coupling strength (further details in Methods "Hamiltonian")
\begin{equation}
\label{eq:dispersive_hamiltonian}
\begin{aligned}
\mathrm{\frac{\hat{U}\hat{H}\hat{U}^\dagger}{\hbar} = \frac{\tilde{\omega}_s\hat{\sigma}_s^z}{2} + \tilde{\omega}_r\hat{a}^\dagger \hat{a} + g_{eff}(\hat{\sigma}_s^+ \hat{a} + \hat{\sigma}_s^- \hat{a})} \, .
\end{aligned}
\end{equation}
In this framework, the influence of the tunable coupler is quantitatively expressed by the renormalization of the characteristic frequencies governing the system dynamics. Specifically, $\mathrm{\tilde{\omega}_k = \omega_k + \frac{g_{kc}^2}{\Delta_{kc}}}$ with $\mathrm{k = s,r}$ represent the Lamb-shifted qubit and reservoir frequency, while the effective coupling $\mathrm{g_{eff}}$ between the source and the reservoir is defined by the following expression: 
\begin{equation}
\label{eq:effective_coupling}
\begin{aligned}
\mathrm{g_{eff}(\Phi_c) = g_{sc}g_{rc}\left(\frac{1}{\Delta_{sc}} + \frac{1}{\Delta_{rc}}\right) + g_{sr}} \, .
\end{aligned}
\end{equation}
The effective coupling $\mathrm{g_{eff}}$ is defined by the sum of two contributions arising from the interactions between the eigenstates $\mathrm{\ket{s,c,r}}$ inside the system spectra. The first contribution comes from the virtual interaction between the first excited state of the coupler 
$\mathrm{\ket{g,1,n_{ph}}}$ with the states $\mathrm{\ket{e,0,n_{ph}}}$ and $\mathrm{\ket{g,0,n_{ph}+1}}$ responsible for the photon transport from the source to the reservoir. When the source and the reservoir are on resonance, the eigenstates $\mathrm{\ket{e,0,n_{ph}-1}}$ and $\mathrm{\ket{g,0,n_{ph}}}$ are degenerate and form a pair of dressed states split by an amount $\mathrm{\delta_N = 2 g_{sr}\sqrt{N}}$. Similarly, the eigenstate with the coupler in the excited state $\mathrm{\ket{g,1,n_{ph}}}$ forms a pair of dressed states with $\mathrm{\ket{e,1,n_{ph}-1}}$.
The splitting $\mathrm{\delta_N}$ becomes comparable to the minimal detuning between energy levels $\mathrm{\left(\Delta_{m} = min\left[\Delta_{rc},\Delta_{sc}\right]\right)}$ for $\mathrm{N^*=\left( \Delta_{min}/g_{sr} \right)^2}$. Within the dispersive regime, this number $\mathrm{N^*}$ always lies outside the range of photons that can be injected before the dispersive regime ceases to be valid.
As a result, we can reasonably assume that the virtual interaction remains independent of the number of photons in the reservoir throughout the entire injection process (further details in Methods "Functional dependence on the number of photons").
The second term accounts for the nearest neighbor coupling that directly couples the two states $\mathrm{\ket{e,0,n_{ph}-1}}$ and $\mathrm{\ket{g,0,n_{ph}}}$. 
The dependence on $\mathrm{\omega_c}$ embedded within $\mathrm{\Delta_{sc}}$ and $\mathrm{\Delta_{rc}}$ offers a knob to regulate the coupling strength $\mathrm{g_{eff}}$ 
\begin{figure}[t]
\label{fig:setup_c}
\centering
\includegraphics[width=0.8\columnwidth]{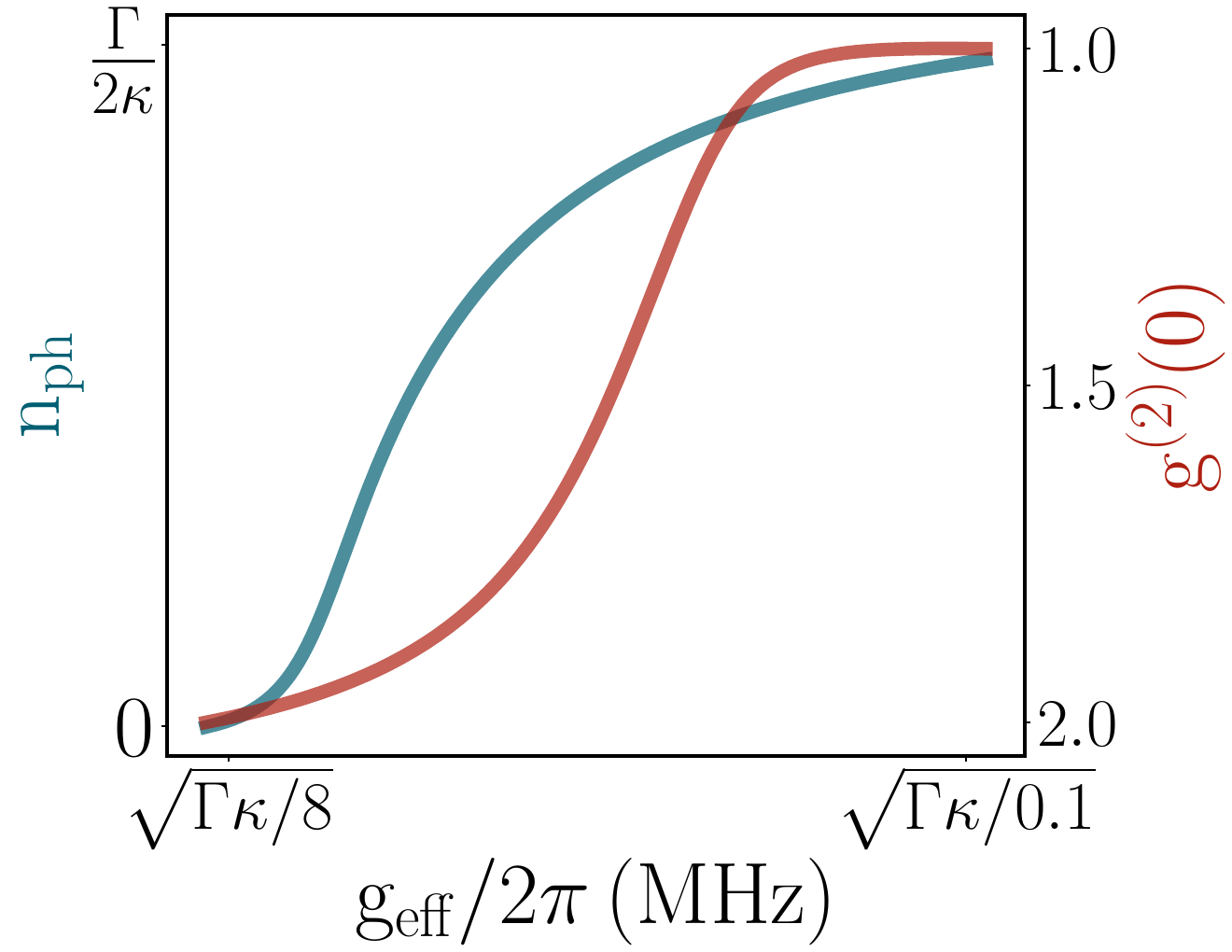}
\caption{ \justifying \textbf{Microwave emission tunability via tunable coupling} The average number of steady-state photons $\mathrm{n_{ph}}$ and the normalized second-order correlation function $\mathrm{g^{(2)}(0)}$ are shown in terms of the effective coupling strength $g_{\text{eff}} $ between the photon source and the reservoir and in terms of the masing ratio $\Gamma \kappa / 4 g_{eff}^2$ where $\kappa$ is the loss rate of the reservoir. The results are obtained by simulating the steady-state dynamics of the Hamiltonian in Eq. \ref{eq:dispersive_hamiltonian} in Qutip varying the effective coupling $\mathrm{g_{eff}}$. The dimension $\mathrm{N}$ of the Hilbert subspace associated with the harmonic reservoir in the simulation is $\mathrm{N=100}$.}
\label{fig:setup_c}
\end{figure}
between the source and the reservoir through an external flux modulating the coupler frequency. 
The result is an intrinsically quantum photon source at cryogenic temperatures, enabling on-demand control of photon flow into a target cavity through an external flux threading the coupler.\\
The practical feasibility of the proposed scheme relies on several key factors. We address them by examining the different engineering steps required for its implementation.
The first step is the selection of the counter-relaxation rate $\mathrm{\Gamma}$ and the loss rate of the target reservoir $\mathrm{\kappa}$ which together define the working point $\mathrm{\left(\Gamma,\kappa \right)}$ of the scheme.
According to Eq. \ref{eq:n_photons}, these parameters determine whether the system can coherently inject photons for a given value of the effective coupling $\mathrm{g_{eff}}$, as well as the maximum emission, which equals to $\mathrm{n_{ph}^{max}=\Gamma/2\kappa}$.
Assuming that the flux-controlled effective coupling allows the masing ratio to vary within a certain range $\mathrm{\left[\lambda_{min}, \lambda_{max}\right]}$, we evaluate the resulting tunability of the emission power, for given $\mathrm{\lambda_{min}, \lambda_{max}}$.   
We perform this analysis by examining the normalized second-order correlation function $\mathrm{g^{(2)}(0) = \braket{\hat{a}^{\dagger}\hat{a}^{\dagger}\hat{a}\hat{a}}/\braket{\hat{a}^{\dagger}\hat{a}}^2}$ \cite{loudon2000quantum}, a common metric for distinguishing quantum effects from classical phenomena in quantum optical systems \cite{loudon2000quantum,gu2017microwave}. Within this scheme, $\mathrm{g^{(2)}(0)}$ serves as an indicator of the system operational regime \cite{ashhab2009single}, as illustrated in Fig. \ref{fig:setup_c}. When $\mathrm{g^{(2)}(0) = 1}$, the photon statistics follow a Poissonian distribution \cite{loudon2000quantum} and the emissions reach a maximum $\mathrm{\left(n_{ph}\sim \Gamma/2\kappa \right)}$, which denotes that the system is in the masing regime. Reducing the coupling between the photon source and the reservoir can suppress the masing dynamics, bringing the photon loss rate closer to the emissions rate. In this intermediate regime, the second-order correlation function assumes values between $\mathrm{1<g^{(2)}(0)<2}$ and the photon distribution follows a super-Poissonian statistics (more details in Methods "From Poissonian to super-Poissonian"). When the losses dominate over the emissions and the masing is fully suppressed, the system is in a thermal regime   $\mathrm{g^{(2)}(0) = 2}$ \cite{loudon2000quantum} where the reservoir remains unpopulated   \cite{ashhab2009single}. \\ Control over emission in the proposed scheme is achieved by modulating the external flux $\mathrm{\Phi_c}$, which adjusts the effective coupling and enables a continuous transition between the deep masing regime and the thermal regime, as shown in Fig. \ref{fig:setup_c}. To maximize the control offered by this scheme, the coupling should allow a full transition from $\mathrm{g^{(2)}(0) = 1}$ (deep masing) to $\mathrm{g^{(2)}(0) = 2}$ (thermal regime).  
We compute the normalized second-order correlation function with QuTip \cite{johansson2012qutip} for different working points $\mathrm{\left(\Gamma,\kappa\right)}$ using the model in Eq. \ref{eq:dispersive_hamiltonian} evaluating its variation $\mathrm{g^{(2)}(0)}$  across a specific range of effective coupling values. Specifically, we consider the difference in $\mathrm{g^{(2)}(0)}$ between scenarios where the masing ratio is set to $\mathrm{\lambda_{min} = 0.1}$ (deep masing) and $\mathrm{\lambda_{max} = 2}$ (suppressed masing). According to Eq. \ref{eq:n_photons}, these values correspond to conditions where the system transitions from a strongly masing regime to a partially suppressed masing regime. 
The results of this analysis are reported in Fig. \ref{fig:las_phasespace}(a), which displays the expected variations in $\mathrm{g^{(2)}(0)}$ for different working points inside a range of experimentally reachable parameters of the system  \cite{astafiev2007single,oelsner2013dressed, sokolova2021single}. The red region in the colormap contains all the points $\mathrm{\left(\Gamma,\kappa\right)}$ characterized by the maximum degree of tunability $\mathrm{\Delta g^{(2)}(0)=1}$, allowing the system to span the full range from $\mathrm{g^{(2)}(0) = 1}$ (maximum emission, $\mathrm{n_{ph} \approx \Gamma / 2\kappa}$) to $\mathrm{g^{(2)}(0) = 2}$ (minimal emission, $\mathrm{n_{ph} \approx 0}$) using the external flux $\mathrm{\Phi_c}$. Conversely, the dark blue regions, instead, denote working points where control via the external flux $\mathrm{\Phi_c}$ is not achievable within the considered range of effective coupling.  The diagonals on the colormap denote all the working points characterized by the same emission maximum $\mathrm{n_{ph}^{max}=\Gamma/2\kappa}$. The dashed black line in Fig. \ref{fig:las_phasespace}(a) indicates a specific case for $\mathrm{n_{ph}^{max}=50}$. For the considered variation range of the masing ratio, $\mathrm{\lambda \in \left[0.1, 2\right]}$, the scheme provides full control $\mathrm{\left(\Delta g^{(2)}(0) \approx 1\right)}$ over the emission for working points above an emission threshold of $\mathrm{\Gamma / 2\kappa \geq 45}$.  For validation, we select the working point $\mathrm{\left(\Gamma, \kappa\right) = \left(50, 0.5\right)\, 2\pi \, MHz}$, marked by a white circle in the red zone of Fig. \ref{fig:las_phasespace}(a). This point represents a feasible experimental configuration within the considered range of effective couplings  \cite{astafiev2007single,sokolova2021single}. Considering the chosen working point, we compute $\mathrm{\Delta g^{(2)}(0)}$ up to an emission maximum of $\mathrm{n_{ph}^{max} = 100}$ in Fig. \ref{fig:las_phasespace}(a).\\
Even when the system is well designed to operate at an optimal working point, it is essential to ensure robustness against fabrication errors, which may affect the range of emissions, and against flux noise, which can degrade control over the steady-state photon distribution. To evaluate the impact of design errors on system performance, we consider the previous reference working point $\mathrm{\left(\Gamma,\kappa\right)}$ and introduce errors of up to $\mathrm{\pm 20\%}$ in both parameters. In each case, we compute the variation $\mathrm{\Delta n_{ph}}$, defined as the difference in emission between the cases where the effective coupling $\mathrm{g_{eff}}$ is set to its maximum and minimum values. The maximum and minimum effective coupling values are given by $\mathrm{g_{eff} = \sqrt{\Gamma \kappa / 0.4}}$ and $\mathrm{g_{eff} = \sqrt{\Gamma \kappa / 8}}$, respectively, assuming the coupling is designed for the reference working point $\mathrm{\left(\Gamma,\kappa\right)}$. The results are depicted in Fig. \ref{fig:las_phasespace}(b) and show a fair robustness of the system within $\mathrm{ 5 \%}$ of error in both degrees of freedom.
To prevent the influence of faulty fabrication on the system feasibility, the coupling can be engineered to work within a broader range. This expansion of the coupling range, however, comes at the expense of an increased sensitivity of the tunable coupler to external flux noise. \\
In a practical realization of the system, a trade-off must be found between these two features. To decrease the sensitivity to external flux, we can employ several strategies that allow us to precisely define the range of working points. Differently from the typical tunable coupler scheme in which the system is engineered to have an off-point value where the coupling is cancelled \cite{yan2018tunable}, here we just aim at modulating it within a specific range of couplings between a deep masing regime $\mathrm{\Gamma\kappa/4g_{eff}^2 \ll 1}$ to a suppressed masing regime $\mathrm{\Gamma\kappa/4g_{eff}^2 > 1}$.
One approach involves designing the tunable coupler with a resonance frequency lower than that of the photon source and the reservoir, resulting in an effective coupling strength $\mathrm{g_{eff} \propto 1 / (\omega_s - \omega_c)}$. Additionally, we can use an asymmetric SQUID as the tunable element of the coupler, which exhibits a more gradual response to external fluxes \cite{krantz2019quantum}. These adjustments provide two significant advantages: (1) the coupler never approaches a frequency close to those of the photon source and the reservoir for any threaded 
flux, ensuring that the dispersive condition is steadily met; (2) the asymmetry allows for non-zero minima in the coupler frequency, narrowing the range of effective coupling to well-defined maxima and minima. 
\begin{figure*}[t]
  \includegraphics[width=\textwidth]{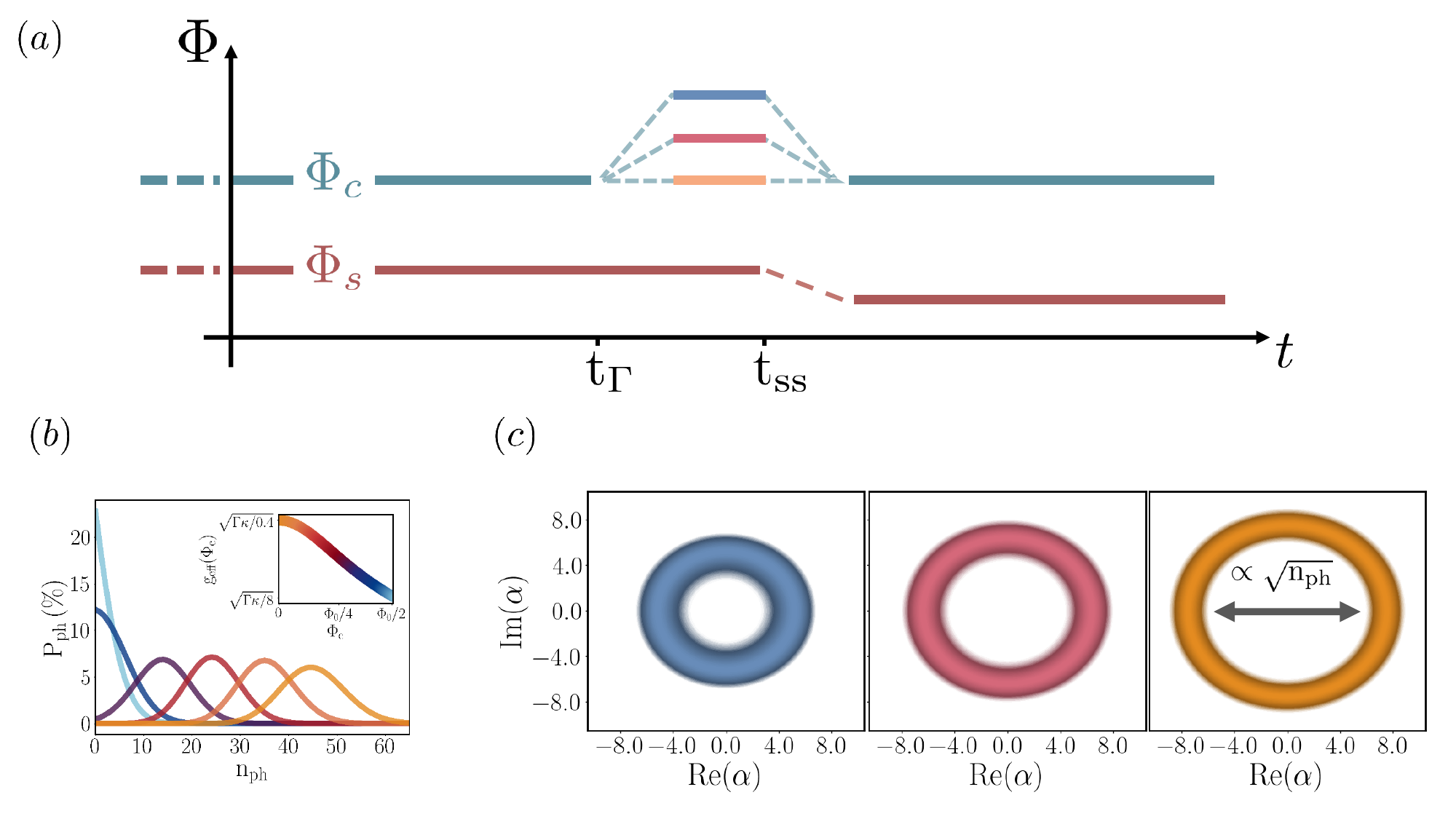}
  \caption{\justifying \textbf{Coherent photon generation through the tunable coupler scheme.} (a) Flux timing diagram illustrating the photon population protocol of a single reservoir with the proposed scheme. The flux $\Phi_s$ sets the emission frequency on resonance with the target reservoir. At the initial time $\mathrm{t_{\Gamma}}$ the counter-relaxation is switched on. Afterwards the flux $\mathrm{\Phi_c}$ on the tunable coupler is chosen according to the wanted steady-state distribution of photons in the reservoir. After the time $t_{ss}$ where the system has reached its steady state, the photon source is set off resonance from the reservoir by changing $\Phi_s$ to avoid further interactions between the two systems. (b) Steady-state photon distributions in the target reservoir under varying flux $\Phi_c$ values threading the coupler. These simulation results are obtained by simulating the steady-state dynamics in Qutip with the Hamiltonian in Eq. \ref{eq:dispersive_hamiltonian} for a device engineered according to the parameters listed at the end of Sec. Control through tunable coupling. Each color represents the distribution of photons in the Fock basis corresponding to specific values of the flux $\mathrm{\Phi_c}$. The dimension $\mathrm{N}$ of the Hilbert subspace associated with the reservoir in the simulation is $\mathrm{N = 100}$. (c) Corresponding steady-state reservoir normalized Wigner functions for different values of the flux $\mathrm{\Phi_c}$. The radius of the Wigner function in the phase space is proportional to the average of the corresponding photon number distribution as in typical IQ representations of spontaneous emission from a photon source \cite{cassidy2017demonstration,liu2015semiconductor,yan2021low}.}
  \label{fig:workflow}
\end{figure*}
 Following these guidelines, the effective coupling can be engineered to have the shape shown in Fig. \ref{fig:workflow}(b). In this configuration, the system passes from a masing regime  $(\lambda=0.1)$ to a suppressed masing regime $(\lambda=2)$ at the two extrema of the flux. Notably, at these extremal points, the sensitivity to flux is suppressed  $(\partial g_{eff}/\partial \Phi \approx 0)$, making the system particularly insensitive to flux noise in these operating regimes. The system can also be operated exclusively at the two extremal points of the flux range, where the photon population can be controlled by adjusting the duration of the flux pulse applied to the tunable coupler, which switches between the points of maximum and minimum effective coupling. Alternative approaches involve the use of ferromagnetic elements, which enable flux tunability without the application of static magnetic fields commonly identified as a primary source of $1/f$ flux noise \cite{ ahmad2023competition}. 
 The design of the tunable coupler must also ensure that its interaction with the other elements always satisfies the dispersive approximation within the operational flux range of the system. To guarantee the general validity of this condition, the tunable coupler and the capacitive coupling to the other components must be engineered such that the number of injected photons $\mathrm{n_{ph}}$ never reach values close to the critical number of photons $\mathrm{n_{cr} = \left( \Delta_{rc}/2g_{rc}\right)^2}$ \cite{blais2021circuit}. The critical number of photons depends on the detuning $\mathrm{\Delta_{rc}}$ between the reservoir and the coupler, and the coupling strength, $\mathrm{g_{rc}}$, between them. In scenarios where the tunable coupler is designed to always have a frequency below the resonance frequencies of the other elements, $\mathrm{n_{cr}}$ reaches its minimum at zero flux $\mathrm{\left(\Phi_c=0\right)}$, where the detuning is minimized. Notably, the zero flux point also corresponds to the regime of maximum photon emission, making it the point where the dispersive approximation is most likely to break.\\  
The maximum number of photons $\mathrm{n_{ph}^{max}=\Gamma/2\kappa}$ is defined by the working point $ \mathrm{\left( \Gamma , \kappa \right)}$. Therefore, the dispersive approximation holds if the ratio $\mathrm{n_{ph}^{max}/n_{cr}=2\Gamma g_{rc}^2\left(0\right)/\kappa \Delta_{rc}^2\left(0\right) \ll 1}$ in any flux point. Moreover, since by definition $n_{cr}/N^* = \left(\Delta_{rc}/\Delta_{min}\right)^2\left(g_{sr}/2g_{rc}\right)^2 < 1$, this assumption also ensures that the splitting between dressed states can be fairly neglected for any flux $\mathrm{\Phi_c}$ threading the tunable coupler.  
Given the typical weak anharmonicity $\mathrm{\alpha_c}$ of the transmon qubits used as tunable coupler \cite{koch2007charge,yan2018tunable}, it is important to account for the effects of higher-energy states on the virtual interaction between the reservoir and the source within the scheme. The two eigenstates $\mathrm{\ket{e,0,n_{ph}}}$ and $\mathrm{\ket{g,0,n_{ph}+1}}$, which govern photon transport from the source to the reservoir, can couple to higher-order states such as $\mathrm{\ket{g,2,n_{ph}-1}}$, originating from the higher excited states of the tunable coupler. Assuming the photon source and reservoir are on resonance $\mathrm{\left(\omega = \omega_s = \omega_r\right)}$, the energy separation from the second excited state of the tunable coupler is approximately $\mathrm{\hbar \delta_{\ket{g,2,n_{ph}-1}}\approx 2\hbar\left(\omega - \omega_c \right) - \hbar\alpha_c}$, where we neglect the dependence on the photon number $\mathrm{n_{ph}}$ embedded in the dressed state splitting of the eigenestates. This expression suggests that a potential mitigation strategy is to design tunable couplers with higher anharmonicity. \\
When engineering the anharmonicity of the tunable coupler, it is also important to consider the self-Kerr effect induced by the coupler on the reservoir dynamics. The nonlinearity of the tunable coupler introduces a term $\mathrm{\hat{H}_K = K_r \hat{a}^\dagger\hat{a}^\dagger\hat{a}\hat{a}}$ in the reservoir Hamiltonian, where $\mathrm{K_r}$ represents the self-Kerr of the reservoir. In this scheme, the only contribution to $\mathrm{K_r}$ arises from the interaction with the tunable coupler and is defined by $\mathrm{K_r = \chi_{rc}^2/4\alpha_c}$ \cite{kirchmair2013observation,he2023fast}, where $\mathrm{\chi_{rc} = - g_{rc}^2 |\alpha_c|/\Delta_{rc}\left( \Delta_{rc} - |\alpha_c| \right)}$ is the dispersive shift associated to the virtual interaction between the reservoir and the coupler \cite{blais2021circuit}.
To ensure that the Kerr term does not impact the time evolution of the generated state in the reservoir, the ratio $\mathrm{K_r/\kappa}$ must satisfy $\mathrm{K_r/\kappa \ll 1}$ \cite{kirchmair2013observation}. Under this condition, the Kerr evolution is sufficiently slow to be negligible compared to the reservoir state collapse, allowing it to be effectively treated as harmonic. The implementation of the scheme for situations where Kerr dynamics dominate over the loss rate is under investigation. \\
To demonstrate the feasibility of the proposed scheme, we consider the dynamics of the following example scheme: a flux-tunable photon source with a counter-relaxation rate  $\Gamma = 50 \, \mathrm{MHz}$, self-capacity of $C_s = 100 \, \mathrm{fF}$, and a transition frequency $\omega_s/2\pi = 7 \, \mathrm{GHz}$ that is resonant with a reservoir having a resonance frequency $\omega_r = \omega_s$, self-capacity of $C_r = 500 \, \mathrm{fF}$, and loss rate $\kappa/2\pi = 0.5 \, \mathrm{MHz}$. The two are connected through a tunable coupler corresponding to an asymmetric flux-tunable qubit with Josephson energies $E_{J1}^c/h = 10.02  \, \mathrm{GHz}$, $E_{J2}^c/h = 8.532 \, \mathrm{GHz}$, and charging energy $E_C^c/h = 200 \, \mathrm{MHz}$, resulting in a ratio $\mathrm{E_{j_2}^c/E_C^c \approx 43}$. The coupling capacities between the photon source and the reservoir with the tunable coupler are $C_{sc} = C_{rc} = 3.5 \, \mathrm{fF}$.
A design with these parameters is perfectly feasible for the state-of-the-art nanofabrication process \cite{krinner2019engineering,yan2018tunable,sung2021tunablecoupler} and the value of $\mathrm{\Gamma}$ is near the one used in other works \cite{sokolova2021single,oelsner2013dressed}.\\A viable candidate that can be used as the photon source in this scheme is an aluminum voltage-biased charge qubit \cite{astafiev2007single}. Designing the chip with a Josephson junction of energy $E_j^s/h \sim 14.66 GHz$ and a shunt resistance on the electrode of $R_b=5 \, M\Omega$ we achieve the target $\mathrm{\Gamma/2\pi \sim 50 \, MHz}$.
With the chosen parameters, the ratio $\mathrm{n_{ph}^{max}/n_{cr}\left(\Phi_c=0\right) = 0.15}$ which ensures that the system is well described by the Hamiltonian in Eq. \ref{eq:dispersive_hamiltonian} in any flux point. The maximum value of the self-Kerr induced by the tunable coupler on the reservoir appears at zero flux point where $\mathrm{|K_r| \sim 0.0365 \, kHZ \ll \kappa }$. The small value of the self-Kerr $\mathrm{K_r}$ compared to the loss rate $\mathrm{\kappa}$ guarantees that Kerr evolution is negligible over the lifetime of the generated state in the reservoir of the proposed scheme.
We analyze the system dynamics for various values of the flux $\mathrm{\Phi_c}$ threading the coupler.
By activating the pumping process and exploring the system across different flux values, we obtain the results shown in Fig. \ref{fig:workflow}(b). Different values of the flux threading the coupler yield distinct values of the effective coupling $\mathrm{g_{eff}}$, which ultimately lead to different photon distributions within the reservoir at the steady state.
Assuming typical flux fluctuations on the order of $10 \, \mu \Phi_0$ \cite{yoshihara2006decoherence,yan2018tunable}, the resulting variation in photon number is negligible, with deviations of the order of $\sim 0.1 \,  \%$ across the entire flux range. Similarly, the variation in the emission linewidth converges to $\sim 0.03 \, \%$ for a Hilbert space dimension of $N=70$, at a representative flux of $\Phi = \Phi_0 /4 $, suggesting no significant impact on the overall system performance (further details can be found in the Supplementary Note 1). 
The Wigner functions defining the steady states of the reservoir for different values of the effective coupling $\mathrm{g_{eff}}$ are shown in Fig. \ref{fig:workflow}(c). We observe the typical "ring" shape that characterizes coherent photon sources, where the radius of the Wigner function is proportional to the average of the steady-state photon distribution \cite{yan2021low,liu2015semiconductor,cassidy2017demonstration}.
After the reservoir has reached its steady state, the pumping process can be stopped and the flux $\mathrm{\Phi_c}$ set to zero. At this stage, to avoid further interactions between the photon source and the reservoir we can change the transition frequency of the photon source through the external flux $\mathrm{\Phi_s}$. 
\subsection{Multiple resonator generalization}
Another advantage offered by the proposed scheme is the ability to address multiple cavities by exploiting the tunability of the photon source. 
Under the same assumptions as the single-reservoir case, we can extend the analysis to an $\mathrm{N_r}$-reservoir system. The system is described by a Hamiltonian that includes additional terms related to the self-energy of different reservoirs, the coupling of each reservoir to the tunable coupler, and the second nearest neighbor cross-couplings between each resonator. In this discussion, we consider the simple case of a two-reservoir setup as a starting point, where the case of $\mathrm{N_r}$ reservoirs will be a direct generalization (calculations of the general $\mathrm{N_r}$ case are provided in Methods "Schrieffer-Wolff transformation" ).
We investigate the ability to individually address each reservoir through the external fluxes, threading the tunable coupler and assuming that the photon source can be tuned on resonance with both resonators separately. As in the single-reservoir case, a single coaxial line coupled to the tunable coupler can modulate the coupler frequency, enabling controlled emission to different target reservoirs. Assuming that both the coupler and the photon source behave as perfect two-level systems and neglecting second-order interactions, the Hamiltonian is written as:
\begin{equation}
\begin{aligned}
\label{eq:full_hamiltonian_n}
\mathrm{\hat{H}_N = \hat{H}_s + \hat{H}_c + \hat{H}_{r_1} + \hat{H}_{r_2} + \sum_{j=s,r_1,r_2} \hat{H}_{jc} + \hat{H}_{r_1r_2}} \, .
\end{aligned}
\end{equation}
To remove the terms related to the coupler from the Hamiltonian and establish a feasible and controllable scheme, the coupler must be engineered to satisfy the conditions $\mathrm{g_{jc}/\Delta_{jc} \ll 1}$ for $\mathrm{j=s,r_1,r_2}$. Under this assumption, we can transform the Hamiltonian by applying the Schrieffer-Wolff transformation:
\begin{equation}
\begin{aligned}
\label{eq:schrieffer_wolff}
\mathrm{\hat{U}} &\mathrm{= \exp\left[\frac{g_{sc}}{\Delta_{sc}}(\hat{\sigma}_s^+ \hat{\sigma}_c^- - \hat{\sigma}_s^- \hat{\sigma}_c^+)\right.} + \\
&\mathrm{\left.\phantom{=} + \frac{g_{r_1c}}{\Delta_{r_1c}}(\hat{a}_{r_1}^\dagger \hat{\sigma}_c - \hat{a}_{r_1} \hat{\sigma}_c^+) + \frac{g_{r_2c}}{\Delta_{r_2c}}(\hat{a}_{r_2}^\dagger \hat{\sigma}_c - \hat{a}_{r_2} \hat{\sigma}_c^+)  \right]} \, .
\end{aligned}
\end{equation}
This transformation leads to the following Hamiltonian:
\begin{equation}
\begin{aligned}
\label{eq:double_cavity}
\mathrm{\frac{\hat{H}}{\hbar}} &= \mathrm{\frac{\tilde{\omega}_s\hat{\sigma}_s^z}{2} + \tilde{\omega}_{r_1}\hat{a}_{r_1}^\dagger \hat{a}_{r_1} + \tilde{\omega}_{r_2}\hat{a}_{r_2}^\dagger \hat{a}_{r_2}} \\
&\mathrm{\phantom{=} + \sum_{j=r_1,r_2}\tilde{g}_{jc}(\hat{\sigma}_s^+ \hat{a}_j + \hat{\sigma}_s^- \hat{a}_j^+)} \\
&\mathrm{\phantom{=} + \tilde{g}_{r_1r_2}(\hat{a}_{r_1}\hat{a}^\dagger_{r_2} + \hat{a}_{r_1}^\dagger\hat{a}_{r_2})} \, .
\end{aligned}
\end{equation}
\begin{figure}[t]
\includegraphics[width=\columnwidth]{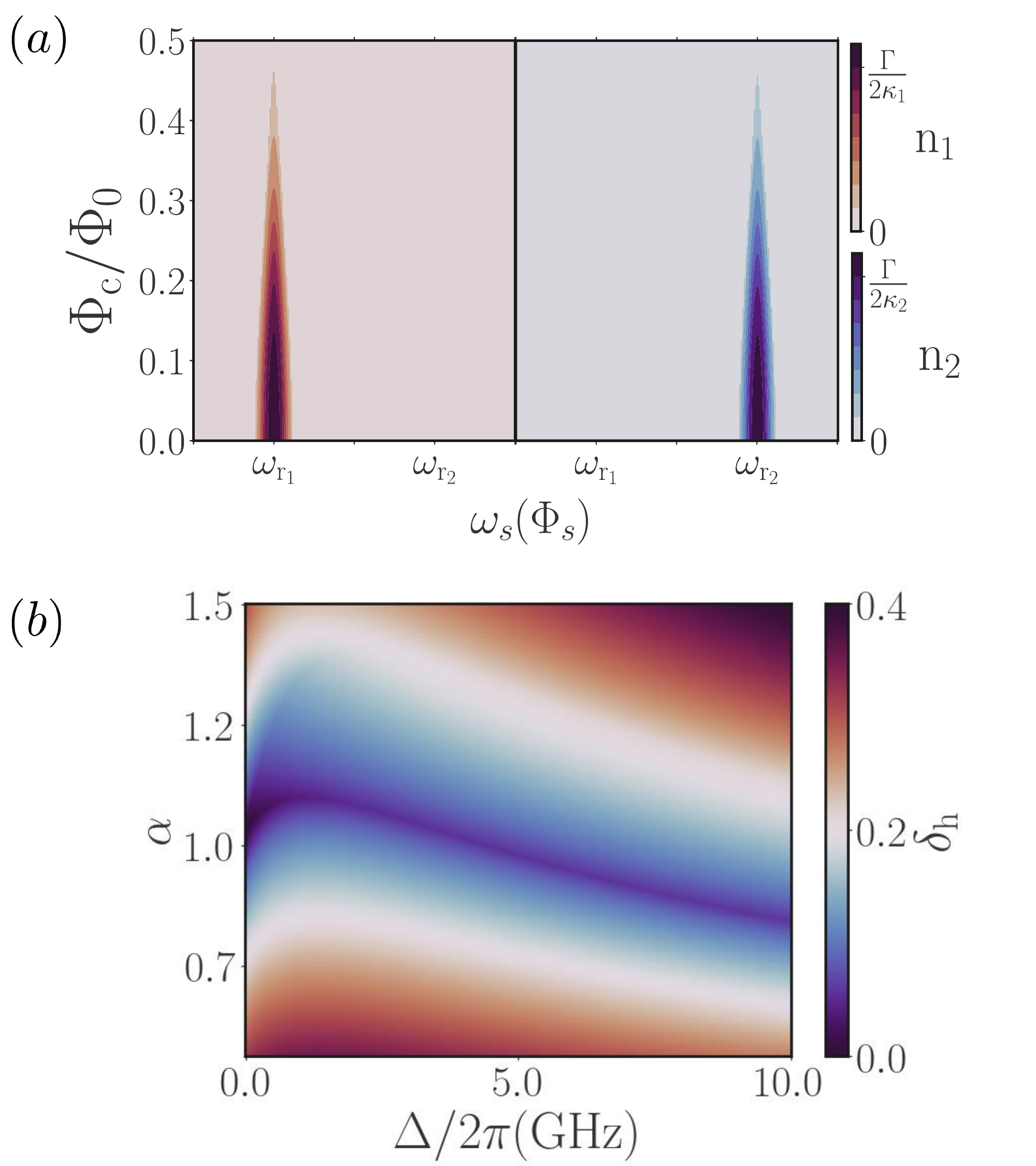}
  \caption{\justifying \textbf{Multiple reservoirs case} (a) Steady-state photon number in different cavities with resonance frequencies $\mathrm{\omega_{r_1}/2\pi = 7 \, GHz}$ and $\mathrm{\omega_{r_2}/2\pi = 7.005 \, GHz}$. The red and blue points on the two plots represent the average number of injected photons in each reservoir for specific choices of flux $\mathrm{\Phi_c}$ and frequency $\mathrm{\omega_s/2\pi}$ of the photon source. The absence of colored points in correspondence to the frequency of the other resonator in each colorplot shows the absence of cross-population phenomena. The circuital and simulation parameters are the same as the one considered in Figs. \ref{fig:setup_c} and \ref{fig:workflow}(b). (b) The Hausdorff distance between the effective coupling strength curve $\mathrm{g_{eff}(\Phi_c)}$ of the reference reservoir and the second reservoir detuned by $\mathrm{\Delta}$ and coupled with a coupling capacity $\mathrm{\alpha}$ times that of the reference reservoir. The dark blue zone indicates a range of coupling capacities that maintain consistent effective coupling dependencies while altering the resonance frequency of the reservoir relative to the reference reservoir.}
\label{fig:multi_res}
\end{figure}
In this expression, the renormalizations of various frequency terms are identical to those observed in the single-reservoir case in Eq. \ref{eq:dispersive_hamiltonian}, except for the new cross-coupling term $\mathrm{\tilde{g}_{r_1r_2}}$ between the two reservoirs. The second-order coupling term $\mathrm{g_{r_1r_2}}$ is summed with another term arising from the coupling between the reservoirs via the tunable coupler, leading to the following expression of the effective cross-coupling strength:
\begin{equation}
    \mathrm{\tilde{g}_{r_1r_2} = g_{r_1c}g_{r_2c} \left(\frac{1}{\Delta_{r_1c}} + \frac{1}{\Delta_{r_2c}} \right) + g_{r_1r_2}} \,.
\end{equation}
It is, therefore, crucial to understand how much detuning is required between the different reservoirs at the fan-out of the system to avoid resonant interactions that lead to cross-population between the cavities.
We consider a setup similar to the single-reservoir scheme analyzed in the previous section, with the addition of a second reservoir at a frequency $\mathrm{\omega_{r_2} = \omega_{r_1} + \Delta}$ and linked to the tunable coupler with the same capacitive coupling used for the target reservoir. We observe the system behavior for different values of $\mathrm{\Delta}$ while sweeping both the photon source and the coupler frequencies, simulating the effect of the external fluxes $\mathrm{\Phi_s}$ and $\Phi_c$ individually threading the two elements, respectively. Assuming that both reservoirs have the same linewidth as the reservoir considered in the previous single reservoir case $\mathrm{\kappa_1 = \kappa_2 = 0.5 \, MHz}$. To avoid cross-population between the two reservoirs the detuning $\mathrm{\Delta = \omega_{r_1} - \omega_{r_2}}$ must be much higher than the spectral linewidth of the two reservoirs $\mathrm{(\Delta \gg \kappa)}$. In this case, we chose a detuning of $\mathrm{\Delta/2\pi \ge 5 \, MHz}$, that allows to individually address and control the steady-state population of each resonator with the proposed scheme.\\
In Fig. \ref{fig:multi_res}(a), we show the behavior of the system, where by setting the photon source on resonance with one of the cavities, it is possible to control their population without populating also the other resonator.
For small detuning, the effective coupling between the source and the reservoir has approximately the same dependence on the flux, even for a fixed value of the capacitive coupling between the reservoir and the tunable coupler. When engineered systems are characterized by high detuning between the different reservoirs on the output, the respective capacitive links to the tunable coupler must be adequately designed. To optimize the device, the couplings must be chosen to preserve the shape of $\mathrm{g_{eff}(\Phi_c)}$,  because the effective coupling depends on the resonance frequency of the target reservoir $\mathrm{\omega_t}$. We thus need to take into account the variation of this frequency by changing the coupling between the central coupler and the reservoir under consideration. We use the same setup employed before, taking a reference reservoir, with a resonance frequency $\mathrm{\omega_{r_1}}$ and a capacity $\mathrm{C_{r_1c}}$ linking the reservoir to the central coupler. We consider a second reservoir with a detuned resonance frequency of $\mathrm{\omega_{r_2} = \omega_{r_1} + \Delta}$ and with a coupling capacitor $\mathrm{C_{r_2c} = \alpha C_{r_1c}}$, as depicted in Fig. \ref{fig:setup}(a). We assume that the self-capacity of the reservoir does not change with its resonance frequency. If there is a change it will result in an additional factor on the expression of the coupling strength that can be absorbed by the factor $\alpha$, allowing the same type of analysis. Through a sweep on $\mathrm{\Delta}$ and $\mathrm{\alpha}$, we study how the shape of $\mathrm{g_{eff}(\Phi_c)}$ changes by analyzing the Hausdorff distance relative to the reference curve obtained for the reference reservoir. The results of this analysis are shown in Fig. \ref{fig:multi_res}(b), where we can clearly distinguish a blue region corresponding to the optimal values of the coupling capacity. For small values of the detuning $\mathrm{\Delta}$, the ideal coupling correction $\mathrm{\alpha}$ is near unity and, in general, follows the shape $\mathrm{\alpha=\sqrt{\omega_r/(\omega_r + \Delta)}}$. 

We have proposed a simple and versatile scheme for an on-demand coherent photon source on-chip. The tunability inherent to this scheme enables precise control over the average of the steady-state photon distribution generated in a target reservoir while also facilitating the addressing of different cavities capacitively linked to the coupler, each characterized by distinct resonance frequencies. 
Initially, we consider the case of a single resonator, where after identifying a set of parameters for a practical design we have analyzed the system performance for various fluxes applied across the coupler, demonstrating the capability of the proposed setup to control the photon distribution generated in a target reservoir. Subsequently, we extend our analysis to a setup comprising an additional reservoir, showcasing the capability to individually address different reservoirs at different frequencies by adequately designing the detunings and the couplings in the system. 
The device acts as a tunable and finely controllable cryogenic micromaser, which can be an important tool in many quantum architectures based on bosonic quantum systems, \cite{cai2021bosonic,ofek2016extending, sivak2023real}, like continuous variable quantum computing \cite{gu2009quantum}, quantum memories \cite{naik2017random} and quantum metrology \cite{wang2019heisenberg}. The fine degree of control over the emission also enables the implementation of this device to flip the state of superconducting qubits \cite{simon2018theory,yan2021low} or to act as a master clock for microwave operations at the cryogenic stage \cite{ball2016role}. Besides providing an external knob to control the emission from the photon source, the circuit defining the tunable coupler can be designed to enhance the emission linewidth \cite{wiseman1999light,Liu_2021}. While linewidth optimization via coupling engineering falls beyond the scope of this work, it represents a promising direction for future studies, focused on engineering the coupling between the atom and the reservoir by incorporating non-linear elements \cite{wiseman1999light,Liu_2021}.
Moreover, the demultiplexing capability of the proposed scheme offers the advantage of decoupling multiple reservoirs from direct connections to room-temperature devices, leading to significant enhancements in terms of scalability. The scheme is also well-suited for 3D integrated systems, further expanding its domain of potential applications.

\section{Methods}

\subsection{Hamiltonian}
\label{sec:hamiltonian_appendix}
The photon source scheme in the most general case, with N resonators in output, can be described by the following Hamiltonian:

\begin{equation}
\begin{aligned}
\label{eq:full_hamiltonian_n_res}
\hat{H}_N/\hbar = &\left(\hat{H}_s + \hat{H}_c + \hat{H}_{sc}\right)/\hbar \\
&+ \sum_{n=1}^{N} (\hat{H}_{r_n c} + \hat{H}_{r_n} + \hat{H}_{sr_n}) \\
&+ \sum_{n \neq m = 1}^{N} \hat{H}_{r_n r_m} \, .
\end{aligned}
\end{equation}
Assuming that the coupler is in the transmon regime and considering just the two-level transitions on resonance with the reservoir for the photon source and the first transition level of the coupler, the Hamiltonian terms can be rewritten as: 
\begin{equation}
\begin{aligned}
\hat{H}_{s,c} = \hbar\frac{\omega_{s,c} \sigma_{s,c}^z }{2} \, , 
\end{aligned}
\end{equation}
\begin{equation}
\begin{aligned}
\hat{H}_{r_n} = \hbar \omega_{r} \hat{a}_n^+ \hat{a}_n \, ,
\end{aligned}
\end{equation}
\begin{equation}
\begin{aligned}
\hat{H}_{r_n c} = \hbar g_{nc}(\hat{a}_n \hat{\sigma}_c^+ + \hat{a}_n^+ \hat{\sigma}_c^-) \, ,
\end{aligned}
\end{equation}
\begin{equation}
\begin{aligned}
\hat{H}_{sc} = \hbar g_{sc}(\hat{\sigma}_s \hat{\sigma}_c^+ + \hat{\sigma}_s \hat{\sigma}_c^-) \, ,
\end{aligned}
\end{equation}
\begin{equation}
\begin{aligned}
\hat{H}_{r_nr_m} = \hbar g_{nm}(\hat{a}_n \hat{a}_m^+ + \hat{a}_n^+ \hat{a}_m) \, .
\end{aligned}
\end{equation}
\subsection{Schrieffer-Wolff transformation}
\label{sec:schrieffer_wolff_transformation}
To decouple the coupler from the other elements of the scheme we can use a Schrieffer-Wolff transformation. Before introducing the transformation for the general case we first consider the situation with a single reservoir. The case with a larger fan-out will be a simple generalization of the single reservoir case.

\subsubsection{Single reservoir}

The Schrieffer-Wolff transformation that we used in the single reservoir case is: 
\begin{equation}
\begin{aligned}
\hat{U} = \exp\Biggl[&\frac{g_{sc}}{\Delta_{sc}}(\hat{\sigma}_s^+\hat{\sigma}_c^- - \hat{\sigma}_s^-\hat{\sigma}_c^+) \\
&+ \frac{g_{rc}}{\Delta_{rc}}(\hat{a}^+\hat{\sigma}_c^- - \hat{a}\hat{\sigma}_c^+)\Biggr] \, ,
\end{aligned}
\end{equation}
where $\Delta_{sc} = \omega_s - \omega_c$ and $\Delta_{rc} = \omega_r - \omega_c$ are respectively the detuning between the coupler with the main transition of the photon source and the reservoir. By applying this transformation to the single reservoir Hamiltonian we pass from a three-body system to the following two-body system Hamiltonian: 
\begin{equation}
\begin{aligned}
\frac{\hat{U}\hat{H_1}\hat{U}\dagger}{\hbar} =  \tilde{\omega}_s \hat{\sigma}_s^z + \tilde{\omega}_r a^+ a + \tilde{g}_{sr}(\hat{a}\hat{\sigma}_c^+ + \hat{a}^+\hat{\sigma}_c^-)  \, .
\end{aligned}
\end{equation}
The characteristic frequencies of the system are renormalized: 
\begin{equation}
\begin{aligned}
\tilde{\omega}_s = \omega_s + g_{sc}^2/\Delta_{sc} \, ,
\end{aligned}
\end{equation}
\begin{equation}
\begin{aligned}
\tilde{\omega}_r = \omega_r + g_{rc}^2/\Delta_{rc} \, ,
\end{aligned}
\end{equation}
\begin{equation}
\begin{aligned}
\tilde{g}_{sr} = \omega_r + g_{rc}^2/\Delta_{rc} \, .
\end{aligned}
\end{equation}
For this calculation, we have assumed that the coupler is always at the ground state. 
\subsubsection{Multiple reservoirs}
To decouple the coupler from the source and the resonator on the fan-out of the scheme we can use a Schrieffer-Wolff transformation which is defined by the sum of various single resonator Schrieffer-Wolff transformations. We use the following unitary operator: 
\begin{equation}
\begin{aligned}
\hat{U} &= \exp\left[\frac{g_{sc}}{\Delta_{sc}}(\hat{\sigma}_s^\dagger \hat{\sigma}_c^- - \hat{\sigma}_s^- \hat{\sigma}_c^\dagger) \right. \\
&\quad + \left. \sum_{n=1}^{N} \frac{g_{nc}}{\Delta_{nc}}(\hat{a}_n^\dagger \hat{\sigma}_c - \hat{a}_n \hat{\sigma}_c^\dagger)\right] \, .
\end{aligned}
\end{equation}
By applying this transformation to the Hamiltonian in Eq. \ref{eq:full_hamiltonian_n_res} and keeping the terms up to the second order the Hamiltonian is rewritten as follows:
\begin{equation}
\begin{aligned}
\label{eq:full_hamiltonian_disp_n_res}
\frac{\hat{U}\hat{H_N}\hat{U^\dagger}}{\hbar} &= \frac{\tilde{\omega}_s \hat{\sigma}_s^z}{2} + \sum_{n=1}^{N} \left[\tilde{\omega}_n \hat{a}_n^\dagger \hat{a}_n + \tilde{g}_{sn}(\hat{a}_n^\dagger\hat{\sigma}_s^- + \hat{a}_n\hat{\sigma}_s^+)\right] \\
&\quad + \sum_{n \neq m}^{N} \tilde{g}_{nm}(\hat{a}_n^\dagger \hat{a}_m + \hat{a}_n\hat{a}_m^\dagger) \, .
\end{aligned}
\end{equation}
As before the effect of the coupler on the whole system is to renormalize the characteristic frequencies of the system as follows: 
\begin{equation}
\begin{aligned}
\tilde{\omega}_s = \omega_s + g_{sc}^2/\Delta_s \, ,
\end{aligned}
\end{equation}
\begin{equation}
\begin{aligned}
\tilde{\omega}_n = \omega_n + g_{nc}^2/\Delta_n \, ,
\end{aligned}
\end{equation}
\begin{equation}
\begin{aligned}
\tilde{g}_{sn} = \frac{g_{sc}g_{nc}}{2}(\frac{1}{\Delta_{nc}} + \frac{1}{\Delta_{sc}}) + g_{ns} \, ,
\end{aligned}
\end{equation}
\begin{equation}
\begin{aligned}
\tilde{g}_{nm} = \frac{g_{nc}g_{mc}}{2}(\frac{1}{\Delta_{nc}} + \frac{1}{\Delta_{mc}}) + g_{nm} \, .
\end{aligned}
\end{equation}
It turns out that the coupler also mediates the interaction between the various pairs of resonators but with respect to the interaction between the target reservoir and the photon source, it is an off-resonance interaction that does not lead to any cross-population phenomena if the reservoirs are adequately detuned between each other.

\subsection{From Poissonian to super-Poissonian}
\label{sec:poissonian_to_super_poissonian}
The tunability of the scheme on the average of the photon number distribution is achieved by passing from a deep masing regime to a thermal regime. This continuous transition, controlled by the flux $\mathrm{\Phi_c}$ threading the tunable coupler, is accompanied by a change in the photon distribution represented by a variation of the second-order correlation function $g^{(2)} (0)$.
In the deep masing regime, where $\mathrm{\Gamma \kappa / 4 g_{eff}^2 \ll 1}$, the second-order correlation is equal to $\mathrm{g^{(2)} (0) = 1}$, indicative of a coherent photon source \cite{loudon2000quantum}. In this regime, the generated photon distribution follows a Poissonian profile, as expected for an ideal coherent source. As the masing ratio increases and approaches the value of $\mathrm{\Gamma \kappa / 4 g_{eff}^2}=1$, the system transitions into a partially suppressed masing regime, characterized by $\mathrm{g^{(2)}(0) > 1}$. In these intermediate working points, the photon distribution deviates from Poissonian statistics and exhibits a super-Poissonian profile with increased variance. At higher values of the ratio $\mathrm{\Gamma \kappa / 4 g_{eff}^2 \rightarrow \infty}$, the source gets closer to a thermal regime characterized by a second-order correlation function $\mathrm{g^{(2)}(0)=2}$ \cite{loudon2000quantum}.
To quantitatively assess how the photon distribution changes for different average photon numbers $\mathrm{n_{ph}}$, we consider the distributions generated by the circuital scheme described in the main text. We compute the variance of the photon distributions generated at different flux $\mathrm{\Phi_c}$, resulting in a $\mathrm{g_{eff}}$ that leads to a masing ratio that goes from $\mathrm{\Gamma \kappa / 4 g_{eff}^2 = 0.1}$ to $\mathrm{\Gamma \kappa / 4 g_{eff}^2 = 2}$, as reported in Fig. \ref{fig:setup}(c). The results, displayed in Fig. \ref{fig:variance}, before reaching the maximum power emission from the device, the variance of the photon distribution exceeds the variance of a Poissonian profile, confirming the super-Poissonian regime
\begin{figure}[t]
\label{fig:variance}
\includegraphics[width=0.94\columnwidth]{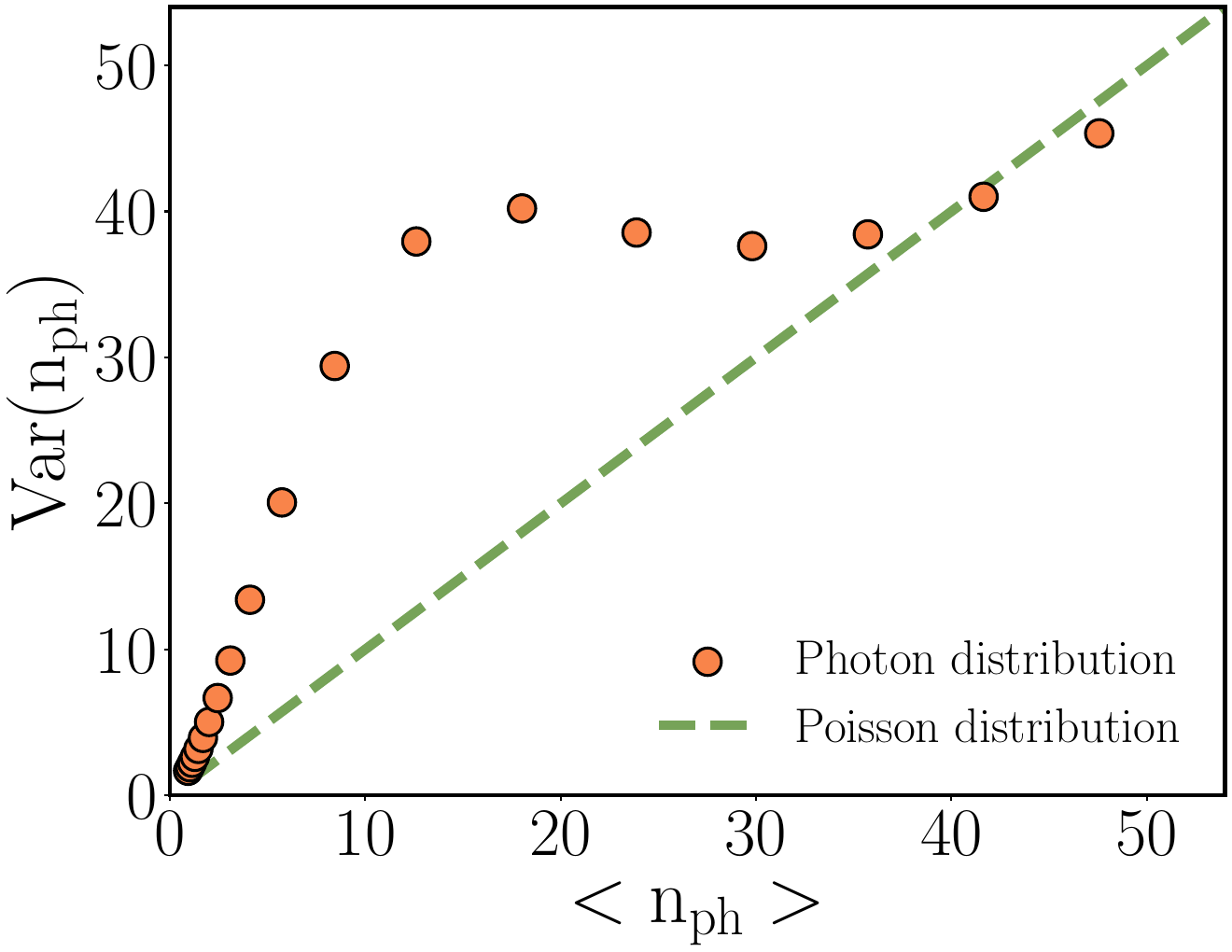}
\caption{ \justifying \textbf{Photon statistics analysis} Variance of the photon distribution generated by the circuit described in Sec. Control through tunable coupling for various average photon numbers (orange scatter plot), compared with the variance of a Poissonian distribution with the same average (green dashed line). The photon distributions exhibit a super-Poissonian variance exceeding that of a Poissonian distribution for average photon numbers $\mathrm{\braket{n} > 5}$, characteristic of a thermal regime, according to Fig. \ref{fig:setup}(c). As the average photon number approaches the maximum value $\mathrm{\braket{n_{ph}} \sim \Gamma / 2\kappa}$, the distributions acquire a profile closer to a Poissonian shape, demonstrating that the device functions as a coherent photon source in the deep masing regime. The results are obtained by simulating the steady-state dynamics of the system described in Eq. \ref{eq:dispersive_hamiltonian} for different values of the effective coupling. In the simulation, the dimension of the Hilbert subspace associated with the reservoir is equal to $\mathrm{N=100}$.} 
\label{fig:variance}
\end{figure}
in the suppressed masing regime. As the average photon number approaches the maximum value of $\mathrm{n_{ph}^{max}=\Gamma/2\kappa=50}$, the shape of the distribution also approaches a Poissonian profile, indicating that the source acts as a coherent photon source.

\subsection{Functional dependence on the number of photons}
\label{sec:large_photon_number}

The scheme proposed in the manuscript introduces stable control over photon emission into a reservoir on resonance with a photon source defined by a single atom maser. Photon injection occurs via the interaction between the state $\mathrm{\ket{e,0,n_{ph}}}$ and $\mathrm{\ket{g,0,n_{ph}+1}}$ in the spectrum of the system. The coupling strength between these states arises from a combination of virtual coupling via the intermediate state $\mathrm{\ket{g,1,n_{ph}}}$ and a direct nearest-neighbor interaction. The position of eigenstates within the system spectrum depends on the frequencies and anharmonicities of the three elements in the scheme, as well as the photon number inside the reservoir.
The photons $\mathrm{n_{ph}}$ within the reservoir influence the relative distance between the eigenstates and thus impact also the coupling between them. It is thus critical to study how the coupling strength changes throughout the population of the reservoir. 
The eignestates $\mathrm{\ket{e,0,n_{ph}}}$ and $\mathrm{\ket{g,0,n_{ph}+1}}$ directly interacting with a coupling strength $\mathrm{g_{sr}}$ form a pair of dressed states with eigenenergies:
\begin{equation}
\begin{aligned}
\label{eq:first_pair_dress_state}
\mathrm{E_{\ket{g,0,n_{ph}+1}}} &=  \hbar \left(n_{ph}+1\right) \omega_r - \frac{\hbar}{2} \sqrt{\Delta_{sr}^2 + 4 g_{sr}^2 \left(n_{ph}+1\right)} \, , \\
\mathrm{E_{\ket{e,0,n_{ph}}}} &=  \hbar \left(n_{ph}+1\right) \omega_r + \frac{\hbar}{2} \sqrt{\Delta_{sr}^2 + 4 g_{sr}^2 \left(n_{ph}+1\right)} \, ,
\end{aligned}
\end{equation}
where $\mathrm{\Delta_{sr} = \omega_r - \omega_s}$ is the detuning between the source and the reservoir. These eigenstates virtually interact through the coupling with the eigenstate $\mathrm{\ket{g,1,n_{ph}}}$ that with $\mathrm{\ket{e,1,n_{ph}-1}}$ forms a pair of dressed state with eigenenergies:

\begin{equation}
\begin{aligned}
\label{eq:second_pair_dress_state}
\mathrm{E_{\ket{g,1,n_{ph}}}} &= \hbar n_{ph} \omega_r + \hbar \omega_c - \frac{\hbar}{2} \sqrt{\Delta_{sr}^2 + 4 g_{sr}^2 n_{ph}} \, , \\
\mathrm{E_{\ket{e,1,n_{ph}-1}}} &= \hbar n_{ph} \omega_r + \hbar \omega_c + \frac{\hbar}{2} \sqrt{\Delta_{sr}^2 + 4 g_{sr}^2 n_{ph}} \, ,
\end{aligned}
\end{equation}

where $\mathrm{\omega_c}$ is the frequency of the tunable coupler. When the source is on resonance with the reservoir $\mathrm{\left(\Delta_{sr} = 0\right) \leftrightarrow \omega = \omega_r = \omega_s}$ the energy separations between the first pair of dressed states in Eqs. \ref{eq:first_pair_dress_state} and the second pair of dressed states in Eqs. \ref{eq:second_pair_dress_state} are defined:

\begin{equation}
\begin{aligned}
\label{eq:energy_distance_eigenenergies}
\mathrm{\delta E_{g,g/e}^{(1)}} &= \hbar \Delta - \hbar g_{sr} \left( \sqrt{n_{ph}+1} \mp \sqrt{n_{ph}} \right) \, , \\
\mathrm{\delta E_{e,g/e}^{(1)}} &= \hbar \Delta + \hbar g_{sr} \left( \sqrt{n_{ph}+1} \pm \sqrt{n_{ph}} \right) \, ,
\end{aligned}
\end{equation}

where $\mathrm{\Delta = \omega - \omega_c}$. Here, $\mathrm{\delta E_{i,j}^{(1)}}$ for $\mathrm{i,j = g,e}$ represent the energy differences between the eigenstate with the tunable coupler in the ground state while the source is in the $\mathrm{\ket{i}}$ state and the eigenstate with the tunable coupler in the excited state while the source in the $\mathrm{\ket{j}}$ state. These expressions indicate that the distances between eigenstates in the spectrum depend differently on the photon number $\mathrm{n_{ph}}$. In particular, for the diagonal terms $\mathrm{\delta E_{i,i}}$ (with $\mathrm{i=g,e}$) the splitting decreases with the number of photons, while in the off-diagonal terms $\mathrm{\Delta E_{i,j}}$ (with $\mathrm{i \neq j = g,e}$) it increases as the reservoir is more populated.
To evaluate the impact of this splitting, we consider the circuit configuration described in the main text and calculate the number of photons $\mathrm{n_{ph}}$ required for the splitting to become comparable to the detuning $\mathrm{\Delta}$. The results, displayed in Fig. \ref{fig:n_ph_dep},
\begin{figure}[t]
\label{fig:n_ph_dep}
\includegraphics[width=0.94\columnwidth]{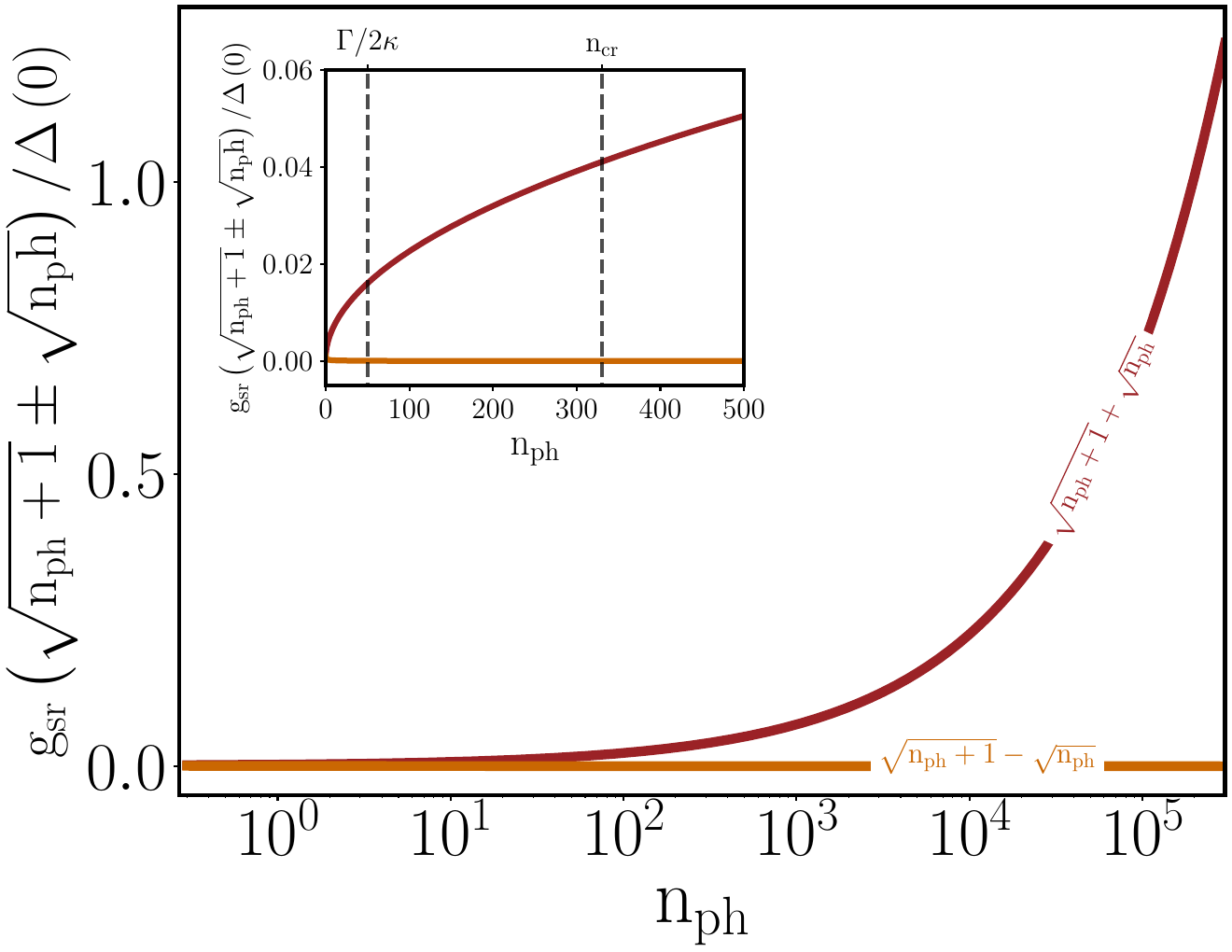}
\caption{ \justifying \textbf{Dressed-state splitting versus photon number} Ratio of the dressed-state splitting, $\mathrm{g_{sr}\left(\sqrt{n_{ph}+1}\pm\sqrt{n_{ph}}\right)}$, induced by the degeneracy between coupled eigenstates, to the detuning between the reservoir/source and the coupler at zero flux, $\mathrm{\Delta(0)}$, as a function of the average photon number $\mathrm{n_{ph}}$. The orange and red curves represent the ratio for different dependencies on the photon number in the reservoir. The inset highlights the ratio within the working range of the scheme, $\mathrm{n_{ph}<\Gamma/2\kappa}$, as described in Sec. Control through tunable coupling.}
\label{fig:n_ph_dep}
\end{figure}
demonstrate that in the proposed scheme, using typical circuit parameters of circuit quantum electrodynamics, the number of photons must reach values well above the critical number of photons imposed by the dispersive regime to become comparable with $\mathrm{\Delta}$. In particular, within the accessible power emission $\mathrm{n_{ph} \leq \Gamma / 2\kappa}$, the splitting caused by the degeneracy introduces a maximum correction which is less than the $\mathrm{2 \%}$ of $\mathrm{\Delta}$ to the relative distance between the eigenstates. Consequently, it is fair to assume that the relative distances between energy levels are independent of the number of photons in the reservoir. In the context of the scheme, this means that the effective coupling $\mathrm{g_{eff}\left(\Phi_c\right)}$ for a fixed $\mathrm{\Phi_c}$ remains approximately unaltered during the whole injection process. 

We follow the same approach to evaluate the influence of the eigenstates originating from the high energy levels of the tunable coupler. In particular, we consider the influence of the eigenstate $\mathrm{\ket{g,2,n_{ph}-1}}$, which is the closest to the eigenstate $\mathrm{\ket{g,1,n_{ph}}}$ mediating the virtual interaction between the source and the reservoir. The state $\mathrm{\ket{g,2,n_{ph}-1}}$ is degenerate with $\mathrm{\ket{e,2,n_{ph}-2}}$, forming a dressed pair with eigenenergies:

\begin{equation}
\begin{aligned}
\label{eq:energy_high_coupler_eigenstate}
\mathrm{E_{\ket{g,2,n_{ph}-1}}} &= \hbar \left( n_{ph} - 1 \right) \omega_r - \hbar g_{sr} \sqrt{n_{ph} - 1} + \\ &+ \hbar \left( 2\omega_c + \alpha_c \right) \, , \\
\mathrm{E_{\ket{e,2,n_{ph}-2}}} &= \hbar \left( n_{ph} - 1 \right) \omega_r + \hbar g_{sr} \sqrt{n_{ph} - 1} + \\ &+ \hbar \left( 2\omega_c + \alpha_c \right) \, ,
\end{aligned}
\end{equation}
where $\mathrm{\alpha_c}$ is the anharmonicity of the coupler. The energy separations between these states and those in Eqs. \ref{eq:first_pair_dress_state} are given by:

\begin{equation}
\begin{aligned}
\label{eq:second_energy_distance_eigenenergies}
\mathrm{\delta E_{g,g/e}^{(2)}} &= 2\hbar \left( \Delta - \alpha_c/2 \right) - \hbar g_{sr} \left( \sqrt{n_{ph}+1} \mp \sqrt{n_{ph}} \right) \, , \\
\mathrm{\delta E_{e,g/e}^{(2)}} &= 2\hbar \left( \Delta - \alpha_c/2 \right) + \hbar g_{sr} \left( \sqrt{n_{ph}+1} \pm \sqrt{n_{ph}} \right) \, .
\end{aligned}
\end{equation}

The dependence over the number of photons $\mathrm{n_{ph}}$ is the same as the energy distances in Eqs. \ref{eq:energy_distance_eigenenergies}. Thus, the effect of the higher energy state on the masing dynamics remains the same within the whole emission range of the system. To evaluate the influence of these states on the injection dynamics we consider the distance between the states in Eqs. \ref{eq:second_energy_distance_eigenenergies} mediating the virtual interaction and the higher order states in Eqs. \ref{eq:energy_high_coupler_eigenstate} which results in:
\begin{equation}
\begin{aligned}
\label{eq:energy_distance_coupler_eigenstates}
\mathrm{\delta E ^{(2)} - \delta E^{(1)} \approx \Delta - \alpha_c} \, . \\
\end{aligned}
\end{equation}
As $\mathrm{\Delta > 0}$ and $\mathrm{\alpha_c < 0}$ in this scheme, the effect of higher energy state coupler states can be mitigated by increasing $\mathrm{\Delta}$ or the anharmonicity $\mathrm{\alpha_c}$ of the tunable coupler.

\section{Data availability}
All the generated data supporting this study can be reproduced using the publicly available simulation codes. 

\section{Code availability}
The source codes for the numerical simulations presented in the paper are available at \url{https://github.com/Pask97/on-chip-photon-source}

\section{Acknowledgments}
The authors thank F. Nori for the fruitful discussion and would also like to thank G. Di Bello and M. Vizzuso for their help in the work development. The work was supported by the  Pathfinder EIC 2023 project “FERROMON-Ferrotransmons and Ferrogatemons for Scalable Superconducting Quantum Computers", by the project “Superconducting quantum-classical linked computing systems (SuperLink)”, in the frame of QuantERA2 ERANET COFUND in Quantum Technologies, and by the Project PRIN 2022-Advanced Control and Readout of scalable Superconducting NISQ Architectures (SuperNISQ)-CUP E53D23001910006. The research activities were also supported by the PNRR MUR project PE0000023-NQSTI and the PNRR MUR project CN\textunderscore00000013—ICSC.

\section{Author Contributions}

P.M. conceived and developed the theoretical framework, performed the numerical simulations and wrote the manuscript. H.G.A., A.P., M.E., D.M. and F.T. contributed to the physical interpretation of the numerical results and to the writing of the manuscript.  

\section{Competing Interests}
The authors declare no competing interests.

\bibliography{bibl}

\providecommand{\noopsort}[1]{}\providecommand{\singleletter}[1]{#1}%
\begin{thebibliography}{82}%
\makeatletter
\providecommand \@ifxundefined [1]{%
 \@ifx{#1\undefined}
}%
\providecommand \@ifnum [1]{%
 \ifnum #1\expandafter \@firstoftwo
 \else \expandafter \@secondoftwo
 \fi
}%
\providecommand \@ifx [1]{%
 \ifx #1\expandafter \@firstoftwo
 \else \expandafter \@secondoftwo
 \fi
}%
\providecommand \natexlab [1]{#1}%
\providecommand \enquote  [1]{``#1''}%
\providecommand \bibnamefont  [1]{#1}%
\providecommand \bibfnamefont [1]{#1}%
\providecommand \citenamefont [1]{#1}%
\providecommand \href@noop [0]{\@secondoftwo}%
\providecommand \href [0]{\begingroup \@sanitize@url \@href}%
\providecommand \@href[1]{\@@startlink{#1}\@@href}%
\providecommand \@@href[1]{\endgroup#1\@@endlink}%
\providecommand \@sanitize@url [0]{\catcode `\\12\catcode `\$12\catcode `\&12\catcode `\#12\catcode `\^12\catcode `\_12\catcode `\%12\relax}%
\providecommand \@@startlink[1]{}%
\providecommand \@@endlink[0]{}%
\providecommand \url  [0]{\begingroup\@sanitize@url \@url }%
\providecommand \@url [1]{\endgroup\@href {#1}{\urlprefix }}%
\providecommand \urlprefix  [0]{URL }%
\providecommand \Eprint [0]{\href }%
\providecommand \doibase [0]{https://doi.org/}%
\providecommand \selectlanguage [0]{\@gobble}%
\providecommand \bibinfo  [0]{\@secondoftwo}%
\providecommand \bibfield  [0]{\@secondoftwo}%
\providecommand \translation [1]{[#1]}%
\providecommand \BibitemOpen [0]{}%
\providecommand \bibitemStop [0]{}%
\providecommand \bibitemNoStop [0]{.\EOS\space}%
\providecommand \EOS [0]{\spacefactor3000\relax}%
\providecommand \BibitemShut  [1]{\csname bibitem#1\endcsname}%
\let\auto@bib@innerbib\@empty
\bibitem [{\citenamefont {Preskill}(2018)}]{preskill2018quantum}%
  \BibitemOpen
  \bibfield  {author} {\bibinfo {author} {\bibfnamefont {J.}~\bibnamefont {Preskill}},\ }\bibfield  {title} {\bibinfo {title} {Quantum computing in the nisq era and beyond},\ }\href@noop {} {\bibfield  {journal} {\bibinfo  {journal} {Quantum}\ }\textbf {\bibinfo {volume} {2}},\ \bibinfo {pages} {79} (\bibinfo {year} {2018})}\BibitemShut {NoStop}%
\bibitem [{\citenamefont {Saffman}(2019)}]{saffman2019quantum}%
  \BibitemOpen
  \bibfield  {author} {\bibinfo {author} {\bibfnamefont {M.}~\bibnamefont {Saffman}},\ }\bibfield  {title} {\bibinfo {title} {Quantum computing with neutral atoms},\ }\href@noop {} {\bibfield  {journal} {\bibinfo  {journal} {National Science Review}\ }\textbf {\bibinfo {volume} {6}},\ \bibinfo {pages} {24} (\bibinfo {year} {2019})}\BibitemShut {NoStop}%
\bibitem [{cir(1995)}]{cirac1995quantum}%
  \BibitemOpen
  \bibfield  {title} {\bibinfo {title} {Quantum computations with cold trapped ions},\ }\href@noop {} {\bibfield  {journal} {\bibinfo  {journal} {Physical review letters}\ }\textbf {\bibinfo {volume} {74}},\ \bibinfo {pages} {4091} (\bibinfo {year} {1995})}\BibitemShut {NoStop}%
\bibitem [{\citenamefont {Gottesman}\ \emph {et~al.}(2001)\citenamefont {Gottesman}, \citenamefont {Kitaev},\ and\ \citenamefont {Preskill}}]{gottesman2001encoding}%
  \BibitemOpen
  \bibfield  {author} {\bibinfo {author} {\bibfnamefont {D.}~\bibnamefont {Gottesman}}, \bibinfo {author} {\bibfnamefont {A.}~\bibnamefont {Kitaev}},\ and\ \bibinfo {author} {\bibfnamefont {J.}~\bibnamefont {Preskill}},\ }\bibfield  {title} {\bibinfo {title} {Encoding a qubit in an oscillator},\ }\href@noop {} {\bibfield  {journal} {\bibinfo  {journal} {Physical Review A}\ }\textbf {\bibinfo {volume} {64}},\ \bibinfo {pages} {012310} (\bibinfo {year} {2001})}\BibitemShut {NoStop}%
\bibitem [{\citenamefont {Gershenfeld}\ and\ \citenamefont {Chuang}(1998)}]{gershenfeld1998quantum}%
  \BibitemOpen
  \bibfield  {author} {\bibinfo {author} {\bibfnamefont {N.}~\bibnamefont {Gershenfeld}}\ and\ \bibinfo {author} {\bibfnamefont {I.~L.}\ \bibnamefont {Chuang}},\ }\bibfield  {title} {\bibinfo {title} {Quantum computing with molecules},\ }\href@noop {} {\bibfield  {journal} {\bibinfo  {journal} {Scientific American}\ }\textbf {\bibinfo {volume} {278}},\ \bibinfo {pages} {66} (\bibinfo {year} {1998})}\BibitemShut {NoStop}%
\bibitem [{\citenamefont {Houck}\ \emph {et~al.}(2007)\citenamefont {Houck}, \citenamefont {Schuster}, \citenamefont {Gambetta}, \citenamefont {Schreier}, \citenamefont {Johnson}, \citenamefont {Chow}, \citenamefont {Frunzio}, \citenamefont {Majer}, \citenamefont {Devoret}, \citenamefont {Girvin} \emph {et~al.}}]{houck2007generating}%
  \BibitemOpen
  \bibfield  {author} {\bibinfo {author} {\bibfnamefont {A.~A.}\ \bibnamefont {Houck}}, \bibinfo {author} {\bibfnamefont {D.}~\bibnamefont {Schuster}}, \bibinfo {author} {\bibfnamefont {J.}~\bibnamefont {Gambetta}}, \bibinfo {author} {\bibfnamefont {J.}~\bibnamefont {Schreier}}, \bibinfo {author} {\bibfnamefont {B.}~\bibnamefont {Johnson}}, \bibinfo {author} {\bibfnamefont {J.}~\bibnamefont {Chow}}, \bibinfo {author} {\bibfnamefont {L.}~\bibnamefont {Frunzio}}, \bibinfo {author} {\bibfnamefont {J.}~\bibnamefont {Majer}}, \bibinfo {author} {\bibfnamefont {M.}~\bibnamefont {Devoret}}, \bibinfo {author} {\bibfnamefont {S.}~\bibnamefont {Girvin}}, \emph {et~al.},\ }\bibfield  {title} {\bibinfo {title} {Generating single microwave photons in a circuit},\ }\href@noop {} {\bibfield  {journal} {\bibinfo  {journal} {Nature}\ }\textbf {\bibinfo {volume} {449}},\ \bibinfo {pages} {328} (\bibinfo {year} {2007})}\BibitemShut {NoStop}%
\bibitem [{\citenamefont {Hofheinz}\ \emph {et~al.}(2009)\citenamefont {Hofheinz}, \citenamefont {Wang}, \citenamefont {Ansmann}, \citenamefont {Bialczak}, \citenamefont {Lucero}, \citenamefont {Neeley}, \citenamefont {O'connell}, \citenamefont {Sank}, \citenamefont {Wenner}, \citenamefont {Martinis} \emph {et~al.}}]{hofheinz2009synthesizing}%
  \BibitemOpen
  \bibfield  {author} {\bibinfo {author} {\bibfnamefont {M.}~\bibnamefont {Hofheinz}}, \bibinfo {author} {\bibfnamefont {H.}~\bibnamefont {Wang}}, \bibinfo {author} {\bibfnamefont {M.}~\bibnamefont {Ansmann}}, \bibinfo {author} {\bibfnamefont {R.~C.}\ \bibnamefont {Bialczak}}, \bibinfo {author} {\bibfnamefont {E.}~\bibnamefont {Lucero}}, \bibinfo {author} {\bibfnamefont {M.}~\bibnamefont {Neeley}}, \bibinfo {author} {\bibfnamefont {A.}~\bibnamefont {O'connell}}, \bibinfo {author} {\bibfnamefont {D.}~\bibnamefont {Sank}}, \bibinfo {author} {\bibfnamefont {J.}~\bibnamefont {Wenner}}, \bibinfo {author} {\bibfnamefont {J.~M.}\ \bibnamefont {Martinis}}, \emph {et~al.},\ }\bibfield  {title} {\bibinfo {title} {Synthesizing arbitrary quantum states in a superconducting resonator},\ }\href@noop {} {\bibfield  {journal} {\bibinfo  {journal} {Nature}\ }\textbf {\bibinfo {volume} {459}},\ \bibinfo {pages} {546} (\bibinfo {year} {2009})}\BibitemShut {NoStop}%
\bibitem [{\citenamefont {Clerk}\ \emph {et~al.}(2010)\citenamefont {Clerk}, \citenamefont {Devoret}, \citenamefont {Girvin}, \citenamefont {Marquardt},\ and\ \citenamefont {Schoelkopf}}]{clerk2010introduction}%
  \BibitemOpen
  \bibfield  {author} {\bibinfo {author} {\bibfnamefont {A.~A.}\ \bibnamefont {Clerk}}, \bibinfo {author} {\bibfnamefont {M.~H.}\ \bibnamefont {Devoret}}, \bibinfo {author} {\bibfnamefont {S.~M.}\ \bibnamefont {Girvin}}, \bibinfo {author} {\bibfnamefont {F.}~\bibnamefont {Marquardt}},\ and\ \bibinfo {author} {\bibfnamefont {R.~J.}\ \bibnamefont {Schoelkopf}},\ }\bibfield  {title} {\bibinfo {title} {Introduction to quantum noise, measurement, and amplification},\ }\href@noop {} {\bibfield  {journal} {\bibinfo  {journal} {Reviews of Modern Physics}\ }\textbf {\bibinfo {volume} {82}},\ \bibinfo {pages} {1155} (\bibinfo {year} {2010})}\BibitemShut {NoStop}%
\bibitem [{\citenamefont {Albert}\ \emph {et~al.}(2016)\citenamefont {Albert}, \citenamefont {Shu}, \citenamefont {Krastanov}, \citenamefont {Shen}, \citenamefont {Liu}, \citenamefont {Yang}, \citenamefont {Schoelkopf}, \citenamefont {Mirrahimi}, \citenamefont {Devoret},\ and\ \citenamefont {Jiang}}]{albert2016holonomic}%
  \BibitemOpen
  \bibfield  {author} {\bibinfo {author} {\bibfnamefont {V.~V.}\ \bibnamefont {Albert}}, \bibinfo {author} {\bibfnamefont {C.}~\bibnamefont {Shu}}, \bibinfo {author} {\bibfnamefont {S.}~\bibnamefont {Krastanov}}, \bibinfo {author} {\bibfnamefont {C.}~\bibnamefont {Shen}}, \bibinfo {author} {\bibfnamefont {R.-B.}\ \bibnamefont {Liu}}, \bibinfo {author} {\bibfnamefont {Z.-B.}\ \bibnamefont {Yang}}, \bibinfo {author} {\bibfnamefont {R.~J.}\ \bibnamefont {Schoelkopf}}, \bibinfo {author} {\bibfnamefont {M.}~\bibnamefont {Mirrahimi}}, \bibinfo {author} {\bibfnamefont {M.~H.}\ \bibnamefont {Devoret}},\ and\ \bibinfo {author} {\bibfnamefont {L.}~\bibnamefont {Jiang}},\ }\bibfield  {title} {\bibinfo {title} {Holonomic quantum control with continuous variable systems},\ }\href@noop {} {\bibfield  {journal} {\bibinfo  {journal} {Physical review letters}\ }\textbf {\bibinfo {volume} {116}},\ \bibinfo {pages} {140502} (\bibinfo {year} {2016})}\BibitemShut {NoStop}%
\bibitem [{\citenamefont {Degen}\ \emph {et~al.}(2017)\citenamefont {Degen}, \citenamefont {Reinhard},\ and\ \citenamefont {Cappellaro}}]{degen2017quantum}%
  \BibitemOpen
  \bibfield  {author} {\bibinfo {author} {\bibfnamefont {C.~L.}\ \bibnamefont {Degen}}, \bibinfo {author} {\bibfnamefont {F.}~\bibnamefont {Reinhard}},\ and\ \bibinfo {author} {\bibfnamefont {P.}~\bibnamefont {Cappellaro}},\ }\bibfield  {title} {\bibinfo {title} {Quantum sensing},\ }\href@noop {} {\bibfield  {journal} {\bibinfo  {journal} {Reviews of modern physics}\ }\textbf {\bibinfo {volume} {89}},\ \bibinfo {pages} {035002} (\bibinfo {year} {2017})}\BibitemShut {NoStop}%
\bibitem [{\citenamefont {Blais}\ \emph {et~al.}(2020)\citenamefont {Blais}, \citenamefont {Girvin},\ and\ \citenamefont {Oliver}}]{blais2020quantum}%
  \BibitemOpen
  \bibfield  {author} {\bibinfo {author} {\bibfnamefont {A.}~\bibnamefont {Blais}}, \bibinfo {author} {\bibfnamefont {S.~M.}\ \bibnamefont {Girvin}},\ and\ \bibinfo {author} {\bibfnamefont {W.~D.}\ \bibnamefont {Oliver}},\ }\bibfield  {title} {\bibinfo {title} {Quantum information processing and quantum optics with circuit quantum electrodynamics},\ }\href@noop {} {\bibfield  {journal} {\bibinfo  {journal} {Nature Physics}\ }\textbf {\bibinfo {volume} {16}},\ \bibinfo {pages} {247} (\bibinfo {year} {2020})}\BibitemShut {NoStop}%
\bibitem [{\citenamefont {Krinner}\ \emph {et~al.}(2022)\citenamefont {Krinner}, \citenamefont {Lacroix}, \citenamefont {Remm}, \citenamefont {Di~Paolo}, \citenamefont {Genois}, \citenamefont {Leroux}, \citenamefont {Hellings}, \citenamefont {Lazar}, \citenamefont {Swiadek}, \citenamefont {Herrmann} \emph {et~al.}}]{krinner2022realizing}%
  \BibitemOpen
  \bibfield  {author} {\bibinfo {author} {\bibfnamefont {S.}~\bibnamefont {Krinner}}, \bibinfo {author} {\bibfnamefont {N.}~\bibnamefont {Lacroix}}, \bibinfo {author} {\bibfnamefont {A.}~\bibnamefont {Remm}}, \bibinfo {author} {\bibfnamefont {A.}~\bibnamefont {Di~Paolo}}, \bibinfo {author} {\bibfnamefont {E.}~\bibnamefont {Genois}}, \bibinfo {author} {\bibfnamefont {C.}~\bibnamefont {Leroux}}, \bibinfo {author} {\bibfnamefont {C.}~\bibnamefont {Hellings}}, \bibinfo {author} {\bibfnamefont {S.}~\bibnamefont {Lazar}}, \bibinfo {author} {\bibfnamefont {F.}~\bibnamefont {Swiadek}}, \bibinfo {author} {\bibfnamefont {J.}~\bibnamefont {Herrmann}}, \emph {et~al.},\ }\bibfield  {title} {\bibinfo {title} {Realizing repeated quantum error correction in a distance-three surface code},\ }\href@noop {} {\bibfield  {journal} {\bibinfo  {journal} {Nature}\ }\textbf {\bibinfo {volume} {605}},\ \bibinfo {pages} {669} (\bibinfo {year} {2022})}\BibitemShut {NoStop}%
\bibitem [{\citenamefont {Arute}\ \emph {et~al.}(2019)\citenamefont {Arute}, \citenamefont {Arya}, \citenamefont {Babbush}, \citenamefont {Bacon}, \citenamefont {Bardin}, \citenamefont {Barends}, \citenamefont {Biswas}, \citenamefont {Boixo}, \citenamefont {Brandao}, \citenamefont {Buell} \emph {et~al.}}]{arute2019quantum}%
  \BibitemOpen
  \bibfield  {author} {\bibinfo {author} {\bibfnamefont {F.}~\bibnamefont {Arute}}, \bibinfo {author} {\bibfnamefont {K.}~\bibnamefont {Arya}}, \bibinfo {author} {\bibfnamefont {R.}~\bibnamefont {Babbush}}, \bibinfo {author} {\bibfnamefont {D.}~\bibnamefont {Bacon}}, \bibinfo {author} {\bibfnamefont {J.~C.}\ \bibnamefont {Bardin}}, \bibinfo {author} {\bibfnamefont {R.}~\bibnamefont {Barends}}, \bibinfo {author} {\bibfnamefont {R.}~\bibnamefont {Biswas}}, \bibinfo {author} {\bibfnamefont {S.}~\bibnamefont {Boixo}}, \bibinfo {author} {\bibfnamefont {F.~G.}\ \bibnamefont {Brandao}}, \bibinfo {author} {\bibfnamefont {D.~A.}\ \bibnamefont {Buell}}, \emph {et~al.},\ }\bibfield  {title} {\bibinfo {title} {Quantum supremacy using a programmable superconducting processor},\ }\href@noop {} {\bibfield  {journal} {\bibinfo  {journal} {Nature}\ }\textbf {\bibinfo {volume} {574}},\ \bibinfo {pages} {505} (\bibinfo {year} {2019})}\BibitemShut {NoStop}%
\bibitem [{\citenamefont {Krinner}\ \emph {et~al.}(2019)\citenamefont {Krinner}, \citenamefont {Storz}, \citenamefont {Kurpiers}, \citenamefont {Magnard}, \citenamefont {Heinsoo}, \citenamefont {Keller}, \citenamefont {Luetolf}, \citenamefont {Eichler},\ and\ \citenamefont {Wallraff}}]{krinner2019engineering}%
  \BibitemOpen
  \bibfield  {author} {\bibinfo {author} {\bibfnamefont {S.}~\bibnamefont {Krinner}}, \bibinfo {author} {\bibfnamefont {S.}~\bibnamefont {Storz}}, \bibinfo {author} {\bibfnamefont {P.}~\bibnamefont {Kurpiers}}, \bibinfo {author} {\bibfnamefont {P.}~\bibnamefont {Magnard}}, \bibinfo {author} {\bibfnamefont {J.}~\bibnamefont {Heinsoo}}, \bibinfo {author} {\bibfnamefont {R.}~\bibnamefont {Keller}}, \bibinfo {author} {\bibfnamefont {J.}~\bibnamefont {Luetolf}}, \bibinfo {author} {\bibfnamefont {C.}~\bibnamefont {Eichler}},\ and\ \bibinfo {author} {\bibfnamefont {A.}~\bibnamefont {Wallraff}},\ }\bibfield  {title} {\bibinfo {title} {Engineering cryogenic setups for 100-qubit scale superconducting circuit systems},\ }\href@noop {} {\bibfield  {journal} {\bibinfo  {journal} {EPJ Quantum Technology}\ }\textbf {\bibinfo {volume} {6}},\ \bibinfo {pages} {2} (\bibinfo {year} {2019})}\BibitemShut {NoStop}%
\bibitem [{\citenamefont {Bardin}\ \emph {et~al.}(2021)\citenamefont {Bardin}, \citenamefont {Slichter},\ and\ \citenamefont {Reilly}}]{bardin2021microwaves}%
  \BibitemOpen
  \bibfield  {author} {\bibinfo {author} {\bibfnamefont {J.~C.}\ \bibnamefont {Bardin}}, \bibinfo {author} {\bibfnamefont {D.~H.}\ \bibnamefont {Slichter}},\ and\ \bibinfo {author} {\bibfnamefont {D.~J.}\ \bibnamefont {Reilly}},\ }\bibfield  {title} {\bibinfo {title} {Microwaves in quantum computing},\ }\href@noop {} {\bibfield  {journal} {\bibinfo  {journal} {IEEE journal of microwaves}\ }\textbf {\bibinfo {volume} {1}},\ \bibinfo {pages} {403} (\bibinfo {year} {2021})}\BibitemShut {NoStop}%
\bibitem [{\citenamefont {Houck}\ \emph {et~al.}(2008)\citenamefont {Houck}, \citenamefont {Schreier}, \citenamefont {Johnson}, \citenamefont {Chow}, \citenamefont {Koch}, \citenamefont {Gambetta}, \citenamefont {Schuster}, \citenamefont {Frunzio}, \citenamefont {Devoret}, \citenamefont {Girvin} \emph {et~al.}}]{houck2008controlling}%
  \BibitemOpen
  \bibfield  {author} {\bibinfo {author} {\bibfnamefont {A.~A.}\ \bibnamefont {Houck}}, \bibinfo {author} {\bibfnamefont {J.}~\bibnamefont {Schreier}}, \bibinfo {author} {\bibfnamefont {B.}~\bibnamefont {Johnson}}, \bibinfo {author} {\bibfnamefont {J.}~\bibnamefont {Chow}}, \bibinfo {author} {\bibfnamefont {J.}~\bibnamefont {Koch}}, \bibinfo {author} {\bibfnamefont {J.}~\bibnamefont {Gambetta}}, \bibinfo {author} {\bibfnamefont {D.}~\bibnamefont {Schuster}}, \bibinfo {author} {\bibfnamefont {L.}~\bibnamefont {Frunzio}}, \bibinfo {author} {\bibfnamefont {M.}~\bibnamefont {Devoret}}, \bibinfo {author} {\bibfnamefont {S.}~\bibnamefont {Girvin}}, \emph {et~al.},\ }\bibfield  {title} {\bibinfo {title} {Controlling the spontaneous emission of a superconducting transmon qubit},\ }\href@noop {} {\bibfield  {journal} {\bibinfo  {journal} {Physical review letters}\ }\textbf {\bibinfo {volume} {101}},\ \bibinfo {pages} {080502} (\bibinfo {year} {2008})}\BibitemShut {NoStop}%
\bibitem [{\citenamefont {Reilly}(2019)}]{reilly2019challenges}%
  \BibitemOpen
  \bibfield  {author} {\bibinfo {author} {\bibfnamefont {D.}~\bibnamefont {Reilly}},\ }\bibfield  {title} {\bibinfo {title} {Challenges in scaling-up the control interface of a quantum computer},\ }in\ \href@noop {} {\emph {\bibinfo {booktitle} {2019 IEEE International Electron Devices Meeting (IEDM)}}}\ (\bibinfo {organization} {IEEE},\ \bibinfo {year} {2019})\ pp.\ \bibinfo {pages} {31--7}\BibitemShut {NoStop}%
\bibitem [{\citenamefont {McDermott}\ \emph {et~al.}(2018)\citenamefont {McDermott}, \citenamefont {Vavilov}, \citenamefont {Plourde}, \citenamefont {Wilhelm}, \citenamefont {Liebermann}, \citenamefont {Mukhanov},\ and\ \citenamefont {Ohki}}]{mcdermott2018quantum}%
  \BibitemOpen
  \bibfield  {author} {\bibinfo {author} {\bibfnamefont {R.}~\bibnamefont {McDermott}}, \bibinfo {author} {\bibfnamefont {M.~G.}\ \bibnamefont {Vavilov}}, \bibinfo {author} {\bibfnamefont {B.}~\bibnamefont {Plourde}}, \bibinfo {author} {\bibfnamefont {F.~K.}\ \bibnamefont {Wilhelm}}, \bibinfo {author} {\bibfnamefont {P.~J.}\ \bibnamefont {Liebermann}}, \bibinfo {author} {\bibfnamefont {O.~A.}\ \bibnamefont {Mukhanov}},\ and\ \bibinfo {author} {\bibfnamefont {T.~A.}\ \bibnamefont {Ohki}},\ }\bibfield  {title} {\bibinfo {title} {Quantum--classical interface based on single flux quantum digital logic},\ }\href@noop {} {\bibfield  {journal} {\bibinfo  {journal} {Quantum science and technology}\ }\textbf {\bibinfo {volume} {3}},\ \bibinfo {pages} {024004} (\bibinfo {year} {2018})}\BibitemShut {NoStop}%
\bibitem [{\citenamefont {Di~Palma}\ \emph {et~al.}(2023)\citenamefont {Di~Palma}, \citenamefont {Miano}, \citenamefont {Mastrovito}, \citenamefont {Massarotti}, \citenamefont {Arzeo}, \citenamefont {Pepe}, \citenamefont {Tafuri},\ and\ \citenamefont {Mukhanov}}]{DiPalma2023}%
  \BibitemOpen
  \bibfield  {author} {\bibinfo {author} {\bibfnamefont {L.}~\bibnamefont {Di~Palma}}, \bibinfo {author} {\bibfnamefont {A.}~\bibnamefont {Miano}}, \bibinfo {author} {\bibfnamefont {P.}~\bibnamefont {Mastrovito}}, \bibinfo {author} {\bibfnamefont {D.}~\bibnamefont {Massarotti}}, \bibinfo {author} {\bibfnamefont {M.}~\bibnamefont {Arzeo}}, \bibinfo {author} {\bibfnamefont {G.}~\bibnamefont {Pepe}}, \bibinfo {author} {\bibfnamefont {F.}~\bibnamefont {Tafuri}},\ and\ \bibinfo {author} {\bibfnamefont {O.}~\bibnamefont {Mukhanov}},\ }\bibfield  {title} {\bibinfo {title} {Discriminating the phase of a coherent tone with a flux-switchable superconducting circuit},\ }\href {https://doi.org/10.1103/PhysRevApplied.19.064025} {\bibfield  {journal} {\bibinfo  {journal} {Phys. Rev. Appl.}\ }\textbf {\bibinfo {volume} {19}},\ \bibinfo {pages} {064025} (\bibinfo {year} {2023})}\BibitemShut {NoStop}%
\bibitem [{\citenamefont {Joshi}\ and\ \citenamefont {Moazeni}(2023)}]{joshi2023scaling}%
  \BibitemOpen
  \bibfield  {author} {\bibinfo {author} {\bibfnamefont {S.}~\bibnamefont {Joshi}}\ and\ \bibinfo {author} {\bibfnamefont {S.}~\bibnamefont {Moazeni}},\ }\bibfield  {title} {\bibinfo {title} {Scaling up superconducting quantum computers with cryogenic rf-photonics},\ }\href@noop {} {\bibfield  {journal} {\bibinfo  {journal} {Journal of Lightwave Technology}\ } (\bibinfo {year} {2023})}\BibitemShut {NoStop}%
\bibitem [{\citenamefont {Yan}\ \emph {et~al.}(2021)\citenamefont {Yan}, \citenamefont {Hassel}, \citenamefont {Vesterinen}, \citenamefont {Zhang}, \citenamefont {Ikonen}, \citenamefont {Gr{\"o}nberg}, \citenamefont {Goetz},\ and\ \citenamefont {M{\"o}tt{\"o}nen}}]{yan2021low}%
  \BibitemOpen
  \bibfield  {author} {\bibinfo {author} {\bibfnamefont {C.}~\bibnamefont {Yan}}, \bibinfo {author} {\bibfnamefont {J.}~\bibnamefont {Hassel}}, \bibinfo {author} {\bibfnamefont {V.}~\bibnamefont {Vesterinen}}, \bibinfo {author} {\bibfnamefont {J.}~\bibnamefont {Zhang}}, \bibinfo {author} {\bibfnamefont {J.}~\bibnamefont {Ikonen}}, \bibinfo {author} {\bibfnamefont {L.}~\bibnamefont {Gr{\"o}nberg}}, \bibinfo {author} {\bibfnamefont {J.}~\bibnamefont {Goetz}},\ and\ \bibinfo {author} {\bibfnamefont {M.}~\bibnamefont {M{\"o}tt{\"o}nen}},\ }\bibfield  {title} {\bibinfo {title} {A low-noise on-chip coherent microwave source},\ }\href@noop {} {\bibfield  {journal} {\bibinfo  {journal} {Nature Electronics}\ }\textbf {\bibinfo {volume} {4}},\ \bibinfo {pages} {885} (\bibinfo {year} {2021})}\BibitemShut {NoStop}%
\bibitem [{\citenamefont {Leonard~Jr}\ \emph {et~al.}(2019)\citenamefont {Leonard~Jr}, \citenamefont {Beck}, \citenamefont {Nelson}, \citenamefont {Christensen}, \citenamefont {Thorbeck}, \citenamefont {Howington}, \citenamefont {Opremcak}, \citenamefont {Pechenezhskiy}, \citenamefont {Dodge}, \citenamefont {Dupuis} \emph {et~al.}}]{leonard2019digital}%
  \BibitemOpen
  \bibfield  {author} {\bibinfo {author} {\bibfnamefont {E.}~\bibnamefont {Leonard~Jr}}, \bibinfo {author} {\bibfnamefont {M.~A.}\ \bibnamefont {Beck}}, \bibinfo {author} {\bibfnamefont {J.}~\bibnamefont {Nelson}}, \bibinfo {author} {\bibfnamefont {B.~G.}\ \bibnamefont {Christensen}}, \bibinfo {author} {\bibfnamefont {T.}~\bibnamefont {Thorbeck}}, \bibinfo {author} {\bibfnamefont {C.}~\bibnamefont {Howington}}, \bibinfo {author} {\bibfnamefont {A.}~\bibnamefont {Opremcak}}, \bibinfo {author} {\bibfnamefont {I.~V.}\ \bibnamefont {Pechenezhskiy}}, \bibinfo {author} {\bibfnamefont {K.}~\bibnamefont {Dodge}}, \bibinfo {author} {\bibfnamefont {N.~P.}\ \bibnamefont {Dupuis}}, \emph {et~al.},\ }\bibfield  {title} {\bibinfo {title} {Digital coherent control of a superconducting qubit},\ }\href@noop {} {\bibfield  {journal} {\bibinfo  {journal} {Physical Review Applied}\ }\textbf {\bibinfo {volume} {11}},\ \bibinfo {pages} {014009} (\bibinfo {year} {2019})}\BibitemShut {NoStop}%
\bibitem [{\citenamefont {Xue}\ \emph {et~al.}(2021)\citenamefont {Xue}, \citenamefont {Patra}, \citenamefont {van Dijk}, \citenamefont {Samkharadze}, \citenamefont {Subramanian}, \citenamefont {Corna}, \citenamefont {Paquelet~Wuetz}, \citenamefont {Jeon}, \citenamefont {Sheikh}, \citenamefont {Juarez-Hernandez} \emph {et~al.}}]{xue2021cmos}%
  \BibitemOpen
  \bibfield  {author} {\bibinfo {author} {\bibfnamefont {X.}~\bibnamefont {Xue}}, \bibinfo {author} {\bibfnamefont {B.}~\bibnamefont {Patra}}, \bibinfo {author} {\bibfnamefont {J.~P.}\ \bibnamefont {van Dijk}}, \bibinfo {author} {\bibfnamefont {N.}~\bibnamefont {Samkharadze}}, \bibinfo {author} {\bibfnamefont {S.}~\bibnamefont {Subramanian}}, \bibinfo {author} {\bibfnamefont {A.}~\bibnamefont {Corna}}, \bibinfo {author} {\bibfnamefont {B.}~\bibnamefont {Paquelet~Wuetz}}, \bibinfo {author} {\bibfnamefont {C.}~\bibnamefont {Jeon}}, \bibinfo {author} {\bibfnamefont {F.}~\bibnamefont {Sheikh}}, \bibinfo {author} {\bibfnamefont {E.}~\bibnamefont {Juarez-Hernandez}}, \emph {et~al.},\ }\bibfield  {title} {\bibinfo {title} {Cmos-based cryogenic control of silicon quantum circuits},\ }\href@noop {} {\bibfield  {journal} {\bibinfo  {journal} {Nature}\ }\textbf {\bibinfo {volume} {593}},\ \bibinfo {pages} {205} (\bibinfo {year} {2021})}\BibitemShut {NoStop}%
\bibitem [{\citenamefont {Meschede}\ \emph {et~al.}(1985)\citenamefont {Meschede}, \citenamefont {Walther},\ and\ \citenamefont {M{\"u}ller}}]{meschede1985one}%
  \BibitemOpen
  \bibfield  {author} {\bibinfo {author} {\bibfnamefont {D.}~\bibnamefont {Meschede}}, \bibinfo {author} {\bibfnamefont {H.}~\bibnamefont {Walther}},\ and\ \bibinfo {author} {\bibfnamefont {G.}~\bibnamefont {M{\"u}ller}},\ }\bibfield  {title} {\bibinfo {title} {One-atom maser},\ }\href@noop {} {\bibfield  {journal} {\bibinfo  {journal} {Physical review letters}\ }\textbf {\bibinfo {volume} {54}},\ \bibinfo {pages} {551} (\bibinfo {year} {1985})}\BibitemShut {NoStop}%
\bibitem [{\citenamefont {Mu}\ and\ \citenamefont {Savage}(1992)}]{mu1992one}%
  \BibitemOpen
  \bibfield  {author} {\bibinfo {author} {\bibfnamefont {Y.}~\bibnamefont {Mu}}\ and\ \bibinfo {author} {\bibfnamefont {C.}~\bibnamefont {Savage}},\ }\bibfield  {title} {\bibinfo {title} {One-atom lasers},\ }\href@noop {} {\bibfield  {journal} {\bibinfo  {journal} {Physical Review A}\ }\textbf {\bibinfo {volume} {46}},\ \bibinfo {pages} {5944} (\bibinfo {year} {1992})}\BibitemShut {NoStop}%
\bibitem [{\citenamefont {McKeever}\ \emph {et~al.}(2003)\citenamefont {McKeever}, \citenamefont {Boca}, \citenamefont {Boozer}, \citenamefont {Buck},\ and\ \citenamefont {Kimble}}]{mckeever2003experimental}%
  \BibitemOpen
  \bibfield  {author} {\bibinfo {author} {\bibfnamefont {J.}~\bibnamefont {McKeever}}, \bibinfo {author} {\bibfnamefont {A.}~\bibnamefont {Boca}}, \bibinfo {author} {\bibfnamefont {A.~D.}\ \bibnamefont {Boozer}}, \bibinfo {author} {\bibfnamefont {J.~R.}\ \bibnamefont {Buck}},\ and\ \bibinfo {author} {\bibfnamefont {H.~J.}\ \bibnamefont {Kimble}},\ }\bibfield  {title} {\bibinfo {title} {Experimental realization of a one-atom laser in the regime of strong coupling},\ }\href@noop {} {\bibfield  {journal} {\bibinfo  {journal} {Nature}\ }\textbf {\bibinfo {volume} {425}},\ \bibinfo {pages} {268} (\bibinfo {year} {2003})}\BibitemShut {NoStop}%
\bibitem [{\citenamefont {Pellizzari}\ and\ \citenamefont {Ritsch}(1994{\natexlab{a}})}]{pellizzari1994preparation}%
  \BibitemOpen
  \bibfield  {author} {\bibinfo {author} {\bibfnamefont {T.}~\bibnamefont {Pellizzari}}\ and\ \bibinfo {author} {\bibfnamefont {H.}~\bibnamefont {Ritsch}},\ }\bibfield  {title} {\bibinfo {title} {Preparation of stationary fock states in a one-atom raman laser},\ }\href@noop {} {\bibfield  {journal} {\bibinfo  {journal} {Physical review letters}\ }\textbf {\bibinfo {volume} {72}},\ \bibinfo {pages} {3973} (\bibinfo {year} {1994}{\natexlab{a}})}\BibitemShut {NoStop}%
\bibitem [{\citenamefont {Pellizzari}\ and\ \citenamefont {Ritsch}(1994{\natexlab{b}})}]{pellizzari1994photon}%
  \BibitemOpen
  \bibfield  {author} {\bibinfo {author} {\bibfnamefont {T.}~\bibnamefont {Pellizzari}}\ and\ \bibinfo {author} {\bibfnamefont {H.}~\bibnamefont {Ritsch}},\ }\bibfield  {title} {\bibinfo {title} {Photon statistics of the three-level one-atom laser},\ }\href@noop {} {\bibfield  {journal} {\bibinfo  {journal} {Journal of Modern Optics}\ }\textbf {\bibinfo {volume} {41}},\ \bibinfo {pages} {609} (\bibinfo {year} {1994}{\natexlab{b}})}\BibitemShut {NoStop}%
\bibitem [{\citenamefont {Lugiato}\ \emph {et~al.}(1987)\citenamefont {Lugiato}, \citenamefont {Scully},\ and\ \citenamefont {Walther}}]{lugiato1987connection}%
  \BibitemOpen
  \bibfield  {author} {\bibinfo {author} {\bibfnamefont {L.}~\bibnamefont {Lugiato}}, \bibinfo {author} {\bibfnamefont {M.}~\bibnamefont {Scully}},\ and\ \bibinfo {author} {\bibfnamefont {H.}~\bibnamefont {Walther}},\ }\bibfield  {title} {\bibinfo {title} {Connection between microscopic and macroscopic maser theory},\ }\href@noop {} {\bibfield  {journal} {\bibinfo  {journal} {Physical Review A}\ }\textbf {\bibinfo {volume} {36}},\ \bibinfo {pages} {740} (\bibinfo {year} {1987})}\BibitemShut {NoStop}%
\bibitem [{\citenamefont {Bj{\"o}rk}\ \emph {et~al.}(1994)\citenamefont {Bj{\"o}rk}, \citenamefont {Karlsson},\ and\ \citenamefont {Yamamoto}}]{bjork1994definition}%
  \BibitemOpen
  \bibfield  {author} {\bibinfo {author} {\bibfnamefont {G.}~\bibnamefont {Bj{\"o}rk}}, \bibinfo {author} {\bibfnamefont {A.}~\bibnamefont {Karlsson}},\ and\ \bibinfo {author} {\bibfnamefont {Y.}~\bibnamefont {Yamamoto}},\ }\bibfield  {title} {\bibinfo {title} {Definition of a laser threshold},\ }\href@noop {} {\bibfield  {journal} {\bibinfo  {journal} {Physical Review A}\ }\textbf {\bibinfo {volume} {50}},\ \bibinfo {pages} {1675} (\bibinfo {year} {1994})}\BibitemShut {NoStop}%
\bibitem [{\citenamefont {Dubin}\ \emph {et~al.}(2010)\citenamefont {Dubin}, \citenamefont {Russo}, \citenamefont {Barros}, \citenamefont {Stute}, \citenamefont {Becher}, \citenamefont {Schmidt},\ and\ \citenamefont {Blatt}}]{dubin2010quantum}%
  \BibitemOpen
  \bibfield  {author} {\bibinfo {author} {\bibfnamefont {F.}~\bibnamefont {Dubin}}, \bibinfo {author} {\bibfnamefont {C.}~\bibnamefont {Russo}}, \bibinfo {author} {\bibfnamefont {H.~G.}\ \bibnamefont {Barros}}, \bibinfo {author} {\bibfnamefont {A.}~\bibnamefont {Stute}}, \bibinfo {author} {\bibfnamefont {C.}~\bibnamefont {Becher}}, \bibinfo {author} {\bibfnamefont {P.~O.}\ \bibnamefont {Schmidt}},\ and\ \bibinfo {author} {\bibfnamefont {R.}~\bibnamefont {Blatt}},\ }\bibfield  {title} {\bibinfo {title} {Quantum to classical transition in a single-ion laser},\ }\href@noop {} {\bibfield  {journal} {\bibinfo  {journal} {Nature Physics}\ }\textbf {\bibinfo {volume} {6}},\ \bibinfo {pages} {350} (\bibinfo {year} {2010})}\BibitemShut {NoStop}%
\bibitem [{\citenamefont {Astafiev}\ \emph {et~al.}(2007)\citenamefont {Astafiev}, \citenamefont {Inomata}, \citenamefont {Niskanen}, \citenamefont {Yamamoto}, \citenamefont {Pashkin}, \citenamefont {Nakamura},\ and\ \citenamefont {Tsai}}]{astafiev2007single}%
  \BibitemOpen
  \bibfield  {author} {\bibinfo {author} {\bibfnamefont {O.}~\bibnamefont {Astafiev}}, \bibinfo {author} {\bibfnamefont {K.}~\bibnamefont {Inomata}}, \bibinfo {author} {\bibfnamefont {A.}~\bibnamefont {Niskanen}}, \bibinfo {author} {\bibfnamefont {T.}~\bibnamefont {Yamamoto}}, \bibinfo {author} {\bibfnamefont {Y.~A.}\ \bibnamefont {Pashkin}}, \bibinfo {author} {\bibfnamefont {Y.}~\bibnamefont {Nakamura}},\ and\ \bibinfo {author} {\bibfnamefont {J.}~\bibnamefont {Tsai}},\ }\bibfield  {title} {\bibinfo {title} {Single artificial-atom lasing},\ }\href@noop {} {\bibfield  {journal} {\bibinfo  {journal} {Nature}\ }\textbf {\bibinfo {volume} {449}},\ \bibinfo {pages} {588} (\bibinfo {year} {2007})}\BibitemShut {NoStop}%
\bibitem [{\citenamefont {Cassidy}\ \emph {et~al.}(2017)\citenamefont {Cassidy}, \citenamefont {Bruno}, \citenamefont {Rubbert}, \citenamefont {Irfan}, \citenamefont {Kammhuber}, \citenamefont {Schouten}, \citenamefont {Akhmerov},\ and\ \citenamefont {Kouwenhoven}}]{cassidy2017demonstration}%
  \BibitemOpen
  \bibfield  {author} {\bibinfo {author} {\bibfnamefont {M.}~\bibnamefont {Cassidy}}, \bibinfo {author} {\bibfnamefont {A.}~\bibnamefont {Bruno}}, \bibinfo {author} {\bibfnamefont {S.}~\bibnamefont {Rubbert}}, \bibinfo {author} {\bibfnamefont {M.}~\bibnamefont {Irfan}}, \bibinfo {author} {\bibfnamefont {J.}~\bibnamefont {Kammhuber}}, \bibinfo {author} {\bibfnamefont {R.}~\bibnamefont {Schouten}}, \bibinfo {author} {\bibfnamefont {A.}~\bibnamefont {Akhmerov}},\ and\ \bibinfo {author} {\bibfnamefont {L.}~\bibnamefont {Kouwenhoven}},\ }\bibfield  {title} {\bibinfo {title} {Demonstration of an ac josephson junction laser},\ }\href@noop {} {\bibfield  {journal} {\bibinfo  {journal} {Science}\ }\textbf {\bibinfo {volume} {355}},\ \bibinfo {pages} {939} (\bibinfo {year} {2017})}\BibitemShut {NoStop}%
\bibitem [{\citenamefont {Schawlow}\ and\ \citenamefont {Townes}(1958)}]{schawlow1958infrared}%
  \BibitemOpen
  \bibfield  {author} {\bibinfo {author} {\bibfnamefont {A.~L.}\ \bibnamefont {Schawlow}}\ and\ \bibinfo {author} {\bibfnamefont {C.~H.}\ \bibnamefont {Townes}},\ }\bibfield  {title} {\bibinfo {title} {Infrared and optical masers},\ }\href@noop {} {\bibfield  {journal} {\bibinfo  {journal} {Physical review}\ }\textbf {\bibinfo {volume} {112}},\ \bibinfo {pages} {1940} (\bibinfo {year} {1958})}\BibitemShut {NoStop}%
\bibitem [{\citenamefont {Bartalini}\ \emph {et~al.}(2010)\citenamefont {Bartalini}, \citenamefont {Borri}, \citenamefont {Cancio}, \citenamefont {Castrillo}, \citenamefont {Galli}, \citenamefont {Giusfredi}, \citenamefont {Mazzotti}, \citenamefont {Gianfrani},\ and\ \citenamefont {De~Natale}}]{bartalini2010observing}%
  \BibitemOpen
  \bibfield  {author} {\bibinfo {author} {\bibfnamefont {S.}~\bibnamefont {Bartalini}}, \bibinfo {author} {\bibfnamefont {S.}~\bibnamefont {Borri}}, \bibinfo {author} {\bibfnamefont {P.}~\bibnamefont {Cancio}}, \bibinfo {author} {\bibfnamefont {A.}~\bibnamefont {Castrillo}}, \bibinfo {author} {\bibfnamefont {I.}~\bibnamefont {Galli}}, \bibinfo {author} {\bibfnamefont {G.}~\bibnamefont {Giusfredi}}, \bibinfo {author} {\bibfnamefont {D.}~\bibnamefont {Mazzotti}}, \bibinfo {author} {\bibfnamefont {.~f.~L.}\ \bibnamefont {Gianfrani}},\ and\ \bibinfo {author} {\bibfnamefont {P.}~\bibnamefont {De~Natale}},\ }\bibfield  {title} {\bibinfo {title} {Observing the intrinsic linewidth of a quantum-cascade laser: Beyond the schawlow-townes limit},\ }\href@noop {} {\bibfield  {journal} {\bibinfo  {journal} {Physical review letters}\ }\textbf {\bibinfo {volume} {104}},\ \bibinfo {pages} {083904} (\bibinfo {year} {2010})}\BibitemShut {NoStop}%
\bibitem [{\citenamefont {Wiseman}(1999)}]{wiseman1999light}%
  \BibitemOpen
  \bibfield  {author} {\bibinfo {author} {\bibfnamefont {H.}~\bibnamefont {Wiseman}},\ }\bibfield  {title} {\bibinfo {title} {Light amplification without stimulated emission: beyond the standard quantum limit to the laser linewidth},\ }\href@noop {} {\bibfield  {journal} {\bibinfo  {journal} {Physical Review A}\ }\textbf {\bibinfo {volume} {60}},\ \bibinfo {pages} {4083} (\bibinfo {year} {1999})}\BibitemShut {NoStop}%
\bibitem [{\citenamefont {Liu}\ \emph {et~al.}(2015)\citenamefont {Liu}, \citenamefont {Stehlik}, \citenamefont {Eichler}, \citenamefont {Gullans}, \citenamefont {Taylor},\ and\ \citenamefont {Petta}}]{liu2015semiconductor}%
  \BibitemOpen
  \bibfield  {author} {\bibinfo {author} {\bibfnamefont {Y.-Y.}\ \bibnamefont {Liu}}, \bibinfo {author} {\bibfnamefont {J.}~\bibnamefont {Stehlik}}, \bibinfo {author} {\bibfnamefont {C.}~\bibnamefont {Eichler}}, \bibinfo {author} {\bibfnamefont {M.}~\bibnamefont {Gullans}}, \bibinfo {author} {\bibfnamefont {J.~M.}\ \bibnamefont {Taylor}},\ and\ \bibinfo {author} {\bibfnamefont {J.}~\bibnamefont {Petta}},\ }\bibfield  {title} {\bibinfo {title} {Semiconductor double quantum dot micromaser},\ }\href@noop {} {\bibfield  {journal} {\bibinfo  {journal} {Science}\ }\textbf {\bibinfo {volume} {347}},\ \bibinfo {pages} {285} (\bibinfo {year} {2015})}\BibitemShut {NoStop}%
\bibitem [{\citenamefont {Liu}\ \emph {et~al.}(2021)\citenamefont {Liu}, \citenamefont {Mucci}, \citenamefont {Cao}, \citenamefont {Dutt}, \citenamefont {Hatridge},\ and\ \citenamefont {Pekker}}]{Liu_2021}%
  \BibitemOpen
  \bibfield  {author} {\bibinfo {author} {\bibfnamefont {C.}~\bibnamefont {Liu}}, \bibinfo {author} {\bibfnamefont {M.}~\bibnamefont {Mucci}}, \bibinfo {author} {\bibfnamefont {X.}~\bibnamefont {Cao}}, \bibinfo {author} {\bibfnamefont {M.~V.~G.}\ \bibnamefont {Dutt}}, \bibinfo {author} {\bibfnamefont {M.}~\bibnamefont {Hatridge}},\ and\ \bibinfo {author} {\bibfnamefont {D.}~\bibnamefont {Pekker}},\ }\bibfield  {title} {\bibinfo {title} {Proposal for a continuous wave laser with linewidth well below the standard quantum limit},\ }\bibfield  {journal} {\bibinfo  {journal} {Nature Communications}\ }\textbf {\bibinfo {volume} {12}},\ \href {https://doi.org/10.1038/s41467-021-25879-8} {10.1038/s41467-021-25879-8} (\bibinfo {year} {2021})\BibitemShut {NoStop}%
\bibitem [{\citenamefont {Ball}\ \emph {et~al.}(2016)\citenamefont {Ball}, \citenamefont {Oliver},\ and\ \citenamefont {Biercuk}}]{ball2016role}%
  \BibitemOpen
  \bibfield  {author} {\bibinfo {author} {\bibfnamefont {H.}~\bibnamefont {Ball}}, \bibinfo {author} {\bibfnamefont {W.~D.}\ \bibnamefont {Oliver}},\ and\ \bibinfo {author} {\bibfnamefont {M.~J.}\ \bibnamefont {Biercuk}},\ }\bibfield  {title} {\bibinfo {title} {The role of master clock stability in quantum information processing},\ }\href@noop {} {\bibfield  {journal} {\bibinfo  {journal} {npj Quantum Information}\ }\textbf {\bibinfo {volume} {2}},\ \bibinfo {pages} {1} (\bibinfo {year} {2016})}\BibitemShut {NoStop}%
\bibitem [{\citenamefont {Simon}\ and\ \citenamefont {Cooper}(2018)}]{simon2018theory}%
  \BibitemOpen
  \bibfield  {author} {\bibinfo {author} {\bibfnamefont {S.~H.}\ \bibnamefont {Simon}}\ and\ \bibinfo {author} {\bibfnamefont {N.~R.}\ \bibnamefont {Cooper}},\ }\bibfield  {title} {\bibinfo {title} {Theory of the josephson junction laser},\ }\href@noop {} {\bibfield  {journal} {\bibinfo  {journal} {Physical Review Letters}\ }\textbf {\bibinfo {volume} {121}},\ \bibinfo {pages} {027004} (\bibinfo {year} {2018})}\BibitemShut {NoStop}%
\bibitem [{\citenamefont {Yuen}\ and\ \citenamefont {Shapiro}(1978)}]{yuen1978optical}%
  \BibitemOpen
  \bibfield  {author} {\bibinfo {author} {\bibfnamefont {H.}~\bibnamefont {Yuen}}\ and\ \bibinfo {author} {\bibfnamefont {J.}~\bibnamefont {Shapiro}},\ }\bibfield  {title} {\bibinfo {title} {Optical communication with two-photon coherent states--part i: Quantum-state propagation and quantum-noise},\ }\href@noop {} {\bibfield  {journal} {\bibinfo  {journal} {IEEE Transactions on Information Theory}\ }\textbf {\bibinfo {volume} {24}},\ \bibinfo {pages} {657} (\bibinfo {year} {1978})}\BibitemShut {NoStop}%
\bibitem [{\citenamefont {Ma}\ \emph {et~al.}(2017)\citenamefont {Ma}, \citenamefont {Miao}, \citenamefont {Pang}, \citenamefont {Evans}, \citenamefont {Zhao}, \citenamefont {Harms}, \citenamefont {Schnabel},\ and\ \citenamefont {Chen}}]{ma2017proposal}%
  \BibitemOpen
  \bibfield  {author} {\bibinfo {author} {\bibfnamefont {Y.}~\bibnamefont {Ma}}, \bibinfo {author} {\bibfnamefont {H.}~\bibnamefont {Miao}}, \bibinfo {author} {\bibfnamefont {B.~H.}\ \bibnamefont {Pang}}, \bibinfo {author} {\bibfnamefont {M.}~\bibnamefont {Evans}}, \bibinfo {author} {\bibfnamefont {C.}~\bibnamefont {Zhao}}, \bibinfo {author} {\bibfnamefont {J.}~\bibnamefont {Harms}}, \bibinfo {author} {\bibfnamefont {R.}~\bibnamefont {Schnabel}},\ and\ \bibinfo {author} {\bibfnamefont {Y.}~\bibnamefont {Chen}},\ }\bibfield  {title} {\bibinfo {title} {Proposal for gravitational-wave detection beyond the standard quantum limit through epr entanglement},\ }\href@noop {} {\bibfield  {journal} {\bibinfo  {journal} {Nature Physics}\ }\textbf {\bibinfo {volume} {13}},\ \bibinfo {pages} {776} (\bibinfo {year} {2017})}\BibitemShut {NoStop}%
\bibitem [{\citenamefont {Jerger}\ \emph {et~al.}(2019)\citenamefont {Jerger}, \citenamefont {Kulikov}, \citenamefont {Vasselin},\ and\ \citenamefont {Fedorov}}]{jerger2019situ}%
  \BibitemOpen
  \bibfield  {author} {\bibinfo {author} {\bibfnamefont {M.}~\bibnamefont {Jerger}}, \bibinfo {author} {\bibfnamefont {A.}~\bibnamefont {Kulikov}}, \bibinfo {author} {\bibfnamefont {Z.}~\bibnamefont {Vasselin}},\ and\ \bibinfo {author} {\bibfnamefont {A.}~\bibnamefont {Fedorov}},\ }\bibfield  {title} {\bibinfo {title} {In situ characterization of qubit control lines: a qubit as a vector network analyzer},\ }\href@noop {} {\bibfield  {journal} {\bibinfo  {journal} {Physical review letters}\ }\textbf {\bibinfo {volume} {123}},\ \bibinfo {pages} {150501} (\bibinfo {year} {2019})}\BibitemShut {NoStop}%
\bibitem [{\citenamefont {Danilin}\ and\ \citenamefont {Weides}(2021)}]{danilin2021quantum}%
  \BibitemOpen
  \bibfield  {author} {\bibinfo {author} {\bibfnamefont {S.}~\bibnamefont {Danilin}}\ and\ \bibinfo {author} {\bibfnamefont {M.}~\bibnamefont {Weides}},\ }\bibfield  {title} {\bibinfo {title} {Quantum sensing with superconducting circuits},\ }\href@noop {} {\bibfield  {journal} {\bibinfo  {journal} {arXiv preprint arXiv:2103.11022}\ } (\bibinfo {year} {2021})}\BibitemShut {NoStop}%
\bibitem [{\citenamefont {Chen}\ \emph {et~al.}(2014)\citenamefont {Chen}, \citenamefont {Li}, \citenamefont {Armour}, \citenamefont {Brahimi}, \citenamefont {Stettenheim}, \citenamefont {Sirois}, \citenamefont {Simmonds}, \citenamefont {Blencowe},\ and\ \citenamefont {Rimberg}}]{chen2014realization}%
  \BibitemOpen
  \bibfield  {author} {\bibinfo {author} {\bibfnamefont {F.}~\bibnamefont {Chen}}, \bibinfo {author} {\bibfnamefont {J.}~\bibnamefont {Li}}, \bibinfo {author} {\bibfnamefont {A.}~\bibnamefont {Armour}}, \bibinfo {author} {\bibfnamefont {E.}~\bibnamefont {Brahimi}}, \bibinfo {author} {\bibfnamefont {J.}~\bibnamefont {Stettenheim}}, \bibinfo {author} {\bibfnamefont {A.}~\bibnamefont {Sirois}}, \bibinfo {author} {\bibfnamefont {R.}~\bibnamefont {Simmonds}}, \bibinfo {author} {\bibfnamefont {M.}~\bibnamefont {Blencowe}},\ and\ \bibinfo {author} {\bibfnamefont {A.}~\bibnamefont {Rimberg}},\ }\bibfield  {title} {\bibinfo {title} {Realization of a single-cooper-pair josephson laser},\ }\href@noop {} {\bibfield  {journal} {\bibinfo  {journal} {Physical Review B}\ }\textbf {\bibinfo {volume} {90}},\ \bibinfo {pages} {020506} (\bibinfo {year} {2014})}\BibitemShut {NoStop}%
\bibitem [{\citenamefont {Yan}\ \emph {et~al.}(2018)\citenamefont {Yan}, \citenamefont {Krantz}, \citenamefont {Sung}, \citenamefont {Kjaergaard}, \citenamefont {Campbell}, \citenamefont {Orlando}, \citenamefont {Gustavsson},\ and\ \citenamefont {Oliver}}]{yan2018tunable}%
  \BibitemOpen
  \bibfield  {author} {\bibinfo {author} {\bibfnamefont {F.}~\bibnamefont {Yan}}, \bibinfo {author} {\bibfnamefont {P.}~\bibnamefont {Krantz}}, \bibinfo {author} {\bibfnamefont {Y.}~\bibnamefont {Sung}}, \bibinfo {author} {\bibfnamefont {M.}~\bibnamefont {Kjaergaard}}, \bibinfo {author} {\bibfnamefont {D.~L.}\ \bibnamefont {Campbell}}, \bibinfo {author} {\bibfnamefont {T.~P.}\ \bibnamefont {Orlando}}, \bibinfo {author} {\bibfnamefont {S.}~\bibnamefont {Gustavsson}},\ and\ \bibinfo {author} {\bibfnamefont {W.~D.}\ \bibnamefont {Oliver}},\ }\bibfield  {title} {\bibinfo {title} {Tunable coupling scheme for implementing high-fidelity two-qubit gates},\ }\href@noop {} {\bibfield  {journal} {\bibinfo  {journal} {Physical Review Applied}\ }\textbf {\bibinfo {volume} {10}},\ \bibinfo {pages} {054062} (\bibinfo {year} {2018})}\BibitemShut {NoStop}%
\bibitem [{\citenamefont {Bialczak}\ \emph {et~al.}(2011)\citenamefont {Bialczak}, \citenamefont {Ansmann}, \citenamefont {Hofheinz}, \citenamefont {Lenander}, \citenamefont {Lucero}, \citenamefont {Neeley}, \citenamefont {O’Connell}, \citenamefont {Sank}, \citenamefont {Wang}, \citenamefont {Weides} \emph {et~al.}}]{bialczak2011fast}%
  \BibitemOpen
  \bibfield  {author} {\bibinfo {author} {\bibfnamefont {R.}~\bibnamefont {Bialczak}}, \bibinfo {author} {\bibfnamefont {M.}~\bibnamefont {Ansmann}}, \bibinfo {author} {\bibfnamefont {M.}~\bibnamefont {Hofheinz}}, \bibinfo {author} {\bibfnamefont {M.}~\bibnamefont {Lenander}}, \bibinfo {author} {\bibfnamefont {E.}~\bibnamefont {Lucero}}, \bibinfo {author} {\bibfnamefont {M.}~\bibnamefont {Neeley}}, \bibinfo {author} {\bibfnamefont {A.}~\bibnamefont {O’Connell}}, \bibinfo {author} {\bibfnamefont {D.}~\bibnamefont {Sank}}, \bibinfo {author} {\bibfnamefont {H.}~\bibnamefont {Wang}}, \bibinfo {author} {\bibfnamefont {M.}~\bibnamefont {Weides}}, \emph {et~al.},\ }\bibfield  {title} {\bibinfo {title} {Fast tunable coupler for superconducting qubits},\ }\href@noop {} {\bibfield  {journal} {\bibinfo  {journal} {Physical review letters}\ }\textbf {\bibinfo {volume} {106}},\ \bibinfo {pages} {060501} (\bibinfo {year} {2011})}\BibitemShut {NoStop}%
\bibitem [{\citenamefont {Heunisch}\ \emph {et~al.}(2023)\citenamefont {Heunisch}, \citenamefont {Eichler},\ and\ \citenamefont {Hartmann}}]{heunisch2023tunable}%
  \BibitemOpen
  \bibfield  {author} {\bibinfo {author} {\bibfnamefont {L.}~\bibnamefont {Heunisch}}, \bibinfo {author} {\bibfnamefont {C.}~\bibnamefont {Eichler}},\ and\ \bibinfo {author} {\bibfnamefont {M.~J.}\ \bibnamefont {Hartmann}},\ }\bibfield  {title} {\bibinfo {title} {Tunable coupler to fully decouple and maximally localize superconducting qubits},\ }\href@noop {} {\bibfield  {journal} {\bibinfo  {journal} {Physical Review Applied}\ }\textbf {\bibinfo {volume} {20}},\ \bibinfo {pages} {064037} (\bibinfo {year} {2023})}\BibitemShut {NoStop}%
\bibitem [{\citenamefont {Gottesman}\ and\ \citenamefont {Preskill}(2003)}]{gottesman2003secure}%
  \BibitemOpen
  \bibfield  {author} {\bibinfo {author} {\bibfnamefont {D.}~\bibnamefont {Gottesman}}\ and\ \bibinfo {author} {\bibfnamefont {J.}~\bibnamefont {Preskill}},\ }\bibfield  {title} {\bibinfo {title} {Secure quantum key distribution using squeezed states},\ }\href@noop {} {\bibfield  {journal} {\bibinfo  {journal} {Quantum Information with Continuous Variables}\ ,\ \bibinfo {pages} {317}} (\bibinfo {year} {2003})}\BibitemShut {NoStop}%
\bibitem [{\citenamefont {Menicucci}(2014)}]{menicucci2014fault}%
  \BibitemOpen
  \bibfield  {author} {\bibinfo {author} {\bibfnamefont {N.~C.}\ \bibnamefont {Menicucci}},\ }\bibfield  {title} {\bibinfo {title} {Fault-tolerant measurement-based quantum computing with continuous-variable cluster states},\ }\href@noop {} {\bibfield  {journal} {\bibinfo  {journal} {Physical review letters}\ }\textbf {\bibinfo {volume} {112}},\ \bibinfo {pages} {120504} (\bibinfo {year} {2014})}\BibitemShut {NoStop}%
\bibitem [{\citenamefont {Marshall}\ \emph {et~al.}(2016)\citenamefont {Marshall}, \citenamefont {Jacobsen}, \citenamefont {Sch{\"a}fermeier}, \citenamefont {Gehring}, \citenamefont {Weedbrook},\ and\ \citenamefont {Andersen}}]{marshall2016continuous}%
  \BibitemOpen
  \bibfield  {author} {\bibinfo {author} {\bibfnamefont {K.}~\bibnamefont {Marshall}}, \bibinfo {author} {\bibfnamefont {C.~S.}\ \bibnamefont {Jacobsen}}, \bibinfo {author} {\bibfnamefont {C.}~\bibnamefont {Sch{\"a}fermeier}}, \bibinfo {author} {\bibfnamefont {T.}~\bibnamefont {Gehring}}, \bibinfo {author} {\bibfnamefont {C.}~\bibnamefont {Weedbrook}},\ and\ \bibinfo {author} {\bibfnamefont {U.~L.}\ \bibnamefont {Andersen}},\ }\bibfield  {title} {\bibinfo {title} {Continuous-variable quantum computing on encrypted data},\ }\href@noop {} {\bibfield  {journal} {\bibinfo  {journal} {Nature communications}\ }\textbf {\bibinfo {volume} {7}},\ \bibinfo {pages} {13795} (\bibinfo {year} {2016})}\BibitemShut {NoStop}%
\bibitem [{\citenamefont {Pauka}\ \emph {et~al.}(2019)\citenamefont {Pauka}, \citenamefont {Das}, \citenamefont {Kalra}, \citenamefont {Moini}, \citenamefont {Yang}, \citenamefont {Trainer}, \citenamefont {Bousquet}, \citenamefont {Cantaloube}, \citenamefont {Dick}, \citenamefont {Gardner} \emph {et~al.}}]{pauka2019cryogenic}%
  \BibitemOpen
  \bibfield  {author} {\bibinfo {author} {\bibfnamefont {S.}~\bibnamefont {Pauka}}, \bibinfo {author} {\bibfnamefont {K.}~\bibnamefont {Das}}, \bibinfo {author} {\bibfnamefont {R.}~\bibnamefont {Kalra}}, \bibinfo {author} {\bibfnamefont {A.}~\bibnamefont {Moini}}, \bibinfo {author} {\bibfnamefont {Y.}~\bibnamefont {Yang}}, \bibinfo {author} {\bibfnamefont {M.}~\bibnamefont {Trainer}}, \bibinfo {author} {\bibfnamefont {A.}~\bibnamefont {Bousquet}}, \bibinfo {author} {\bibfnamefont {C.}~\bibnamefont {Cantaloube}}, \bibinfo {author} {\bibfnamefont {N.}~\bibnamefont {Dick}}, \bibinfo {author} {\bibfnamefont {G.}~\bibnamefont {Gardner}}, \emph {et~al.},\ }\bibfield  {title} {\bibinfo {title} {A cryogenic interface for controlling many qubits},\ }\href@noop {} {\bibfield  {journal} {\bibinfo  {journal} {arXiv preprint arXiv:1912.01299}\ } (\bibinfo {year} {2019})}\BibitemShut {NoStop}%
\bibitem [{\citenamefont {Cai}\ \emph {et~al.}(2021)\citenamefont {Cai}, \citenamefont {Ma}, \citenamefont {Wang}, \citenamefont {Zou},\ and\ \citenamefont {Sun}}]{cai2021bosonic}%
  \BibitemOpen
  \bibfield  {author} {\bibinfo {author} {\bibfnamefont {W.}~\bibnamefont {Cai}}, \bibinfo {author} {\bibfnamefont {Y.}~\bibnamefont {Ma}}, \bibinfo {author} {\bibfnamefont {W.}~\bibnamefont {Wang}}, \bibinfo {author} {\bibfnamefont {C.-L.}\ \bibnamefont {Zou}},\ and\ \bibinfo {author} {\bibfnamefont {L.}~\bibnamefont {Sun}},\ }\bibfield  {title} {\bibinfo {title} {Bosonic quantum error correction codes in superconducting quantum circuits},\ }\href@noop {} {\bibfield  {journal} {\bibinfo  {journal} {Fundamental Research}\ }\textbf {\bibinfo {volume} {1}},\ \bibinfo {pages} {50} (\bibinfo {year} {2021})}\BibitemShut {NoStop}%
\bibitem [{\citenamefont {Sung}\ \emph {et~al.}(2021)\citenamefont {Sung}, \citenamefont {Ding}, \citenamefont {Braum\"uller}, \citenamefont {Veps\"al\"ainen}, \citenamefont {Kannan}, \citenamefont {Kjaergaard}, \citenamefont {Greene}, \citenamefont {Samach}, \citenamefont {McNally}, \citenamefont {Kim}, \citenamefont {Melville}, \citenamefont {Niedzielski}, \citenamefont {Schwartz}, \citenamefont {Yoder}, \citenamefont {Orlando}, \citenamefont {Gustavsson},\ and\ \citenamefont {Oliver}}]{sung2021tunablecoupler}%
  \BibitemOpen
  \bibfield  {author} {\bibinfo {author} {\bibfnamefont {Y.}~\bibnamefont {Sung}}, \bibinfo {author} {\bibfnamefont {L.}~\bibnamefont {Ding}}, \bibinfo {author} {\bibfnamefont {J.}~\bibnamefont {Braum\"uller}}, \bibinfo {author} {\bibfnamefont {A.}~\bibnamefont {Veps\"al\"ainen}}, \bibinfo {author} {\bibfnamefont {B.}~\bibnamefont {Kannan}}, \bibinfo {author} {\bibfnamefont {M.}~\bibnamefont {Kjaergaard}}, \bibinfo {author} {\bibfnamefont {A.}~\bibnamefont {Greene}}, \bibinfo {author} {\bibfnamefont {G.~O.}\ \bibnamefont {Samach}}, \bibinfo {author} {\bibfnamefont {C.}~\bibnamefont {McNally}}, \bibinfo {author} {\bibfnamefont {D.}~\bibnamefont {Kim}}, \bibinfo {author} {\bibfnamefont {A.}~\bibnamefont {Melville}}, \bibinfo {author} {\bibfnamefont {B.~M.}\ \bibnamefont {Niedzielski}}, \bibinfo {author} {\bibfnamefont {M.~E.}\ \bibnamefont {Schwartz}}, \bibinfo {author} {\bibfnamefont {J.~L.}\ \bibnamefont {Yoder}}, \bibinfo {author} {\bibfnamefont {T.~P.}\ \bibnamefont {Orlando}}, \bibinfo {author}
  {\bibfnamefont {S.}~\bibnamefont {Gustavsson}},\ and\ \bibinfo {author} {\bibfnamefont {W.~D.}\ \bibnamefont {Oliver}},\ }\bibfield  {title} {\bibinfo {title} {Realization of high-fidelity cz and $zz$-free iswap gates with a tunable coupler},\ }\href {https://doi.org/10.1103/PhysRevX.11.021058} {\bibfield  {journal} {\bibinfo  {journal} {Phys. Rev. X}\ }\textbf {\bibinfo {volume} {11}},\ \bibinfo {pages} {021058} (\bibinfo {year} {2021})}\BibitemShut {NoStop}%
\bibitem [{\citenamefont {Fowles}(1989)}]{fowles1989introduction}%
  \BibitemOpen
  \bibfield  {author} {\bibinfo {author} {\bibfnamefont {G.~R.}\ \bibnamefont {Fowles}},\ }\href@noop {} {\emph {\bibinfo {title} {Introduction to modern optics}}}\ (\bibinfo  {publisher} {Courier Corporation},\ \bibinfo {year} {1989})\BibitemShut {NoStop}%
\bibitem [{\citenamefont {Wallraff}\ \emph {et~al.}(2004)\citenamefont {Wallraff}, \citenamefont {Schuster}, \citenamefont {Blais}, \citenamefont {Frunzio}, \citenamefont {Huang}, \citenamefont {Majer}, \citenamefont {Kumar}, \citenamefont {Girvin},\ and\ \citenamefont {Schoelkopf}}]{wallraff2004strong}%
  \BibitemOpen
  \bibfield  {author} {\bibinfo {author} {\bibfnamefont {A.}~\bibnamefont {Wallraff}}, \bibinfo {author} {\bibfnamefont {D.~I.}\ \bibnamefont {Schuster}}, \bibinfo {author} {\bibfnamefont {A.}~\bibnamefont {Blais}}, \bibinfo {author} {\bibfnamefont {L.}~\bibnamefont {Frunzio}}, \bibinfo {author} {\bibfnamefont {R.-S.}\ \bibnamefont {Huang}}, \bibinfo {author} {\bibfnamefont {J.}~\bibnamefont {Majer}}, \bibinfo {author} {\bibfnamefont {S.}~\bibnamefont {Kumar}}, \bibinfo {author} {\bibfnamefont {S.~M.}\ \bibnamefont {Girvin}},\ and\ \bibinfo {author} {\bibfnamefont {R.~J.}\ \bibnamefont {Schoelkopf}},\ }\bibfield  {title} {\bibinfo {title} {Strong coupling of a single photon to a superconducting qubit using circuit quantum electrodynamics},\ }\href@noop {} {\bibfield  {journal} {\bibinfo  {journal} {Nature}\ }\textbf {\bibinfo {volume} {431}},\ \bibinfo {pages} {162} (\bibinfo {year} {2004})}\BibitemShut {NoStop}%
\bibitem [{\citenamefont {Oelsner}\ and\ \citenamefont {Il'ichev}(2018)}]{Oelsner2018}%
  \BibitemOpen
  \bibfield  {author} {\bibinfo {author} {\bibfnamefont {G.}~\bibnamefont {Oelsner}}\ and\ \bibinfo {author} {\bibfnamefont {E.}~\bibnamefont {Il'ichev}},\ }\bibinfo {title} {Lasing in circuit quantum electrodynamics},\ in\ \href {https://doi.org/10.1007/978-3-319-90481-8_9} {\emph {\bibinfo {booktitle} {Functional Nanostructures and Metamaterials for Superconducting Spintronics: From Superconducting Qubits to Self-Organized Nanostructures}}},\ \bibinfo {editor} {edited by\ \bibinfo {editor} {\bibfnamefont {A.}~\bibnamefont {Sidorenko}}}\ (\bibinfo  {publisher} {Springer International Publishing},\ \bibinfo {address} {Cham},\ \bibinfo {year} {2018})\ pp.\ \bibinfo {pages} {175--194}\BibitemShut {NoStop}%
\bibitem [{\citenamefont {Thomas}\ \emph {et~al.}(2020)\citenamefont {Thomas}, \citenamefont {Gubaydullin}, \citenamefont {Golubev},\ and\ \citenamefont {Pekola}}]{thomas2020thermally}%
  \BibitemOpen
  \bibfield  {author} {\bibinfo {author} {\bibfnamefont {G.}~\bibnamefont {Thomas}}, \bibinfo {author} {\bibfnamefont {A.}~\bibnamefont {Gubaydullin}}, \bibinfo {author} {\bibfnamefont {D.~S.}\ \bibnamefont {Golubev}},\ and\ \bibinfo {author} {\bibfnamefont {J.~P.}\ \bibnamefont {Pekola}},\ }\bibfield  {title} {\bibinfo {title} {Thermally pumped on-chip maser},\ }\href@noop {} {\bibfield  {journal} {\bibinfo  {journal} {Physical Review B}\ }\textbf {\bibinfo {volume} {102}},\ \bibinfo {pages} {104503} (\bibinfo {year} {2020})}\BibitemShut {NoStop}%
\bibitem [{\citenamefont {Sokolova}\ \emph {et~al.}(2021)\citenamefont {Sokolova}, \citenamefont {Fedorov}, \citenamefont {Il'ichev},\ and\ \citenamefont {Astafiev}}]{sokolova2021single}%
  \BibitemOpen
  \bibfield  {author} {\bibinfo {author} {\bibfnamefont {A.}~\bibnamefont {Sokolova}}, \bibinfo {author} {\bibfnamefont {G.}~\bibnamefont {Fedorov}}, \bibinfo {author} {\bibfnamefont {E.}~\bibnamefont {Il'ichev}},\ and\ \bibinfo {author} {\bibfnamefont {O.}~\bibnamefont {Astafiev}},\ }\bibfield  {title} {\bibinfo {title} {Single-atom maser with an engineered circuit for population inversion},\ }\href@noop {} {\bibfield  {journal} {\bibinfo  {journal} {Physical Review A}\ }\textbf {\bibinfo {volume} {103}},\ \bibinfo {pages} {013718} (\bibinfo {year} {2021})}\BibitemShut {NoStop}%
\bibitem [{\citenamefont {Joo}\ \emph {et~al.}(2010)\citenamefont {Joo}, \citenamefont {Bourassa}, \citenamefont {Blais},\ and\ \citenamefont {Sanders}}]{jaewoo2010electroinduc}%
  \BibitemOpen
  \bibfield  {author} {\bibinfo {author} {\bibfnamefont {J.}~\bibnamefont {Joo}}, \bibinfo {author} {\bibfnamefont {J.}~\bibnamefont {Bourassa}}, \bibinfo {author} {\bibfnamefont {A.}~\bibnamefont {Blais}},\ and\ \bibinfo {author} {\bibfnamefont {B.~C.}\ \bibnamefont {Sanders}},\ }\bibfield  {title} {\bibinfo {title} {Electromagnetically induced transparency with amplification in superconducting circuits},\ }\href {https://doi.org/10.1103/PhysRevLett.105.073601} {\bibfield  {journal} {\bibinfo  {journal} {Phys. Rev. Lett.}\ }\textbf {\bibinfo {volume} {105}},\ \bibinfo {pages} {073601} (\bibinfo {year} {2010})}\BibitemShut {NoStop}%
\bibitem [{\citenamefont {Marthaler}\ \emph {et~al.}(2011)\citenamefont {Marthaler}, \citenamefont {Utsumi}, \citenamefont {Golubev}, \citenamefont {Shnirman},\ and\ \citenamefont {Sch{\"o}n}}]{marthaler2011lasing}%
  \BibitemOpen
  \bibfield  {author} {\bibinfo {author} {\bibfnamefont {M.}~\bibnamefont {Marthaler}}, \bibinfo {author} {\bibfnamefont {Y.}~\bibnamefont {Utsumi}}, \bibinfo {author} {\bibfnamefont {D.~S.}\ \bibnamefont {Golubev}}, \bibinfo {author} {\bibfnamefont {A.}~\bibnamefont {Shnirman}},\ and\ \bibinfo {author} {\bibfnamefont {G.}~\bibnamefont {Sch{\"o}n}},\ }\bibfield  {title} {\bibinfo {title} {Lasing without inversion in circuit quantum electrodynamics},\ }\href@noop {} {\bibfield  {journal} {\bibinfo  {journal} {Physical Review Letters}\ }\textbf {\bibinfo {volume} {107}},\ \bibinfo {pages} {093901} (\bibinfo {year} {2011})}\BibitemShut {NoStop}%
\bibitem [{\citenamefont {Oelsner}\ \emph {et~al.}(2013)\citenamefont {Oelsner}, \citenamefont {Macha}, \citenamefont {Astafiev}, \citenamefont {Il’ichev}, \citenamefont {Grajcar}, \citenamefont {H{\"u}bner}, \citenamefont {Ivanov}, \citenamefont {Neilinger},\ and\ \citenamefont {Meyer}}]{oelsner2013dressed}%
  \BibitemOpen
  \bibfield  {author} {\bibinfo {author} {\bibfnamefont {G.}~\bibnamefont {Oelsner}}, \bibinfo {author} {\bibfnamefont {P.}~\bibnamefont {Macha}}, \bibinfo {author} {\bibfnamefont {O.}~\bibnamefont {Astafiev}}, \bibinfo {author} {\bibfnamefont {E.}~\bibnamefont {Il’ichev}}, \bibinfo {author} {\bibfnamefont {M.}~\bibnamefont {Grajcar}}, \bibinfo {author} {\bibfnamefont {U.}~\bibnamefont {H{\"u}bner}}, \bibinfo {author} {\bibfnamefont {B.}~\bibnamefont {Ivanov}}, \bibinfo {author} {\bibfnamefont {P.}~\bibnamefont {Neilinger}},\ and\ \bibinfo {author} {\bibfnamefont {H.-G.}\ \bibnamefont {Meyer}},\ }\bibfield  {title} {\bibinfo {title} {Dressed-state amplification by a single superconducting qubit},\ }\href@noop {} {\bibfield  {journal} {\bibinfo  {journal} {Physical review letters}\ }\textbf {\bibinfo {volume} {110}},\ \bibinfo {pages} {053602} (\bibinfo {year} {2013})}\BibitemShut {NoStop}%
\bibitem [{\citenamefont {Blais}\ \emph {et~al.}(2021)\citenamefont {Blais}, \citenamefont {Grimsmo}, \citenamefont {Girvin},\ and\ \citenamefont {Wallraff}}]{blais2021circuit}%
  \BibitemOpen
  \bibfield  {author} {\bibinfo {author} {\bibfnamefont {A.}~\bibnamefont {Blais}}, \bibinfo {author} {\bibfnamefont {A.~L.}\ \bibnamefont {Grimsmo}}, \bibinfo {author} {\bibfnamefont {S.~M.}\ \bibnamefont {Girvin}},\ and\ \bibinfo {author} {\bibfnamefont {A.}~\bibnamefont {Wallraff}},\ }\bibfield  {title} {\bibinfo {title} {Circuit quantum electrodynamics},\ }\href@noop {} {\bibfield  {journal} {\bibinfo  {journal} {Reviews of Modern Physics}\ }\textbf {\bibinfo {volume} {93}},\ \bibinfo {pages} {025005} (\bibinfo {year} {2021})}\BibitemShut {NoStop}%
\bibitem [{\citenamefont {Lang}\ \emph {et~al.}(2011)\citenamefont {Lang}, \citenamefont {Bozyigit}, \citenamefont {Eichler}, \citenamefont {Steffen}, \citenamefont {Fink}, \citenamefont {Abdumalikov~Jr}, \citenamefont {Baur}, \citenamefont {Filipp}, \citenamefont {Da~Silva}, \citenamefont {Blais} \emph {et~al.}}]{lang2011observation}%
  \BibitemOpen
  \bibfield  {author} {\bibinfo {author} {\bibfnamefont {C.}~\bibnamefont {Lang}}, \bibinfo {author} {\bibfnamefont {D.}~\bibnamefont {Bozyigit}}, \bibinfo {author} {\bibfnamefont {C.}~\bibnamefont {Eichler}}, \bibinfo {author} {\bibfnamefont {L.}~\bibnamefont {Steffen}}, \bibinfo {author} {\bibfnamefont {J.}~\bibnamefont {Fink}}, \bibinfo {author} {\bibfnamefont {A.}~\bibnamefont {Abdumalikov~Jr}}, \bibinfo {author} {\bibfnamefont {M.}~\bibnamefont {Baur}}, \bibinfo {author} {\bibfnamefont {S.}~\bibnamefont {Filipp}}, \bibinfo {author} {\bibfnamefont {M.~P.}\ \bibnamefont {Da~Silva}}, \bibinfo {author} {\bibfnamefont {A.}~\bibnamefont {Blais}}, \emph {et~al.},\ }\bibfield  {title} {\bibinfo {title} {Observation of resonant photon blockade at microwave frequencies using correlation function measurements},\ }\href@noop {} {\bibfield  {journal} {\bibinfo  {journal} {Physical review letters}\ }\textbf {\bibinfo {volume} {106}},\ \bibinfo {pages} {243601} (\bibinfo {year} {2011})}\BibitemShut {NoStop}%
\bibitem [{\citenamefont {Itano}\ \emph {et~al.}(1990)\citenamefont {Itano}, \citenamefont {Heinzen}, \citenamefont {Bollinger},\ and\ \citenamefont {Wineland}}]{itano1990quantum}%
  \BibitemOpen
  \bibfield  {author} {\bibinfo {author} {\bibfnamefont {W.~M.}\ \bibnamefont {Itano}}, \bibinfo {author} {\bibfnamefont {D.~J.}\ \bibnamefont {Heinzen}}, \bibinfo {author} {\bibfnamefont {J.~J.}\ \bibnamefont {Bollinger}},\ and\ \bibinfo {author} {\bibfnamefont {D.~J.}\ \bibnamefont {Wineland}},\ }\bibfield  {title} {\bibinfo {title} {Quantum zeno effect},\ }\href@noop {} {\bibfield  {journal} {\bibinfo  {journal} {Physical Review A}\ }\textbf {\bibinfo {volume} {41}},\ \bibinfo {pages} {2295} (\bibinfo {year} {1990})}\BibitemShut {NoStop}%
\bibitem [{\citenamefont {De~Jong}\ \emph {et~al.}(1997)\citenamefont {De~Jong}, \citenamefont {Spreeuw},\ and\ \citenamefont {van~den Heuvell}}]{de1997quantum}%
  \BibitemOpen
  \bibfield  {author} {\bibinfo {author} {\bibfnamefont {F.}~\bibnamefont {De~Jong}}, \bibinfo {author} {\bibfnamefont {R.}~\bibnamefont {Spreeuw}},\ and\ \bibinfo {author} {\bibfnamefont {H.~v.~L.}\ \bibnamefont {van~den Heuvell}},\ }\bibfield  {title} {\bibinfo {title} {Quantum zeno effect and v-scheme lasing without inversion},\ }\href@noop {} {\bibfield  {journal} {\bibinfo  {journal} {Physical Review A}\ }\textbf {\bibinfo {volume} {55}},\ \bibinfo {pages} {3918} (\bibinfo {year} {1997})}\BibitemShut {NoStop}%
\bibitem [{\citenamefont {Ashhab}\ \emph {et~al.}(2009)\citenamefont {Ashhab}, \citenamefont {Johansson}, \citenamefont {Zagoskin},\ and\ \citenamefont {Nori}}]{ashhab2009single}%
  \BibitemOpen
  \bibfield  {author} {\bibinfo {author} {\bibfnamefont {S.}~\bibnamefont {Ashhab}}, \bibinfo {author} {\bibfnamefont {J.}~\bibnamefont {Johansson}}, \bibinfo {author} {\bibfnamefont {A.}~\bibnamefont {Zagoskin}},\ and\ \bibinfo {author} {\bibfnamefont {F.}~\bibnamefont {Nori}},\ }\bibfield  {title} {\bibinfo {title} {Single-artificial-atom lasing using a voltage-biased superconducting charge qubit},\ }\href@noop {} {\bibfield  {journal} {\bibinfo  {journal} {New Journal of Physics}\ }\textbf {\bibinfo {volume} {11}},\ \bibinfo {pages} {023030} (\bibinfo {year} {2009})}\BibitemShut {NoStop}%
\bibitem [{\citenamefont {Mukhanov}(2011)}]{mukhanov2011energy}%
  \BibitemOpen
  \bibfield  {author} {\bibinfo {author} {\bibfnamefont {O.~A.}\ \bibnamefont {Mukhanov}},\ }\bibfield  {title} {\bibinfo {title} {Energy-efficient single flux quantum technology},\ }\href@noop {} {\bibfield  {journal} {\bibinfo  {journal} {IEEE Transactions on Applied Superconductivity}\ }\textbf {\bibinfo {volume} {21}},\ \bibinfo {pages} {760} (\bibinfo {year} {2011})}\BibitemShut {NoStop}%
\bibitem [{\citenamefont {Loudon}(2000)}]{loudon2000quantum}%
  \BibitemOpen
  \bibfield  {author} {\bibinfo {author} {\bibfnamefont {R.}~\bibnamefont {Loudon}},\ }\href@noop {} {\emph {\bibinfo {title} {The quantum theory of light}}}\ (\bibinfo  {publisher} {OUP Oxford},\ \bibinfo {year} {2000})\BibitemShut {NoStop}%
\bibitem [{\citenamefont {Gu}\ \emph {et~al.}(2017)\citenamefont {Gu}, \citenamefont {Kockum}, \citenamefont {Miranowicz}, \citenamefont {Liu},\ and\ \citenamefont {Nori}}]{gu2017microwave}%
  \BibitemOpen
  \bibfield  {author} {\bibinfo {author} {\bibfnamefont {X.}~\bibnamefont {Gu}}, \bibinfo {author} {\bibfnamefont {A.~F.}\ \bibnamefont {Kockum}}, \bibinfo {author} {\bibfnamefont {A.}~\bibnamefont {Miranowicz}}, \bibinfo {author} {\bibfnamefont {Y.-x.}\ \bibnamefont {Liu}},\ and\ \bibinfo {author} {\bibfnamefont {F.}~\bibnamefont {Nori}},\ }\bibfield  {title} {\bibinfo {title} {Microwave photonics with superconducting quantum circuits},\ }\href@noop {} {\bibfield  {journal} {\bibinfo  {journal} {Physics Reports}\ }\textbf {\bibinfo {volume} {718}},\ \bibinfo {pages} {1} (\bibinfo {year} {2017})}\BibitemShut {NoStop}%
\bibitem [{\citenamefont {Johansson}\ \emph {et~al.}(2012)\citenamefont {Johansson}, \citenamefont {Nation},\ and\ \citenamefont {Nori}}]{johansson2012qutip}%
  \BibitemOpen
  \bibfield  {author} {\bibinfo {author} {\bibfnamefont {J.~R.}\ \bibnamefont {Johansson}}, \bibinfo {author} {\bibfnamefont {P.~D.}\ \bibnamefont {Nation}},\ and\ \bibinfo {author} {\bibfnamefont {F.}~\bibnamefont {Nori}},\ }\bibfield  {title} {\bibinfo {title} {Qutip: An open-source python framework for the dynamics of open quantum systems},\ }\href@noop {} {\bibfield  {journal} {\bibinfo  {journal} {Computer Physics Communications}\ }\textbf {\bibinfo {volume} {183}},\ \bibinfo {pages} {1760} (\bibinfo {year} {2012})}\BibitemShut {NoStop}%
\bibitem [{\citenamefont {Krantz}\ \emph {et~al.}(2019)\citenamefont {Krantz}, \citenamefont {Kjaergaard}, \citenamefont {Yan}, \citenamefont {Orlando}, \citenamefont {Gustavsson},\ and\ \citenamefont {Oliver}}]{krantz2019quantum}%
  \BibitemOpen
  \bibfield  {author} {\bibinfo {author} {\bibfnamefont {P.}~\bibnamefont {Krantz}}, \bibinfo {author} {\bibfnamefont {M.}~\bibnamefont {Kjaergaard}}, \bibinfo {author} {\bibfnamefont {F.}~\bibnamefont {Yan}}, \bibinfo {author} {\bibfnamefont {T.~P.}\ \bibnamefont {Orlando}}, \bibinfo {author} {\bibfnamefont {S.}~\bibnamefont {Gustavsson}},\ and\ \bibinfo {author} {\bibfnamefont {W.~D.}\ \bibnamefont {Oliver}},\ }\bibfield  {title} {\bibinfo {title} {A quantum engineer's guide to superconducting qubits},\ }\href@noop {} {\bibfield  {journal} {\bibinfo  {journal} {Applied physics reviews}\ }\textbf {\bibinfo {volume} {6}} (\bibinfo {year} {2019})}\BibitemShut {NoStop}%
\bibitem [{\citenamefont {Ahmad}\ \emph {et~al.}(2023)\citenamefont {Ahmad}, \citenamefont {Brosco}, \citenamefont {Miano}, \citenamefont {Di~Palma}, \citenamefont {Arzeo}, \citenamefont {Satariano}, \citenamefont {Ferraiuolo}, \citenamefont {Lucignano}, \citenamefont {Vettoliere}, \citenamefont {Granata} \emph {et~al.}}]{ahmad2023competition}%
  \BibitemOpen
  \bibfield  {author} {\bibinfo {author} {\bibfnamefont {H.}~\bibnamefont {Ahmad}}, \bibinfo {author} {\bibfnamefont {V.}~\bibnamefont {Brosco}}, \bibinfo {author} {\bibfnamefont {A.}~\bibnamefont {Miano}}, \bibinfo {author} {\bibfnamefont {L.}~\bibnamefont {Di~Palma}}, \bibinfo {author} {\bibfnamefont {M.}~\bibnamefont {Arzeo}}, \bibinfo {author} {\bibfnamefont {R.}~\bibnamefont {Satariano}}, \bibinfo {author} {\bibfnamefont {R.}~\bibnamefont {Ferraiuolo}}, \bibinfo {author} {\bibfnamefont {P.}~\bibnamefont {Lucignano}}, \bibinfo {author} {\bibfnamefont {A.}~\bibnamefont {Vettoliere}}, \bibinfo {author} {\bibfnamefont {C.}~\bibnamefont {Granata}}, \emph {et~al.},\ }\bibfield  {title} {\bibinfo {title} {Competition of quasiparticles and magnetization noise in hybrid ferromagnetic transmon qubits},\ }\href@noop {} {\bibfield  {journal} {\bibinfo  {journal} {IEEE Transactions on Applied Superconductivity}\ }\textbf {\bibinfo {volume} {33}},\ \bibinfo {pages} {1} (\bibinfo {year} {2023})}\BibitemShut {NoStop}%
\bibitem [{\citenamefont {Koch}\ \emph {et~al.}(2007)\citenamefont {Koch}, \citenamefont {Yu}, \citenamefont {Gambetta},\ and\ \citenamefont {Blais}}]{koch2007charge}%
  \BibitemOpen
  \bibfield  {author} {\bibinfo {author} {\bibfnamefont {J.}~\bibnamefont {Koch}}, \bibinfo {author} {\bibfnamefont {T.~M.}\ \bibnamefont {Yu}}, \bibinfo {author} {\bibfnamefont {J.}~\bibnamefont {Gambetta}},\ and\ \bibinfo {author} {\bibfnamefont {A.}~\bibnamefont {Blais}},\ }\bibfield  {title} {\bibinfo {title} {Charge-insensitive qubit design derived from the cooper pair box},\ }\href@noop {} {\bibfield  {journal} {\bibinfo  {journal} {Physical Review A—Atomic, Molecular, and Optical Physics}\ }\textbf {\bibinfo {volume} {76}},\ \bibinfo {pages} {042319} (\bibinfo {year} {2007})}\BibitemShut {NoStop}%
\bibitem [{\citenamefont {Kirchmair}\ \emph {et~al.}(2013)\citenamefont {Kirchmair}, \citenamefont {Vlastakis}, \citenamefont {Leghtas}, \citenamefont {Nigg}, \citenamefont {Paik}, \citenamefont {Ginossar}, \citenamefont {Mirrahimi}, \citenamefont {Frunzio}, \citenamefont {Girvin},\ and\ \citenamefont {Schoelkopf}}]{kirchmair2013observation}%
  \BibitemOpen
  \bibfield  {author} {\bibinfo {author} {\bibfnamefont {G.}~\bibnamefont {Kirchmair}}, \bibinfo {author} {\bibfnamefont {B.}~\bibnamefont {Vlastakis}}, \bibinfo {author} {\bibfnamefont {Z.}~\bibnamefont {Leghtas}}, \bibinfo {author} {\bibfnamefont {S.~E.}\ \bibnamefont {Nigg}}, \bibinfo {author} {\bibfnamefont {H.}~\bibnamefont {Paik}}, \bibinfo {author} {\bibfnamefont {E.}~\bibnamefont {Ginossar}}, \bibinfo {author} {\bibfnamefont {M.}~\bibnamefont {Mirrahimi}}, \bibinfo {author} {\bibfnamefont {L.}~\bibnamefont {Frunzio}}, \bibinfo {author} {\bibfnamefont {S.~M.}\ \bibnamefont {Girvin}},\ and\ \bibinfo {author} {\bibfnamefont {R.~J.}\ \bibnamefont {Schoelkopf}},\ }\bibfield  {title} {\bibinfo {title} {Observation of quantum state collapse and revival due to the single-photon kerr effect},\ }\href@noop {} {\bibfield  {journal} {\bibinfo  {journal} {Nature}\ }\textbf {\bibinfo {volume} {495}},\ \bibinfo {pages} {205} (\bibinfo {year} {2013})}\BibitemShut {NoStop}%
\bibitem [{\citenamefont {He}\ \emph {et~al.}(2023)\citenamefont {He}, \citenamefont {Lu}, \citenamefont {Bao}, \citenamefont {Xue}, \citenamefont {Jiang}, \citenamefont {Wang}, \citenamefont {Roudsari}, \citenamefont {Delsing}, \citenamefont {Tsai},\ and\ \citenamefont {Lin}}]{he2023fast}%
  \BibitemOpen
  \bibfield  {author} {\bibinfo {author} {\bibfnamefont {X.}~\bibnamefont {He}}, \bibinfo {author} {\bibfnamefont {Y.}~\bibnamefont {Lu}}, \bibinfo {author} {\bibfnamefont {D.}~\bibnamefont {Bao}}, \bibinfo {author} {\bibfnamefont {H.}~\bibnamefont {Xue}}, \bibinfo {author} {\bibfnamefont {W.}~\bibnamefont {Jiang}}, \bibinfo {author} {\bibfnamefont {Z.}~\bibnamefont {Wang}}, \bibinfo {author} {\bibfnamefont {A.}~\bibnamefont {Roudsari}}, \bibinfo {author} {\bibfnamefont {P.}~\bibnamefont {Delsing}}, \bibinfo {author} {\bibfnamefont {J.}~\bibnamefont {Tsai}},\ and\ \bibinfo {author} {\bibfnamefont {Z.}~\bibnamefont {Lin}},\ }\bibfield  {title} {\bibinfo {title} {Fast generation of schr{\"o}dinger cat states using a kerr-tunable superconducting resonator},\ }\href@noop {} {\bibfield  {journal} {\bibinfo  {journal} {Nature communications}\ }\textbf {\bibinfo {volume} {14}},\ \bibinfo {pages} {6358} (\bibinfo {year} {2023})}\BibitemShut {NoStop}%
\bibitem [{\citenamefont {Yoshihara}\ \emph {et~al.}(2006)\citenamefont {Yoshihara}, \citenamefont {Harrabi}, \citenamefont {Niskanen}, \citenamefont {Nakamura},\ and\ \citenamefont {Tsai}}]{yoshihara2006decoherence}%
  \BibitemOpen
  \bibfield  {author} {\bibinfo {author} {\bibfnamefont {F.}~\bibnamefont {Yoshihara}}, \bibinfo {author} {\bibfnamefont {K.}~\bibnamefont {Harrabi}}, \bibinfo {author} {\bibfnamefont {A.}~\bibnamefont {Niskanen}}, \bibinfo {author} {\bibfnamefont {Y.}~\bibnamefont {Nakamura}},\ and\ \bibinfo {author} {\bibfnamefont {J.~S.}\ \bibnamefont {Tsai}},\ }\bibfield  {title} {\bibinfo {title} {Decoherence of flux qubits due to 1/f flux noise},\ }\href@noop {} {\bibfield  {journal} {\bibinfo  {journal} {Physical review letters}\ }\textbf {\bibinfo {volume} {97}},\ \bibinfo {pages} {167001} (\bibinfo {year} {2006})}\BibitemShut {NoStop}%
\bibitem [{\citenamefont {Ofek}\ \emph {et~al.}(2016)\citenamefont {Ofek}, \citenamefont {Petrenko}, \citenamefont {Heeres}, \citenamefont {Reinhold}, \citenamefont {Leghtas}, \citenamefont {Vlastakis}, \citenamefont {Liu}, \citenamefont {Frunzio}, \citenamefont {Girvin}, \citenamefont {Jiang} \emph {et~al.}}]{ofek2016extending}%
  \BibitemOpen
  \bibfield  {author} {\bibinfo {author} {\bibfnamefont {N.}~\bibnamefont {Ofek}}, \bibinfo {author} {\bibfnamefont {A.}~\bibnamefont {Petrenko}}, \bibinfo {author} {\bibfnamefont {R.}~\bibnamefont {Heeres}}, \bibinfo {author} {\bibfnamefont {P.}~\bibnamefont {Reinhold}}, \bibinfo {author} {\bibfnamefont {Z.}~\bibnamefont {Leghtas}}, \bibinfo {author} {\bibfnamefont {B.}~\bibnamefont {Vlastakis}}, \bibinfo {author} {\bibfnamefont {Y.}~\bibnamefont {Liu}}, \bibinfo {author} {\bibfnamefont {L.}~\bibnamefont {Frunzio}}, \bibinfo {author} {\bibfnamefont {S.}~\bibnamefont {Girvin}}, \bibinfo {author} {\bibfnamefont {L.}~\bibnamefont {Jiang}}, \emph {et~al.},\ }\bibfield  {title} {\bibinfo {title} {Extending the lifetime of a quantum bit with error correction in superconducting circuits},\ }\href@noop {} {\bibfield  {journal} {\bibinfo  {journal} {Nature}\ }\textbf {\bibinfo {volume} {536}},\ \bibinfo {pages} {441} (\bibinfo {year} {2016})}\BibitemShut {NoStop}%
\bibitem [{\citenamefont {Sivak}\ \emph {et~al.}(2023)\citenamefont {Sivak}, \citenamefont {Eickbusch}, \citenamefont {Royer}, \citenamefont {Singh}, \citenamefont {Tsioutsios}, \citenamefont {Ganjam}, \citenamefont {Miano}, \citenamefont {Brock}, \citenamefont {Ding}, \citenamefont {Frunzio} \emph {et~al.}}]{sivak2023real}%
  \BibitemOpen
  \bibfield  {author} {\bibinfo {author} {\bibfnamefont {V.}~\bibnamefont {Sivak}}, \bibinfo {author} {\bibfnamefont {A.}~\bibnamefont {Eickbusch}}, \bibinfo {author} {\bibfnamefont {B.}~\bibnamefont {Royer}}, \bibinfo {author} {\bibfnamefont {S.}~\bibnamefont {Singh}}, \bibinfo {author} {\bibfnamefont {I.}~\bibnamefont {Tsioutsios}}, \bibinfo {author} {\bibfnamefont {S.}~\bibnamefont {Ganjam}}, \bibinfo {author} {\bibfnamefont {A.}~\bibnamefont {Miano}}, \bibinfo {author} {\bibfnamefont {B.}~\bibnamefont {Brock}}, \bibinfo {author} {\bibfnamefont {A.}~\bibnamefont {Ding}}, \bibinfo {author} {\bibfnamefont {L.}~\bibnamefont {Frunzio}}, \emph {et~al.},\ }\bibfield  {title} {\bibinfo {title} {Real-time quantum error correction beyond break-even},\ }\href@noop {} {\bibfield  {journal} {\bibinfo  {journal} {Nature}\ }\textbf {\bibinfo {volume} {616}},\ \bibinfo {pages} {50} (\bibinfo {year} {2023})}\BibitemShut {NoStop}%
\bibitem [{\citenamefont {Gu}\ \emph {et~al.}(2009)\citenamefont {Gu}, \citenamefont {Weedbrook}, \citenamefont {Menicucci}, \citenamefont {Ralph},\ and\ \citenamefont {van Loock}}]{gu2009quantum}%
  \BibitemOpen
  \bibfield  {author} {\bibinfo {author} {\bibfnamefont {M.}~\bibnamefont {Gu}}, \bibinfo {author} {\bibfnamefont {C.}~\bibnamefont {Weedbrook}}, \bibinfo {author} {\bibfnamefont {N.~C.}\ \bibnamefont {Menicucci}}, \bibinfo {author} {\bibfnamefont {T.~C.}\ \bibnamefont {Ralph}},\ and\ \bibinfo {author} {\bibfnamefont {P.}~\bibnamefont {van Loock}},\ }\bibfield  {title} {\bibinfo {title} {Quantum computing with continuous-variable clusters},\ }\href@noop {} {\bibfield  {journal} {\bibinfo  {journal} {Physical Review A}\ }\textbf {\bibinfo {volume} {79}},\ \bibinfo {pages} {062318} (\bibinfo {year} {2009})}\BibitemShut {NoStop}%
\bibitem [{\citenamefont {Naik}\ \emph {et~al.}(2017)\citenamefont {Naik}, \citenamefont {Leung}, \citenamefont {Chakram}, \citenamefont {Groszkowski}, \citenamefont {Lu}, \citenamefont {Earnest}, \citenamefont {McKay}, \citenamefont {Koch},\ and\ \citenamefont {Schuster}}]{naik2017random}%
  \BibitemOpen
  \bibfield  {author} {\bibinfo {author} {\bibfnamefont {R.}~\bibnamefont {Naik}}, \bibinfo {author} {\bibfnamefont {N.}~\bibnamefont {Leung}}, \bibinfo {author} {\bibfnamefont {S.}~\bibnamefont {Chakram}}, \bibinfo {author} {\bibfnamefont {P.}~\bibnamefont {Groszkowski}}, \bibinfo {author} {\bibfnamefont {Y.}~\bibnamefont {Lu}}, \bibinfo {author} {\bibfnamefont {N.}~\bibnamefont {Earnest}}, \bibinfo {author} {\bibfnamefont {D.}~\bibnamefont {McKay}}, \bibinfo {author} {\bibfnamefont {J.}~\bibnamefont {Koch}},\ and\ \bibinfo {author} {\bibfnamefont {D.~I.}\ \bibnamefont {Schuster}},\ }\bibfield  {title} {\bibinfo {title} {Random access quantum information processors using multimode circuit quantum electrodynamics},\ }\href@noop {} {\bibfield  {journal} {\bibinfo  {journal} {Nature communications}\ }\textbf {\bibinfo {volume} {8}},\ \bibinfo {pages} {1904} (\bibinfo {year} {2017})}\BibitemShut {NoStop}%
\bibitem [{\citenamefont {Wang}\ \emph {et~al.}(2019)\citenamefont {Wang}, \citenamefont {Wu}, \citenamefont {Ma}, \citenamefont {Cai}, \citenamefont {Hu}, \citenamefont {Mu}, \citenamefont {Xu}, \citenamefont {Chen}, \citenamefont {Wang}, \citenamefont {Song} \emph {et~al.}}]{wang2019heisenberg}%
  \BibitemOpen
  \bibfield  {author} {\bibinfo {author} {\bibfnamefont {W.}~\bibnamefont {Wang}}, \bibinfo {author} {\bibfnamefont {Y.}~\bibnamefont {Wu}}, \bibinfo {author} {\bibfnamefont {Y.}~\bibnamefont {Ma}}, \bibinfo {author} {\bibfnamefont {W.}~\bibnamefont {Cai}}, \bibinfo {author} {\bibfnamefont {L.}~\bibnamefont {Hu}}, \bibinfo {author} {\bibfnamefont {X.}~\bibnamefont {Mu}}, \bibinfo {author} {\bibfnamefont {Y.}~\bibnamefont {Xu}}, \bibinfo {author} {\bibfnamefont {Z.-J.}\ \bibnamefont {Chen}}, \bibinfo {author} {\bibfnamefont {H.}~\bibnamefont {Wang}}, \bibinfo {author} {\bibfnamefont {Y.}~\bibnamefont {Song}}, \emph {et~al.},\ }\bibfield  {title} {\bibinfo {title} {Heisenberg-limited single-mode quantum metrology in a superconducting circuit},\ }\href@noop {} {\bibfield  {journal} {\bibinfo  {journal} {Nature communications}\ }\textbf {\bibinfo {volume} {10}},\ \bibinfo {pages} {4382} (\bibinfo {year} {2019})}\BibitemShut {NoStop}%
\end{thebibliography}%
\end{document}